\documentclass[iop]{emulateapj}

\newcommand{\gsim}{\;\lower.6ex\hbox{$\sim$}\kern-7.75pt\raise.65ex\hbox{$>$}\;}
\newcommand{\lsim}{\;\lower.6ex\hbox{$\sim$}\kern-7.75pt\raise.65ex\hbox{$<$}\;}

\newcommand{\MSUN}{$M_{\odot}$}
\newcommand{\Myr}{M$_{\odot}$~yr$^{-1}$}
\newcommand{\Myk}{M$_{\odot}$~yr$^{-1}$~kpc$^{-2}$}

\slugcomment{to appear ...}

\shorttitle{}
\shortauthors{}

\begin{document}


\title{PNe and H II regions in the starburst irregular galaxy NGC~4449 from LBT MODS data}


\author{F. Annibali \altaffilmark{1}, 
 M. Tosi\altaffilmark{1},  D. Romano\altaffilmark{1}, A. Buzzoni\altaffilmark{1},
  F. Cusano\altaffilmark{1}, M. Fumana\altaffilmark{2}, A. Marchetti\altaffilmark{2},    
 M. Mignoli\altaffilmark{1}, A. Pasquali\altaffilmark{3}, A. Aloisi\altaffilmark{4}}

\altaffiltext{1}{INAF-Osservatorio Astronomico di Bologna, 
Via Gobetti 93/3, I-40129 Bologna, Italy; francesca.annibali@oabo.inaf.it}

\altaffiltext{2}{INAF-Istituto di Astrofisica Spaziale e Fisica Cosmica,
Via Bassini 15, I-20133 Milano, Italy}

\altaffiltext{3}{Astronomisches Rechen-Institut, Zentrum fuer Astronomie der Universitaet Heidelberg, Moenchhofstr. 12 - 14,
            69120 Heidelberg, Germany}

\altaffiltext{4}{Space Telescope Science Institute, 3700 San Martin Drive, 
Baltimore, MD 21218, USA}



\begin{abstract}

We present deep 3500$-$10000 \AA \ spectra  of H~II regions and planetary nebulae (PNe) in the starburst irregular galaxy NGC~4449, acquired with the Multi Object Double Spectrograph at the Large Binocular Telescope.  Using the ``direct'' method, we derived the abundance of He, N, O, Ne, Ar, and S in six H~II regions and in four PNe in NGC~4449. This is the first case of PNe studied in a starburst irregular outside the Local Group. 
Our H~II region and PN sample extends over a galacto-centric distance range of $\approx$2 kpc and spans $\approx$0.2 dex in oxygen abundance, with average values of  
 $12+\log(O/H)=8.37 \pm 0.05$ and   $8.3 \pm 0.1$ for H~II regions and PNe, respectively. 
PNe and H~II regions exhibit similar oxygen abundances in the galacto-centric distance range of overlap, while PNe appear more than $\sim$1 dex enhanced in nitrogen with
 respect to H~II regions. The latter result is the natural consequence of N  being mostly synthesized in intermediate-mass stars and brought to the stellar surface during dredge-up episodes. On the other hand, the similarity in O abundance between H~II regions and PNe suggests that NGC~4449' s interstellar medium has been poorly enriched in $\alpha$-elements since the progenitors of the PNe were formed.
Finally, our data reveal the presence of a negative oxygen gradient for both H~II regions and PNe, whilst nitrogen does not exhibit any significant radial trend.  We ascribe 
the (unexpected) nitrogen behaviour  as due to  local N enrichment by the conspicuous Wolf-Rayet population in NGC~4449.

\end{abstract}



\section{Introduction}

Stellar and galaxy evolution are closely coupled: on the one hand, subsequent generations of high- and intermediate-mass stars continuously modify the energy balance and chemical composition of the interstellar medium (ISM) of their host galaxy; on the other hand gas accretion by diffuse or filamentary cold streams  \citep{dekel09} or by gas-rich dwarfs may trigger new star formation and dilute the metallicity of the ISM. It is thus mandatory to combine accurate star formation histories (SFHs) from resolved stellar studies and chemical 
abundance studies in individual systems as key ingredients to reconstruct a coherent and complete picture of how galaxies formed and evolved.

Since the advent of the Hubble Space Telescope (HST), much effort has been done to resolve the individual stars and to derive the SFHs in large galaxy samples within the Local Universe  
\citep[e.g.,][see also  \cite{tht09} for a review]{dalcanton09,mcquinn10,monelli10a,monelli10b,weisz11,dalcanton12,weisz14,legus}. At the same time, the availability of multi-object spectroscopy on 8-10 m telescopes 
has promoted chemical composition studies of  H~II regions and planetary nebulae (PNe) over large galaxy areas 
\citep[e.g.,][]{bresolin05,magrini05,pena07,bresolin09b,magrini09,stasi13,anni15,berg15}. 
In particular, the simultaneous study of chemical abundances in PNe and in H~II regions can provide more stringent constraints on chemical evolution models, since H~II regions probe the present-day composition of the ISM, while PNe, being the  end-product of the evolution of stars with masses 0.8\MSUN $<$ M $<$ 8 \MSUN, offer a view of the ISM composition back to several Gyrs ago.

PNe enrich the ISM mainly in He, C and N, 
while leaving untouched elements such as Ne, S and Ar whose abundance remains the initial one of the PN progenitor. O is also usually thought to be unaffected by
the progenitor reaction processes, although it has sometimes been suggested to be enhanced \citep{marigo01} in metal-poor PNe. 
The important production of He, C and N by PNe is due to dredge-up episodes occurring during the red giant branch (RGB) and asymptotic giant branch (AGB) phases of intermediate- and low-mass stars,  and to hot-bottom burning \citep[HBB, e.g.][]{renzini81} in the most massive AGB stars ($\gsim4 M_{\odot}$, depending on metallicity), that change 
the stellar surface abundances of these elements.

Here we exploited the high performance of the Multi Object Double Spectrograph (MODS) mounted on the Large Binocular Telescope (LBT) to perform the first combined study 
of H~II regions and PNe in the irregular galaxy NGC 4449 ($\alpha_{2000}=12^h~28^m~11^s.9$, $\delta_{2000}=+44^{\circ}~05'~40''$), at a distance of 3.82$\pm$0.27 Mpc from us \citep{anni08}. 
NGC~4449 is of particular interest because it is one of the strongest starburst in the local Universe \citep[star formation rate$\sim$1 \Myr, or $\sim$0.04 \Myk,][]{mcquinn10,anni11}, and moreover it exhibits 
several characteristics suggesting that it has accreted one or possibly several smaller companions: more specifically, i) it has a very extended H~I halo ($\approx$90 kpc in diameter), which is a factor of $\sim$10 larger than the optical diameter of the galaxy and rotates in the opposite direction to the gas in the center \citep{hunter98}; ii) it is one of the very few dwarf galaxies where a stellar tidal stream has been discovered so far \citep{delgado12,rich12}; iii) it hosts an old, $10^6$~\MSUN cluster associated with two tails of young stars, 
potentially the nucleus of an accreted  gas-rich satellite galaxy \citep{anni12}.
Because of these properties, NGC~4449 is a perfect laboratory to test the hypothesis that strong starbursts in dwarf galaxies are caused by 
accretion or merging events, as suggested by recent studies showing that disturbed H~I kinematics, H~I companions, and filamentary H~I structures are more common in starburst dwarfs than in typical star-forming irregulars \citep{lelli14}. 
NGC~4449 was targeted with the Advanced Camera for Survey (ACS) on board of HST a few years ago to resolve its stellar content and to derive its SFH \citep{anni08,mcquinn10,sacchi17}. These analyses indicate that NGC~4449 enhanced its SF activity $\approx$500 Myr ago, while the rate was much lower at earlier epochs; however, the  impossibility to reach, even with HST, the main sequence turnoffs at a distance of $\sim$4 Mpc implies that the SFH of NGC~4449 is very uncertain prior to $1-2$ Gyr ago  \citep[see e.g.][]{sacchi17}.

In this paper we present a study of the H~II region and PN chemical abundances in NGC~4449 with the purpose of providing 
a key complement to previous stellar population studies. Chemical evolution models based on the SFH and on the derived abundances  \citep[see e.g. the approach of][]{romano06}  will be presented in a forthcoming paper and  will provide new insights into the past evolution of NGC~4449. The paper is structured as follows: Section~2 describes the observations and data reduction, Section~3 informs on the procedure for the derivation of the reddening-corrected emission line fluxes, while temperatures, densities, and chemical abundances of H~II regions and PNe are derived in Section~4. Section~5 focuses on the study of the Wolf-Rayet spectral features. In Section~6 and 7 we analyse and discuss the derived abundances, element ratios and spatial abundance distributions, while in Section~8 we compare the properties of our PNe with those of  Local Group star forming dwarfs. Our conclusions are summarised in Section~9.

\section{Observations and data reduction \label{data_reduction}}

 \begin{figure*}
\epsscale{1}
\plotone{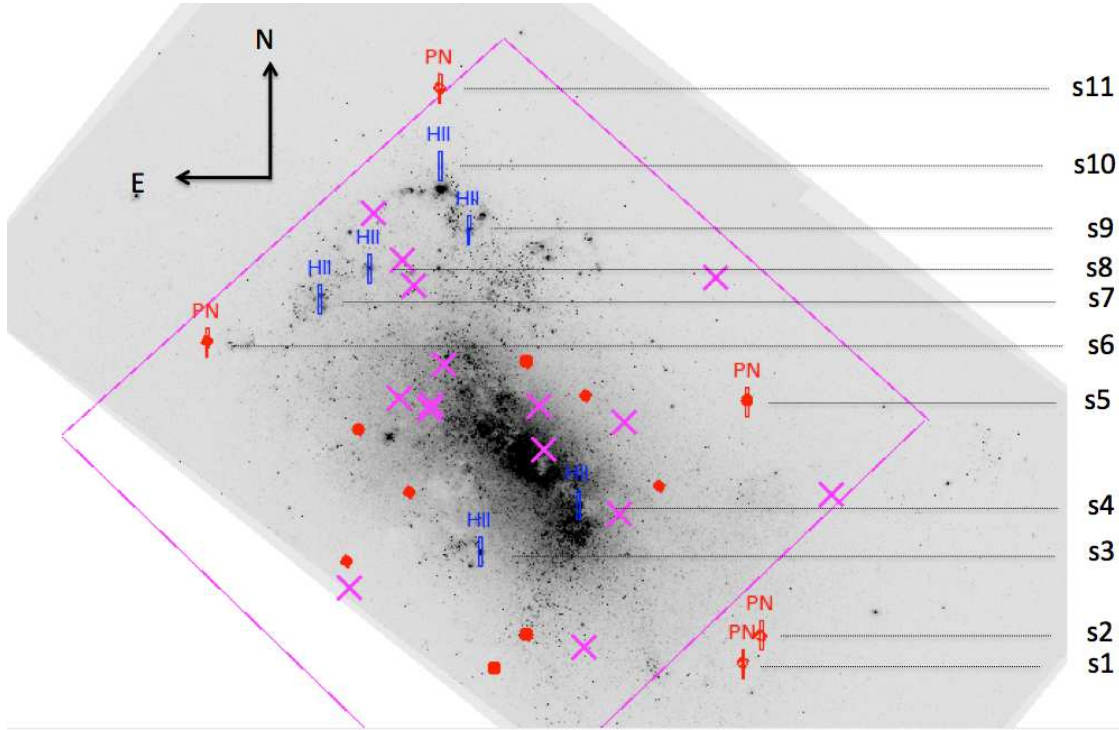}
\caption{HST/ACS image of NGC~4449 in F555W ($\sim$V) with superimposed our H~II region and PN sample. The smaller field of view covered by the
ACS F502N ($\sim$[O~III]) image is also indicated with the magenta line. The small red points and magenta crosses indicate the totality of the 28 PNe identified from the combined 
F435W ($\sim$B)-F502N - F814W (I) image, and then cross-checked in F658N (H$\alpha$). The small red points denote the 13 PNe also identified when using the 
combined B, V, I image. Superimposed on the image are the eleven LBT/MODS   1''$\times$10''  slits at the position of the 6 H~II regions and 5 PNe targeted for spectroscopy.
\label{image}}
\end{figure*}

PN candidates and H~II regions were identified on HST/ACS images in the F435W ($\sim$ B), F555W ($\sim$ V), F814W ($\sim$ I), and F658N (H$\alpha$) filters 
(GO program 10585; PI: Aloisi).  These data cover a field of view as large as $\sim400''\times 200$'' (two ACS pointings) and allow to identify both H~II regions and PNe up to large galacto-centric distances.
In the ACS images, H~II regions are resolved and appear as regions of diffuse H$\alpha$ and V (i.e. [OIII]$\lambda\lambda$4959,5007)  
emission. On the other hand, PN candidates were visually selected from a  B, V, I  color-combined image looking for point-like sources that stand  
out in V compared to B and I because of the [OIII]$\lambda\lambda$4959,5007 emission lines. The 29 selected candidates were then cross-identified on the shallower H$\alpha$ image to eliminate background emission-line sources.  This provided 13 PN candidates in NGC 4449, whose spatial location is shown in Figure~\ref{image}. 
We repeated the PN search using archival ACS images in the narrow-band F502N filter (GO program 10522, PI Calzetti) instead of the F555W image and then cross-checked in  H$\alpha$.  The F502N filter, centered around the [O~III]$\lambda$5007 line, allows for a better contrast of the PNe compared to stars; however,  the smaller $\sim200''\times 200$'' field of view of the available data (corresponding to just one ACS pointing) does not permit an inspection of the NGC~4449 outskirts where PNe can be more easily studied thanks to the lower galaxy background. All the PNe that were identified 
from the F555W image were also found when using the F502N image (however three PNe fall outside the  field of view of the F502N image);  15 additional PNe were identified when adopting the F502N image in place of the  F555W one (see Figure~\ref{image}), for a total sample of 28 PNe.

\begin{figure*}
\epsscale{1}
\plotone{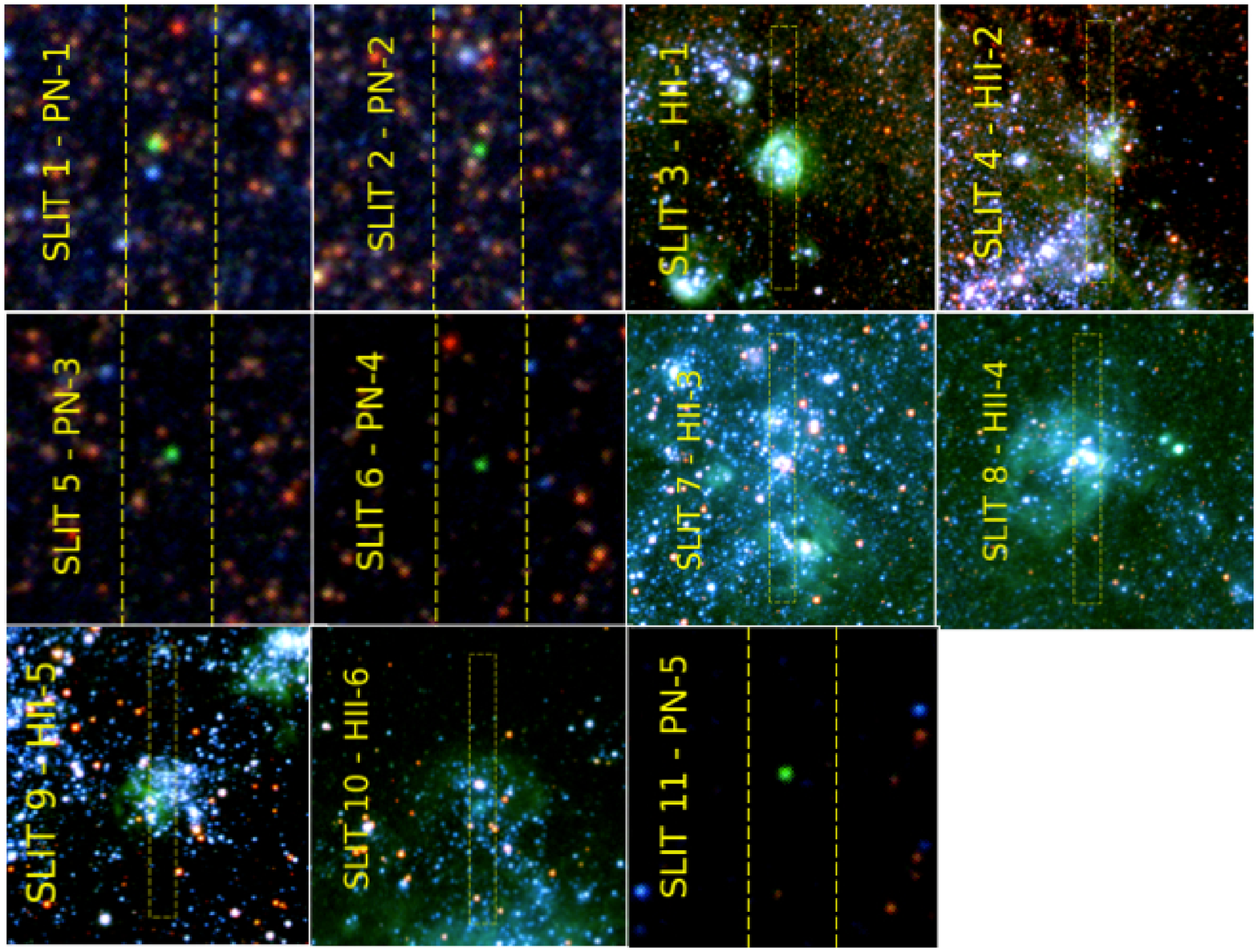}
\caption{HST/ACS color-combined images  (F435W$=$blue, F555W$=$green, F814W$=$red)  of PNe and H~II regions in NGC~4449 targeted for spectroscopy 
with LBT/MODS. The FoV shown is $\sim 3.5'' \times 3.5''$ for the PNe and  $12'' \times 12''$ for the H~II regions. 
\label{mosaic}}
\end{figure*}

PNe and H~II regions were targeted for spectroscopy with LBT/MODS from January 21, 2013 until April 5, 2013 within program 2012B\_23, RUN~A (PI Annibali). 
The 1''$\times$10'' slit mask is shown in Figure~\ref{image}. We were able to accomodate into the MODS slit mask 5 PNe  out of 28, chosen in the most external regions of NGC~4449 to minimize the contamination from the diffuse ionized gas. Six remaining slits were positioned on H~II regions. 
Figure~\ref{mosaic} shows color-composite HST images for the  PNe and H~II regions targeted with LBT, with superimposed the MODS slits. 
We observed our targets using the blue G400L (3200$-$5800 \AA) and the red G670L (5000$-$10000 \AA) gratings on the blue and red channels 
in dichroic mode for a total exposure time of $\sim$10.5 h, organized into 14 sub-exposures of $\sim$2700 s each. The seeing varied between $\sim$0.6'' and $\sim$1.3'', 
and the airmass from $\sim$1.0 to   $\sim$1.3. Typically, the exposures were acquired at hour angles between $\sim$ $-$1 h and $\sim$ $+$1 h to avoid 
significant effects from differential atmospheric refraction \citep[see e.g.][]{filippenko}.
Only  8 sub-exposures  with a seeing $\lesssim 1''$ were retained for our study, for a total integration time of $\sim$ 6 h. 
The journal of the observations is provided in Table~\ref{obs}.

\begin{deluxetable}{lccccc}
\tabletypesize{\scriptsize}
\tablecaption{Journal of LBT/MODS observations \label{obs}}
\tablehead{
\colhead{Exp.N.} & \colhead{Date-obs}  & \colhead{Exptime}   & \colhead{Seeing}    & \colhead{Airmass} & \colhead{Retained?} }  
\startdata
1 &  2013-01-21   & 2700 s  & 0.6'' & 1.0    & {\bf yes}  \\
2 &  2013-01-21   & 2700 s & 0.6'' & 1.0   & {\bf yes}  \\
3 &  2013-01-21   & 2700 s & 0.7'' & 1.0   & {\bf yes} \\
4 & 2013-04-01   & 2700  s & 1.0'' & 1.1   & {\bf yes}  \\
5 & 2013-04-01   & 2700 s  & 0.8'' & 1.1  & {\bf yes } \\
6 & 2013-04-01   & 2700 s  & 0.7'' & 1.0   & {\bf yes} \\
7 & 2013-04-01   & 2700 s  & 0.8'' & 1.0  & {\bf yes } \\
8 & 2013-04-01   & 2700 s  & 1.1'' & 1.0  &  no \\
9 & 2013-04-01    & 2700 s & 1.1'' & 1.1  & no \\
10 & 2013-04-01  & 2700 s & 1.1'' & 1.2  & no \\
11 & 2013-04-01  & 3040 s & 2.0'' & 1.3  & no \\
12 & 2013-04-02   & 2700 s & 0.9''  & 1.0 & {\bf yes }\\
13 & 2013-04-02  & 2700 s & 1.3''  & 1.1 & no \\
14 & 2013-04-05  & 2700 s & 1.3''  & 1.0 & no \\
\enddata
\end{deluxetable}

\begin{figure*}
\epsscale{1}
\plotone{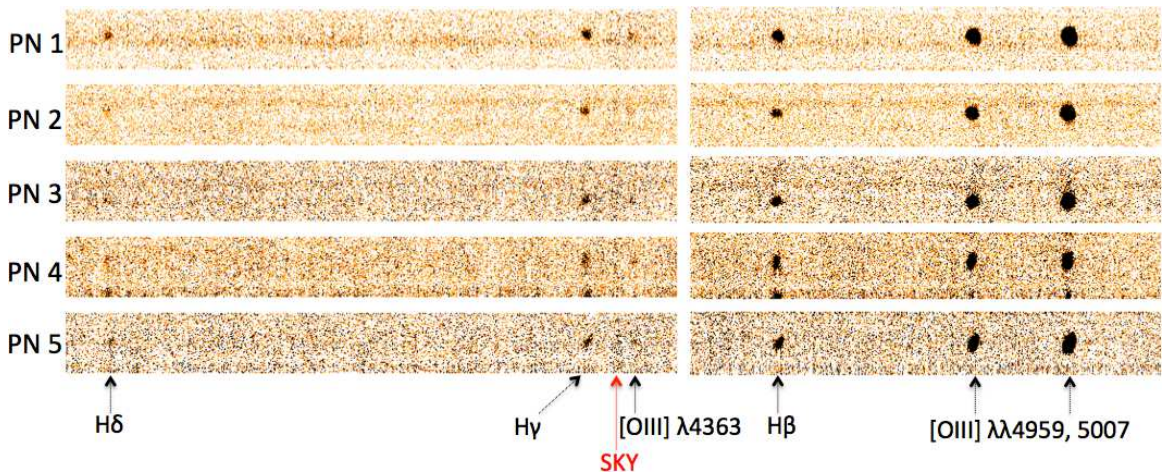}
\caption{Two-dimensional MODS spectra of PNe in NGC~4449 showing the H$\delta$, H$\gamma$, [O~III]$\lambda$4363, H$\beta$ and 
 [O~III]$\lambda\lambda$4959, 5007 lines. The residuals corresponding to the subtraction of the  Hg~I~$\lambda$4358  sky line,  between the  
 H$\gamma$ and [O~III]$\lambda$4363 lines, are visible in the 2D spectra. 
\label{pne_2d}}
\end{figure*}

Bias, flat-field, and wavelength calibrations were performed  with the Italian LBT Spectroscopic reduction Facility at INAF-IASF Milano, 
producing the calibrated two-dimensional (2D) spectra for the individual sub-exposures. Then, the individual sub-exposures were sky-subtracted and combined into 
final 2D blue and red frames. 
The sky subtraction was performed with the {\it{background}} task in IRAF\footnote{ IRAF is distributed by the National Optical Astronomy Observatory, which is operated by the Association of Universities for Research in Astronomy, Inc., under cooperative agreement with the National Science Foundation.}, typically choosing the windows at the two opposite sides of the central source. This procedure removed, together with the sky, also the contribution from the NGC~4449 unresolved background.  
As an example, we show in Figure~\ref{pne_2d} the two-dimensional sky-subtracted combined spectra for our PNe in selected spectral regions. 
The figure shows that we were able to detect the [O~III]$\lambda$4363 line in all PNe thanks both to the good MODS resolution and to the NGC~4449' s systemic 
velocity of $\approx$210 km s$^{-1}$, allowing for sufficient separation with the  Hg~I~$\lambda$4358 sky line. 
For region H~II-3, located in slit~7, it was not possible to evaluate the background, since the slit was entirely filled by gaseous emission. In this case, we adopted  the background derived for PN~5 (slit~11) as a sky template, and subtracted it to region H~II-3. This is a reasonable choice because PN~5 is located at fairly large galactocentric distance and is affected by negligible contribution from the NGC~4449 unresolved background. 
The  one-dimensional (1D) spectra were extracted from the 2D calibrated and sky-subtracted spectra by running the {\it apall} task in the {\it twodspec} IRAF package. 
To derive the effective spectral resolution, we used the combined 1D spectra with no sky subtraction, and measured the FWHM of the most prominent sky lines; 
this resulted into  FWHM$\approx$4.1 \AA \ (or R$\sim$1100 at 4500 \AA) for the blue channel, and  FWHM$\approx$5.8 \AA \ (or R$\sim$1400 at 8000 \AA) for the red channel.

The blue and red 1D spectra were flux calibrated using the sensitivity curves from the Italian LBT spectroscopic reduction pipeline; the curves  were derived using the spectrophotometric standard star Feige~56 observed  in dichroic mode with a 5''-width  slit on April 1, 2013 at an airmass of $\sim$1.4. 
To obtain the red and blue sensitivity curves, the observed standard was compared with reference spectra in the HST CALSPEC database. 
Atmospheric extinction corrections were applied using the average extinction curve available from the MODS calibration webpage at http://www.astronomy.ohio-state.edu/MODS/Calib/. This may introduce some uncertainty in flux calibration, in particular at the bluest 
wavelengths, where the effect of atmospheric extinction is more severe. By comparing the sensitivity curves from Feige~56 with those obtained from another standard, Feige~66,  observed on January 20, 2013 at an airmass of $\sim$1 and with the same setup of Feige~56 as part of 
our 2012B\_23, RUN~B program\footnote{During RUN B we targeted old unresolved stellar clusters in NGC~4449; the results will be presented in another paper (Annibali et al in preparation).},  we found  a $\sim15\%$ difference at wavelegnths below $\sim$ 4000 \AA, while the curves agree within  $\sim1\%$ at redder wavelengths. 
 
\begin{figure*}
\epsscale{1}
\plotone{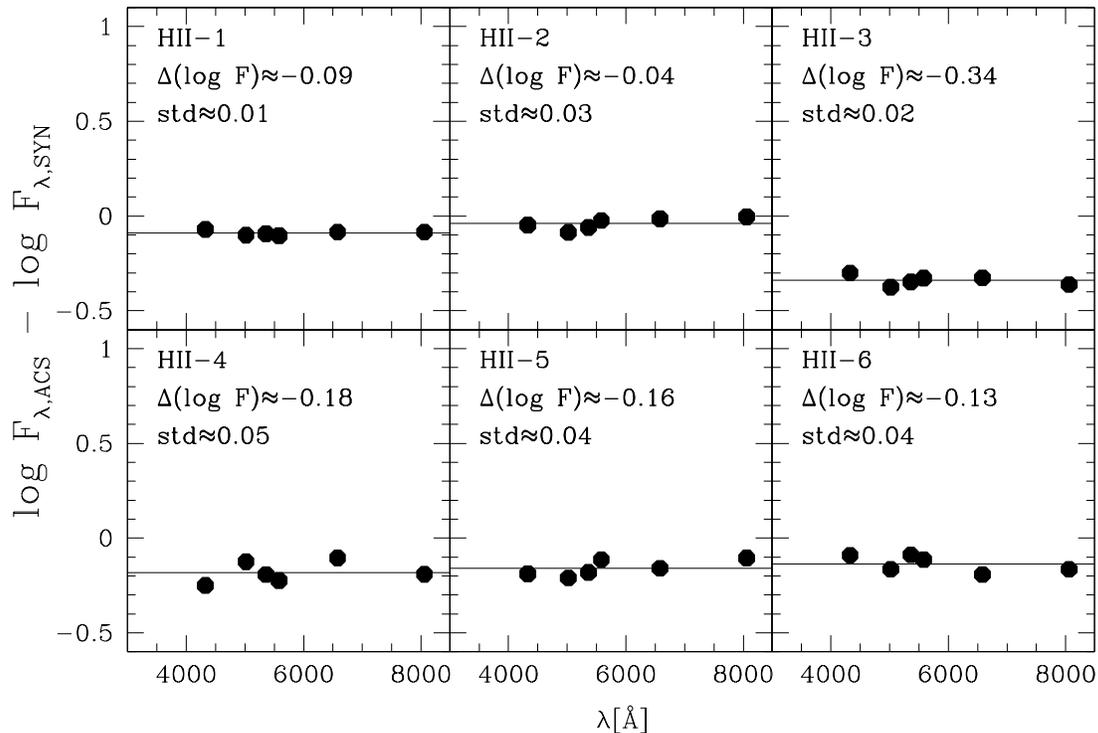}
\caption{Comparison between ACS and  {\it Synphot} fluxes for  H~II regions in NGC~4449 (see Section~2). From blue to red wavelengths, the 
dots correspond to the following ACS bandpasses: F435W, F502N, F555W, F550M, F658N and F814W. 
For each H~II region, the solid horizontal line is the average $log(F_{\lambda,ACS}) - log(F_{\lambda,SYN})$ offset. 
The standard deviation around this value is also indicated within each panel.  
\label{calibration}}
\end{figure*}

 \begin{figure*}
\epsscale{1}
\plotone{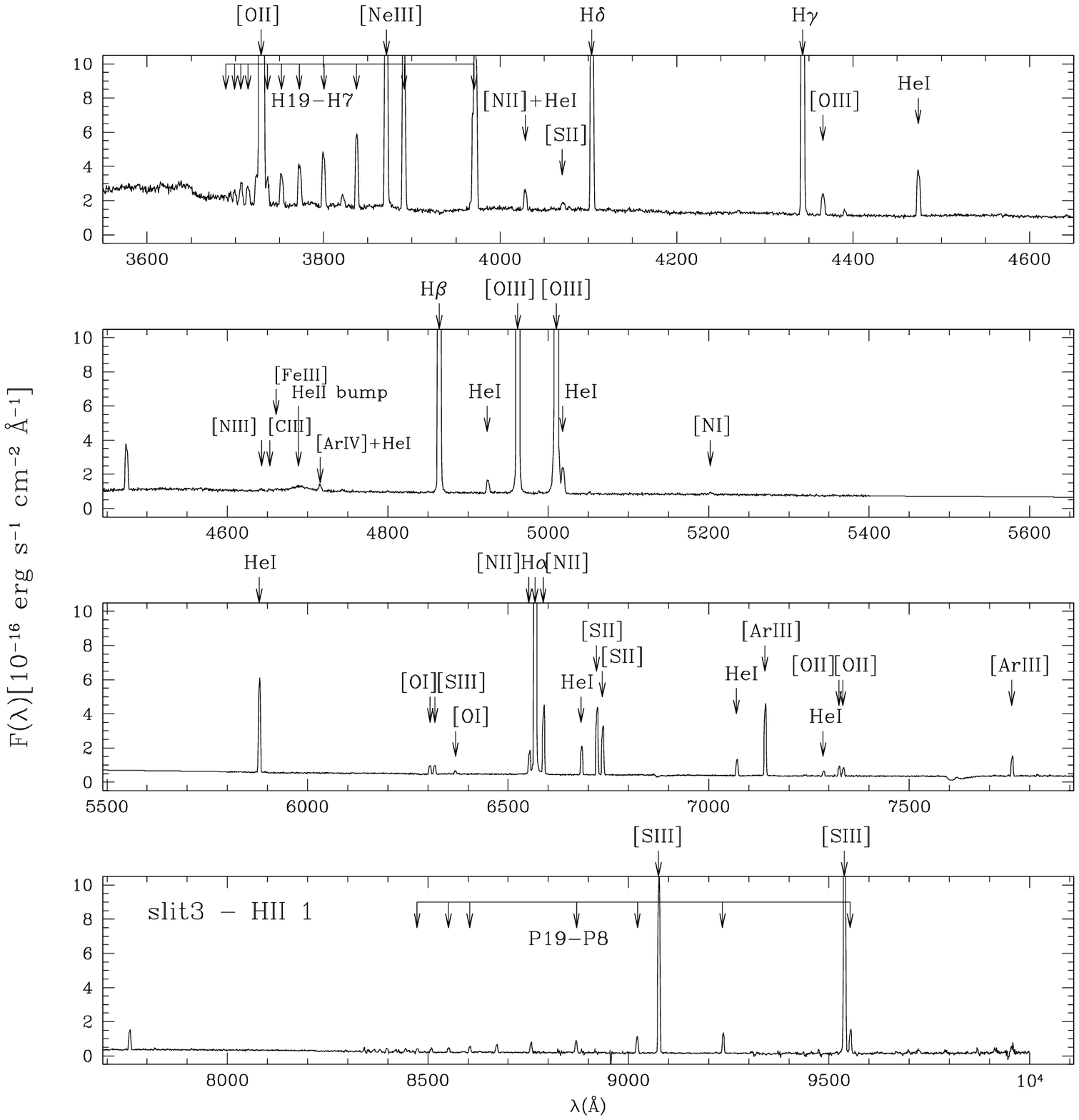}
\caption{ LBT/MODS spectra in the blue and red channels for H~II-1 in NGC~4449 with indicated all the identified emission lines. 
A linear interpolation was performed in the 5400 - 5800 \AA \ region where the sensitivities of the blue and red detectors drop. 
The spectra of the other H~II regions 
(H~II-2, H~II-3, H~II-4, H~II-5, H~II-6) are provided in the Appendix.
 \label{slit3}}
\end{figure*}

 \begin{figure*}
\epsscale{1}
\plotone{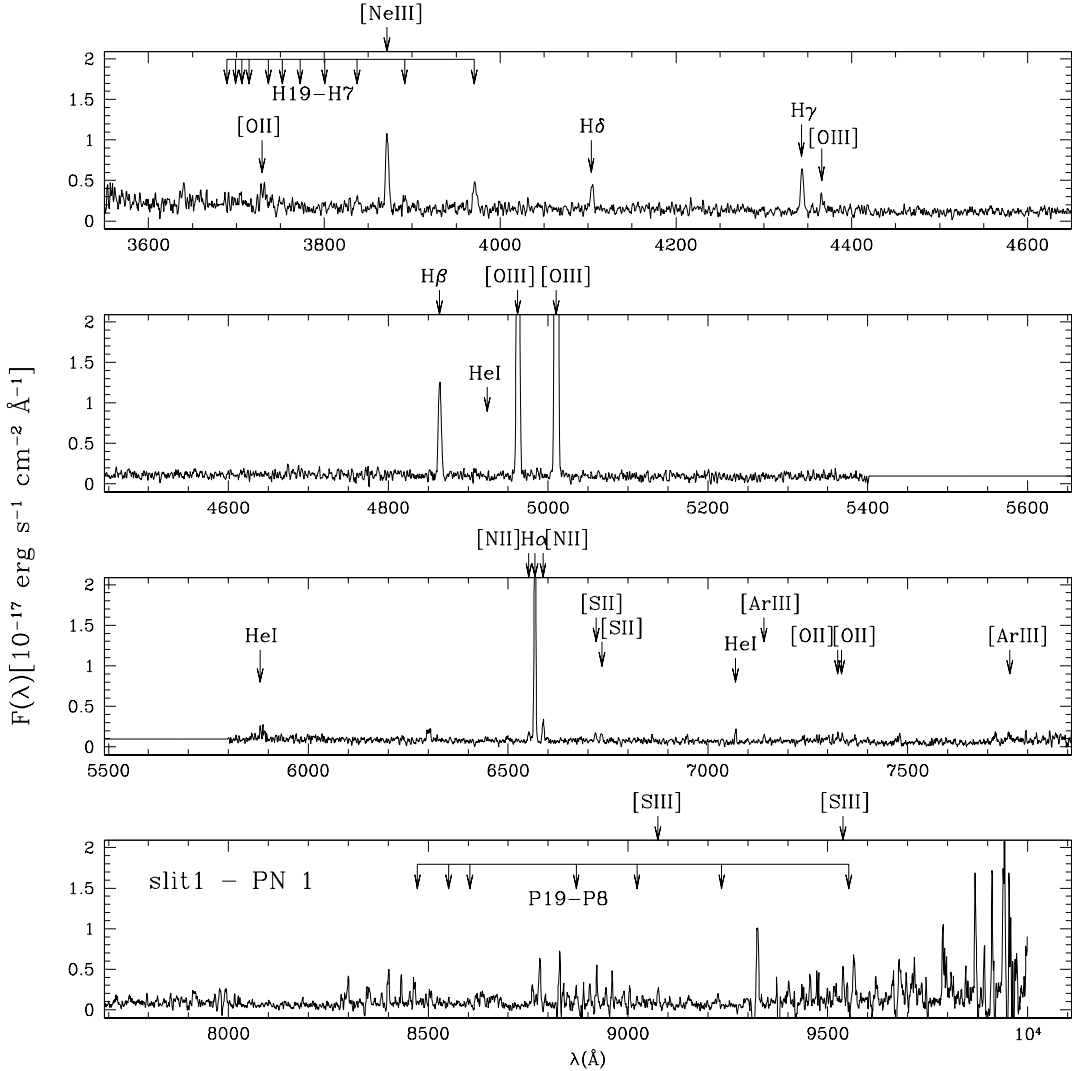}
\caption {LBT/MODS spectra in the blue and red channels for PN-1 in NGC~4449 with indicated all the identified emission lines. 
A linear interpolation was performed in the 5400 - 5800 \AA \ region where the sensitivities of the blue and red detectors drop. 
A  $\sim$1~\AA \ boxcar filter smoothing was applied to the spectrum to better highlight the low singal-to-noise features. 
The spectra of the other PNe (PN-2, PN-3, PN-4, PN-5) are provided in the Appendix.
 \label{slit1}}
\end{figure*}

 \begin{figure*}
\epsscale{1}
\plotone{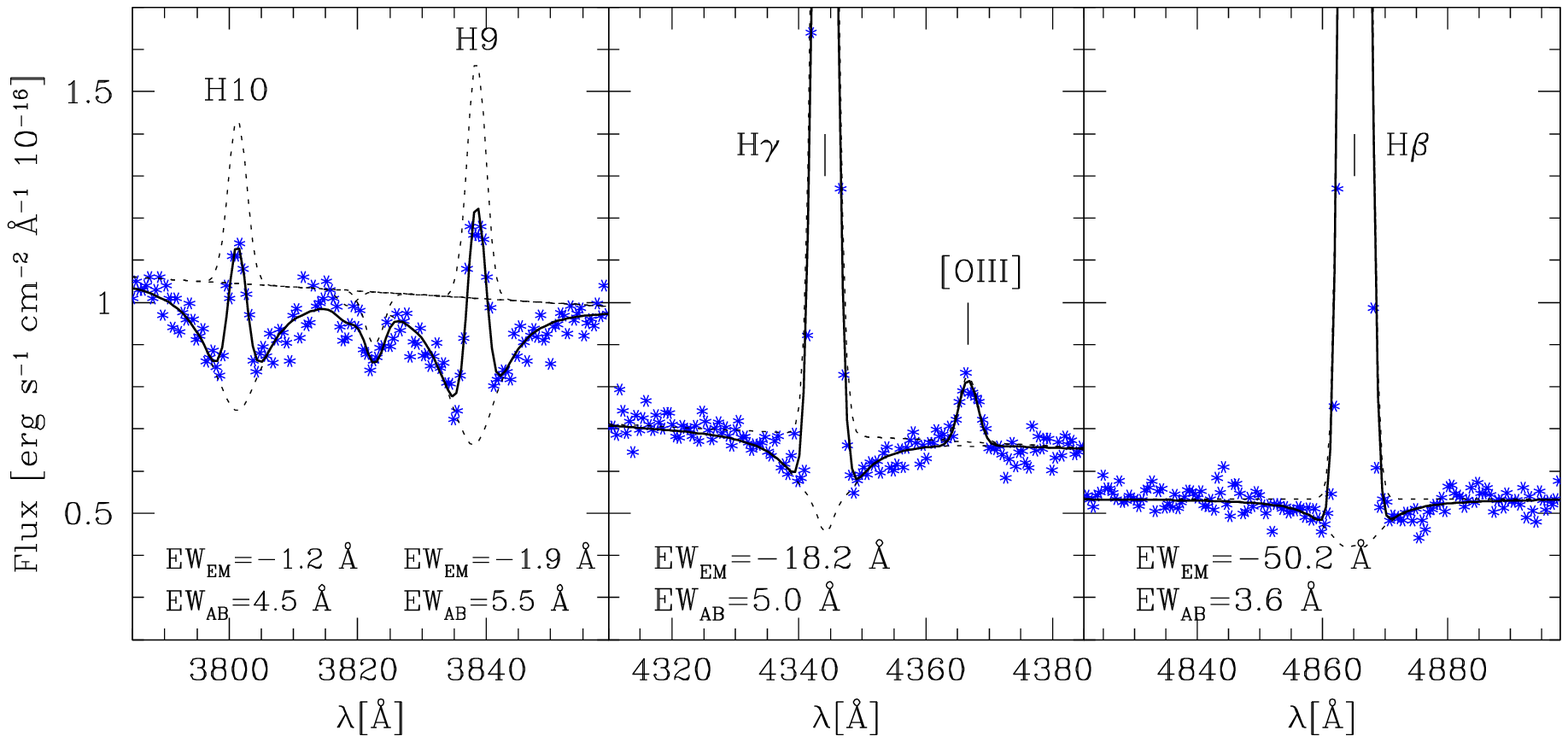}
\caption{ Example (region H~II-6) of spectral fit to the regions around some Balmer lines.  The asterisks are the observed spectrum, while the continuous line is the best fit. The individual components of the fit are plotted with a dashed line: linear continuum, Gaussian profiles for the emission lines, and Voigt profiles for the Balmer absorption lines.
The outcome of fit to the  H10($\lambda$3798) and H9($\lambda$3835) lines, with prominent absorption wings,  was used to fix the Lorentian and   Gaussian FWHMs during the fit to redder Balmer lines. The derived Balmer equivalent widths in absorption and in emission are provided within each panel. 
 The absorption feature between H10 and H9 is due to a  blend of Fe~I lines.
The figure clearly shows that the contribution of absorption with respect to emission becomes increasingly lower toward redder wavelengths (see section~\ref{balmer_abs} for more details.)
\label{balmer_fig}}
\end{figure*}

Eventually, to evaluate the accuracy of the flux calibration for our spectra, we used HST/ACS imaging in F435W, F555W, F814W, and F658N from our GO program 10585 
 and archive HST/ACS imaging in F502N and F550M from GO program 10522 (PI Calzetti),  
with a smaller field of view.  
The six H~II regions, plus PN-3 and PN-4, fall within the field of view of all six images; however, we considered only H~II regions to the purpose of checking the flux calibration, because of their signal-to-noise higher than in PNe. 
Aperture photometry was performed on the images within the same extraction aperture of the 1D spetcra using the {\it Polyphot} task in IRAF.  
The F555W magnitudes in the ACS Vegamag system derived for the H~II regions  and the 5 PNe targeted with MODS are given in Tables~\ref{h2_flux_raw} and \ref{pn_flux_raw} in the Appendix (see instead Table~\ref{pne_total} in Section~8 for a complete list of the  F555W and the F502N magnitudes derived for the total sample of 28 PNe).
For the H~II regions, we also computed synthetic magnitudes in the  F435W, F502N, F550M, F555W, F814W and F658N ACS filters by convolving the MODS spectra with the ACS bandpasses: this was done  by running  the 
{\it Calcphot} task in  the {\it Synphot} package. 
Figure~\ref{calibration} shows, for each H~II region, the  comparison between  the observed  ACS  fluxes and the synthetic ones  (in logarithmic scale) as a function of wavelength.  We notice that the $F_{\lambda,ACS}$ fluxes are smaller than the $F_{\lambda,SYN}$ ones, indicating that the fluxes from the spectra are overestimated. This effect is expected, and its origin is well discussed by \cite{smith07}: flux calibrations are usually tied to reference point source observations, and therefore include an implicit correction for the fraction of the point source light that falls outside the slit; however, in the limit of a perfect uniform, slit-filling extended source, the diffractive losses out of the slit are perfectly balanced by the diffractive gains into the slit from emission beyond its geometric boundary. Therefore, the point-source-based calibration will inevitably cause an  overestimate of the extended source flux.
The  $\log(F_{\lambda,ACS}) - \log(F_{\lambda,SYN})$  zero point offsets of Fig.~\ref{calibration} were used to anchor our  H~II region spectra to the ACS magnitudes;
 this procedure is useful to establish an absolute flux calibration for these objects.  On the other hand, 
no correction was applied to the PN spectra, for two main reasons: 1) these sources are point-like at NGC~4449' s distance (also in the HST images), and therefore we do not expect the spectral fluxes to be overestimated as in the case of extended H~II regions;  2) synthetic fluxes derived by convolving the ACS filter throughputs with the PN spectra are 
highly affected by uncertainties in the background subtraction, and do not permit a reliable comparison with the ACS photometry. 
 The  standard deviation  around the average $\log F_{\lambda,ACS} - log F_{\lambda,SYN}$ value derived for the H~II regions is quite modest,  corresponding to an average 
 flux uncertainty  in the range $\approx$2\% -  $\approx$11\%.
The low dispersion about the mean and the lack of a general trend with wavelength apparently confirms that our observational setup prevented significant flux losses at the bluest wavelengths due to atmospheric differential refraction \citep{filippenko}.

 As an example, we show the final calibrated spectra for H~II region H~II-1 and planetary nebula PN-1 in Figures~\ref{slit3} and ~\ref{slit1}, respectively. 
The spectra of all the other  H~II regions and PNe are provided in Figures~\ref{slit4} to ~\ref{slit11} in the Appendix.

\section{Emission line measurement \label{line_meas}}

Emission line fluxes for H II regions and PNe were obtained with the {\it deblend} function available in the  {\it splot} IRAF task. 
We used this function to fit single lines, groups of lines close in wavelength, or blended lines.
Lines were fitted with Gaussian profiles, treating the centroids and the widths as free parameters;  however, 
when fitting groups of lines close in $\lambda$ (e.g., ${\rm He~I~\lambda6678 + [S~II]\lambda6716 + [S~II]\lambda6731}$) 
or (partially) blended  (${\rm H\alpha+[N~II]\lambda\lambda6548,84}$), we forced the lines to have all the same width. On the other hand, 
no constraint on the line centroids (e.g. fixed separation between the lines) was assumed and we let the {\it deblend} 
function to find  the best-fit line centers independently.  This was a reasonable choice given that  even faint ``key'' lines, such as [O~III]$\lambda4363$ and 
[N~II]$\lambda\lambda6548,84$, are detected with a good signal-to-noise in our spectra; however, had we had worse data, it would have been more appropriate 
to set the wavelengths of the blended  lines and to allow for a common Doppler shift, in order to reduce the noise on the weak lines measurement.
The continuum was defined choosing two continuum windows to the left and to the right of the line or line complex, and fitting with a linear regression. 
 Balmer lines, potentially affected by underlying stellar absorption, were fitted with a combination of Voigt profiles in absorption  plus Gaussian profiles in emission  
(see Section~\ref{balmer_abs} for more details). 

The final emission fluxes were obtained repeating the measurement several times with slightly different continuum choices, and averaging the results. 
To compute the errors, we derived the standard deviation of the different measurements ($\sigma_{F, line}$).  
 The results for the H~II regions and the PNe are provided in Tables~\ref{h2_flux_raw} and ~\ref{pn_flux_raw} of the Appendix, respectively.
From the tables, we notice that the errors on the derived fluxes are very small: for instance, in the case of H~II regions, $\sigma_{F, line}$  is below 1\%
 for the brightest lines such as  [O~III]$\lambda5007$. To get more realistic errors, we added in quadrature to $\sigma_{F, line}$ a 15\% flux error below 
$\sim4000$ \AA \ to account for atmospheric extinction uncertainties (see Section~2), and a  2\% to 11\% flux error,  corresponding to the scatter 
around the average offsets in Figure~\ref{calibration}, at redder wavelengths. 
  For PNe,  the standard deviation from the different line measurements turned out significantly larger than  for H~II regions, 
 typically in the range $\approx$2\% to $\approx$20\%;  an arbitrary additional 15\% error, equalling the flux calibration uncertainty below $\sim4000$ \AA \  and slightly above the largest scatter for the H~II regions in Fig.~\ref{calibration},  was added in quadrature to  $\sigma_{F, line}$ over the entire wavelength range.

\subsection{Absorption from underlying stellar population \label{balmer_abs}}

\setcounter{table}{3}

\begin{center}
\begin{deluxetable*}{lcccccc}
\tabletypesize{\scriptsize}
\tablecaption{Reddening-corrected emission fluxes for H~II regions in NGC~4449 \label{h2_flux}}
\tablewidth{0pt}
\tablehead{
\colhead{Line} & \colhead{H~II-1} &  \colhead{H~II-2}  &  \colhead{H~II-3}  &  \colhead{H~II-4} &  \colhead{H~II-5} &  \colhead{H~II-6} 
}
\startdata
{[O II]} $\lambda$3727 &  110 $\pm$ 20 &  190 $\pm$  30 &  400 $\pm$  70 &  250 $\pm$  50 &  240 $\pm$  40 &  480 $\pm$  90 \\
H10 $\lambda$3978 &    6 $\pm$   1 &    6 $\pm$   1 &    6 $\pm$   1 &    7 $\pm$   1 &    6 $\pm$   1 &    6 $\pm$   1 \\
He I $\lambda$3820  &   0.9 $\pm$  0.1 &   0.7 $\pm$  0.1 & $-$ &   0.6 $\pm$  0.1 & $-$ & $-$ \\
H9$+$He II $\lambda$3835 &    9 $\pm$   1 &    8 $\pm$   1 &    9 $\pm$   1 &    9 $\pm$   2 &    8 $\pm$   1 &    9 $\pm$   2 \\
{[Ne III]} $\lambda$3869 &   32 $\pm$  5 &   23 $\pm$   4 &   14 $\pm$   2 &   23 $\pm$   4 &   18 $\pm$   3 &   35 $\pm$   6 \\
H8$+$He I $\lambda$3889 &   19 $\pm$   3 &   21 $\pm$   4 &   20 $\pm$   3 &   22 $\pm$   4 &   21 $\pm$   4 &   21 $\pm$   4 \\
H$\epsilon$ $+$ He I $+$[Ne III] $\lambda$3970 &   25 $\pm$  4 &   23 $\pm$  4 &   21 $\pm$   3 &   27 $\pm$  5 &   24 $\pm$   4 &   29 $\pm$   5 \\
He I $\lambda$4026 &   2.02 $\pm$  0.06 &   1.8 $\pm$  0.2 &   0.9 $\pm$  0.1 &   1.5 $\pm$  0.2 &   1.0 $\pm$  0.1 & $-$ \\
{[S II]} $\lambda$4068 &   0.73 $\pm$  0.03 &   0.77 $\pm$  0.07 &   2.5 $\pm$  0.2 &   1.0 $\pm$  0.1 &   1.3 $\pm$  0.2 &   3.9 $\pm$  0.5 \\
{[S II]} $\lambda$4076 &   0.21 $\pm$  0.03 &   0.40 $\pm$  0.04 &   0.79 $\pm$  0.06 & $-$ & $-$ &   1.5 $\pm$  0.2 \\
H$\delta$ $\lambda$4101 &   26.3 $\pm$   0.8 &   29 $\pm$   3 &   30 $\pm$   2 &   33 $\pm$   5 &   30 $\pm$  4 &   30 $\pm$  4 \\
H$\gamma$ $\lambda$4340 &   46 $\pm$   1 &   49 $\pm$   4 &   50 $\pm$   4 &   50 $\pm$   7 &   50 $\pm$  6 &   52 $\pm$   6 \\
{[O III]} $\lambda$4363 &   2.1 $\pm$  0.1 &   1.8 $\pm$  0.2 &   1.4 $\pm$  0.1 &   1.6 $\pm$  0.2 &   1.9 $\pm$  0.2 &   2.4 $\pm$  0.3 \\
He I $\lambda$4389 &   0.57 $\pm$  0.02 &   0.38 $\pm$  0.03 & $-$ &   0.48 $\pm$  0.07 & $-$ & $-$ \\
He I $\lambda$4471 &    4.6 $\pm$   0.1 &    4.8 $\pm$   0.4 &    4.1 $\pm$   0.3 &    4.5 $\pm$   0.6 &    4.2 $\pm$   0.4 &    4.2 $\pm$   0.4 \\
{[N III](WR)} $\lambda$4640 &   0.19 $\pm$  0.01 & $-$ & $-$ &   1.1 $\pm$  0.2 & $-$ & $-$ \\
{[C III](WR)} $\lambda$4652 &   0.11 $\pm$  0.03 & $-$ & $-$ &   0.5 $\pm$  0.1 & $-$ & $-$ \\
{[Fe III]} $\lambda$4658 &   0.14 $\pm$  0.02 &   0.28 $\pm$  0.02 &   0.84 $\pm$  0.06 &   0.43 $\pm$  0.07 &   0.50 $\pm$  0.06 &   1.4 $\pm$  0.2 \\
He II (WR) $\lambda$4686 &   3.6 $\pm$  0.2 & $-$ & $-$ &   6 $\pm$ 1 & $-$ & $-$ \\
He II  $\lambda$4686 &   0.4 $\pm$  0.3 & $-$ & $-$ &  $-$   & $-$ &   0.8 $\pm$  0.1 \\
{[Ar IV]}$+$He I $\lambda$4713 &   0.50 $\pm$  0.05 &   0.38 $\pm$  0.03 & $-$ &   0.5 $\pm$  0.1 & $-$ & $-$ \\
{[Ar IV]} $\lambda$4740 &   0.16 $\pm$  0.01 & $-$ & $-$ &   0.30 $\pm$  0.06 & $-$ & $-$ \\
H$\beta$ $\lambda$4861 &  100 $\pm$   3 &  100 $\pm$   8 &  100 $\pm$   7 &  100 $\pm$  14 &  100 $\pm$  11 &  100 $\pm$  11 \\
He I $\lambda$4922 &   1.20 $\pm$  0.04 &   1.2 $\pm$  0.1 &   0.76 $\pm$  0.06 &   1.0 $\pm$  0.2 &   0.8 $\pm$  0.1 & $-$ \\
{[O III]} $\lambda$4959 &  143 $\pm$   4 &  129 $\pm$  11 &   71 $\pm$   5 &  100 $\pm$  20 &  110 $\pm$  10 &  100 $\pm$  10 \\
{[Fe III]} $\lambda$4986 &   0.19 $\pm$  0.02 &   0.37 $\pm$  0.06 &   1.16 $\pm$  0.09 &   0.38 $\pm$  0.05 &   0.57 $\pm$  0.07 &   1.9 $\pm$  0.2 \\
{[O III]} $\lambda$5007 &  420 $\pm$  10 &  380 $\pm$  30 &  210 $\pm$  10 &  300 $\pm$  40 &  320 $\pm$  40 &  300 $\pm$  30 \\
He I $\lambda$5015  &   2.46 $\pm$  0.07 &   2.7 $\pm$  0.2 &   2.2 $\pm$  0.2 &   2.3 $\pm$  0.3 &   2.2 $\pm$  0.2 &   2.2 $\pm$  0.3 \\
{[N I]} $\lambda$5199 &   0.21 $\pm$  0.01 &   0.13 $\pm$  0.01 &   1.01 $\pm$  0.08 &   0.26 $\pm$  0.04 &   0.34 $\pm$  0.04 &   2.2 $\pm$  0.2 \\
He I $\lambda$5876  &   11.8 $\pm$   0.3 &   12 $\pm$   1 &   10.8 $\pm$   0.8 &   12 $\pm$   2 &   12 $\pm$  1 &   13 $\pm$   1 \\
{[OI]}  $\lambda$6302 &   1.17 $\pm$  0.04 &   0.64 $\pm$  0.06 & $-$ &   1.1 $\pm$  0.2 &   1.7 $\pm$  0.2 &   9.4 $\pm$  1.1 \\
{[S III]} $\lambda$6314 &   1.12 $\pm$  0.03 &   1.3 $\pm$  0.1 &   1.2 $\pm$  0.1 &   1.2 $\pm$  0.2 &   1.3 $\pm$  0.2 &   1.6 $\pm$  0.2 \\
{[OI]} $\lambda$6365  &   0.37 $\pm$  0.01 &   0.17 $\pm$  0.02 &   2.0 $\pm$  0.2 &   0.39 $\pm$  0.06 &   0.61 $\pm$  0.08 &   3.1 $\pm$  0.4 \\
{[NII]} $\lambda$6548  &    2.9 $\pm$   0.1 &    4.2 $\pm$   0.4 &    8.8 $\pm$   0.9 &    5.3 $\pm$   0.8 &    4.9 $\pm$   0.6 &   12 $\pm$   1 \\
H$\alpha$ $\lambda$6563  &  287 $\pm$  8 &  300 $\pm$  30 &  300 $\pm$  30 &  310 $\pm$  50 &  300 $\pm$  40 &  310 $\pm$  40 \\
{[N II]} $\lambda$6584  &    8.0 $\pm$   0.3 &   12 $\pm$   1 &   25 $\pm$   2 &   16 $\pm$   2 &   14 $\pm$   2 &   35 $\pm$   4 \\
He I $\lambda$6678   &    3.5 $\pm$   0.1 &    3.4 $\pm$   0.3 &    3.4 $\pm$   0.3 &    3.7 $\pm$   0.6 &    3.4 $\pm$   0.4 &    3.8 $\pm$   0.5 \\
{[S II]} $\lambda$6716  &    8.2 $\pm$   0.2 &    7.1 $\pm$   0.7 &   33 $\pm$   3 &   15 $\pm$   2 &   16 $\pm$   2 &   51 $\pm$   6 \\
{[S II]} $\lambda$6731  &    6.0 $\pm$   0.2 &    5.6 $\pm$   0.5 &   24 $\pm$   2 &   11 $\pm$   2 &   12 $\pm$   1 &   37 $\pm$   5 \\
He I $\lambda$7065  &    1.9 $\pm$   0.1 &    2.1 $\pm$   0.2 &    1.9 $\pm$   0.2 &    2.1 $\pm$   0.3 &    1.9 $\pm$   0.2 &    1.8 $\pm$   0.2 \\
{[Ar III]} $\lambda$7136  &    8.4 $\pm$   0.3 &    9.4 $\pm$   0.9 &    7.1 $\pm$   0.7 &    8 $\pm$   1 &    9 $\pm$   1 &    10$\pm$   1 \\
He I $\lambda$7281  &   0.57 $\pm$  0.02 &   0.59 $\pm$  0.06 & $-$ &   0.6 $\pm$  0.1 &   0.63 $\pm$  0.08 & $-$ \\
{[O II]} $\lambda$7320  &   1.14 $\pm$  0.04 &   1.7 $\pm$  0.2 & $-$ &   2.2 $\pm$  0.4 &   2.3 $\pm$  0.3 &   4.1 $\pm$  0.5 \\
{[O II]} $\lambda$7330  &   0.96 $\pm$  0.03 &   1.5 $\pm$  0.1 & $-$ &   1.8 $\pm$  0.3 &   1.9 $\pm$  0.2 &   3.3 $\pm$  0.4 \\
{[Ar III]} $\lambda$7751  &   2.25 $\pm$  0.07 &   2.4 $\pm$  0.2 & $-$ &   1.9 $\pm$  0.3 &   2.4 $\pm$  0.3 &   2.0 $\pm$  0.3 \\
P10 $\lambda$9017  &   19.8 $\pm$   0.5 &   15$\pm$   2 & $-$ &   19 $\pm$   3 &   20 $\pm$   3 &   16 $\pm$   2 \\
{[S III]} $\lambda$9069  &   20.5 $\pm$   0.7 &   21 $\pm$   2 &   20 $\pm$   2 &   21 $\pm$   4 &   23 $\pm$   3 &   22 $\pm$   3 \\
P9 $\lambda$9229   &   2.24 $\pm$  0.08 &   2.1 $\pm$  0.2 &   2.5 $\pm$  0.3 &   2.4 $\pm$  0.5 &   2.5 $\pm$  0.4 & $-$ \\
{[S III]} $\lambda$9532  &   42 $\pm$   1 &   51 $\pm$  6 &   41 $\pm$   5 &   44 $\pm$   9 &   51 $\pm$   8 &   46 $\pm$   7 \\
P8 $\lambda$9547  &    2.7 $\pm$   0.1 &    3.1 $\pm$   0.4 & $-$ &    2.4 $\pm$   0.5 &    2.6 $\pm$   0.4 & $-$ \\
F(H$\beta$)[$10^{-13}$ erg/s/cm$^2$] &  0.35$\pm$ 0.01&  0.64$\pm$ 0.08&  0.41$\pm$ 0.07 &  0.4$\pm$ 0.1&  0.21$\pm$ 0.04&  0.06$\pm$ 0.01\\
E(B$-$V)  &   0.10 $\pm$  0.01 & 0.24  $\pm$  0.03 & 0.18  $\pm$  0.05 &   0.20 $\pm$ 0.07 &   0.16  $\pm$  0.05 &   0.23 $\pm$ 0.05 \\
\enddata
\tablecomments{Fluxes are given on a scale where F(H$\beta$)$=$100. }
\end{deluxetable*}
\end{center}

\begin{center}
\begin{deluxetable*}{lccccc}
\tabletypesize{\scriptsize}
\tablecaption{Reddening-corrected emission fluxes for PNe in NGC~4449 \label{pn_flux}}
\tablehead{
\colhead{Line} & \colhead{PN-1} &  \colhead{PN-2}  &  \colhead{PN-3}  &  \colhead{PN-4} &  \colhead{PN-5} 
}
\startdata
{[O II]} $\lambda$3727 &   25 $\pm$   7 &   80 $\pm$  20 &   90 $\pm$  20 &  400 $\pm$ 100 & $-$ \\
{[Ne III]} $\lambda$3869 &   70 $\pm$  20 &   80 $\pm$  20 &   70 $\pm$  20 &   80 $\pm$  20 &   90 $\pm$  20 \\
H8$+$He I $\lambda$3889 &   10 $\pm$   3 &   23 $\pm$   6 &   20 $\pm$   5 &   24 $\pm$   6 &   18 $\pm$   5 \\
H$\epsilon$ $+$ He I $+$[Ne III] $\lambda$3970 &   30 $\pm$  8 &   40 $\pm$  10 &   40 $\pm$  10 &   30 $\pm$   9 & $-$ \\
H$\delta$ $\lambda$4101 &   23 $\pm$   5 &   31 $\pm$   7 &   20 $\pm$   5 & $-$ & $-$ \\
H$\gamma$ $\lambda$4340 &   40 $\pm$   8 &   60 $\pm$  10 &   50 $\pm$  10 &   60 $\pm$  10 &   50 $\pm$  10 \\
{[O III]} $\lambda$4363 &   10 $\pm$   3 &   20 $\pm$   4 &   20 $\pm$   4 &   20 $\pm$   4 &   20 $\pm$   4 \\
He II  $\lambda$4686 & $-$ &   16 $\pm$   3 &   40 $\pm$   7 & $-$ & $-$ \\
H$\beta$ $\lambda$4861 &  100 $\pm$  20 &  100 $\pm$  20 &  100 $\pm$  20 &  100 $\pm$  20 &  100 $\pm$  20 \\
{[O III]} $\lambda$4959 &  390 $\pm$  70 &  410 $\pm$  70 &  400 $\pm$  70 &  330 $\pm$  60 &  430 $\pm$  80 \\
{[O III]} $\lambda$5007 & 1100 $\pm$ 200 & 1200 $\pm$ 200 & 1100 $\pm$ 200 &  900 $\pm$ 200 & 1200 $\pm$ 200 \\
He I $\lambda$5876  &   11 $\pm$   3 &   11 $\pm$   3 &    7 $\pm$   2 &   11 $\pm$   2 &   14 $\pm$   3 \\
{[NII]} $\lambda$6548  &   10 $\pm$   3 &    9 $\pm$   3 &   24 $\pm$   6 &   11 $\pm$   3 &    5 $\pm$   1 \\
H$\alpha$ $\lambda$6563  &  280 $\pm$  70 &  280 $\pm$  70 &  260 $\pm$  70 &  280 $\pm$  70 &  280 $\pm$  70 \\
{[N II]} $\lambda$6584  &   25 $\pm$   7 &   36 $\pm$   9 &   70 $\pm$  20 &   24 $\pm$   6 &   11 $\pm$   3 \\
{[S II]} $\lambda$6716  &   10 $\pm$   3 &   11 $\pm$   3 &    9 $\pm$   2 &   27 $\pm$   7 & $-$ \\
{[S II]} $\lambda$6731  &   11 $\pm$   3 &   10 $\pm$   3 &   11 $\pm$   3 &   18 $\pm$   5 & $-$ \\
He I $\lambda$7065  &   14 $\pm$   4 & $-$ & $-$ & $-$ &    7 $\pm$   2 \\
{[Ar III]} $\lambda$7136  &    8 $\pm$   2 &   17 $\pm$   5 &    9 $\pm$   3 & $-$ & $-$ \\
{[S III]} $\lambda$9069  &   18 $\pm$   7 &   20 $\pm$   10 &   19 $\pm$   8 & $-$ & $-$ \\
{[S III]} $\lambda$9532  &   30 $\pm$  10 & $-$ & $-$ & $-$ & $-$ \\
F(H$\beta$)[10$^{-16}$erg/s/cm$^2$] &  0.52 $\pm$ 0.07 &  0.4 $\pm$ 0.3 &  0.35 $\pm$ 0.05 &  0.8 $\pm$ 0.5 &  0.4 $\pm$ 0.3 \\
E(B$-$V) &   0.0 $\pm$ 0.2  &   0.1  $\pm$   0.2 &   0.0  $\pm$  0.2 &  0.4 $\pm$ 0.2 &   0.1  $\pm$ 0.2 \\
\enddata
\tablecomments{Fluxes are given on a scale where F(H$\beta$)$=$100.
}
\end{deluxetable*}
\end{center}

To account for underlying stellar absorption in H~II regions, spectral regions around Balmer lines were fitted with a combination of Voigt profiles (in absorption) plus Gaussian profiles (in emission),  as shown in Figure~\ref{balmer_fig}. 
Absorption wings are more prominent for bluer lines, because Balmer emission decreases very rapidly toward bluer wavelength, while the absorption equivalent widths remain roughly constant with wavelength. This implies that the contribution from absorption and emission can be better separated for bluer lines, while fits to red Balmer lines (H$\beta$ and H$\alpha$) tend to be highly affected by a degeneracy between absorption strength and FWHM of the line. To overcome this problem, we fitted the H10($\lambda$3798) and H9($\lambda$3835) lines in the first place, and then adopted the derived  (Lorentian and Gaussian) FWHM values for the fits to all the other Balmer lines up to H$\beta$. We derived average absorption equivalent widths and standard deviations of $3.3\pm0.6$ \AA,  $4.2\pm0.7$ \AA, $4.2\pm0.3$ \AA, $4.2\pm0.6$ \AA, $4.9\pm0.1$ \AA, $5.6\pm0.3$ \AA,  and  $5.6\pm1.2$ \AA \ for H10, H9, H8, H$\epsilon$, H$\delta$, H$\gamma$, and H$\beta$,  respectively, 
compatible with a simple stellar population (SSP) of age $\simeq$10-20 Myr and $Z=0.008$ (about NGC~4449' s metallicity) from the 
Starburst99 models \citep[][hereafter SB99]{sb99}. 
For such an SSP, the models predict  a typical  absorption equivalent width of  $\simeq3.5$ \AA \ for H$\alpha$,  
 to be compared with emission strengths in the range 100$-$2000 \AA.  Therefore, we neglected the effect of underlying absorption on the $\rm {H\alpha+[N~II]\lambda\lambda6548, 84}$ emission complex. In conclusion, our procedure accounts for the presence of underlying Balmer absorptions by simultaneously fitting the emission and absorption components, so that no further correction needs to be applied to the Hydrogen emission lines.

As for Helium, absorption wings are too shallow to allow for a decomposition of absorption and emission through spectral fits. Thus we used the predictions of simple stellar population models to correct for this effect. The SB~99 models provide for a $\simeq$10-20 Myr old population with $Z=0.008$ typical absorption equivalent widths of 
$\simeq$0.5 \AA, $\simeq$0.3 \AA \ and $\simeq$0.3 \AA \ in  the  He~I $\lambda$4471,  He~I $\lambda$5876 and  He~I $\lambda$6678 lines, respectively. These contributions are not negligible when compared with observed emissions in the range $\simeq1-9$ \AA,  $\simeq9-55$ \AA,  and  $\simeq3-23$ \AA \ in  our sample. 
 Therefore, we adopted the absorption EWs provided by SB~99 and  corrected the He~I emission lines according to the formula 
$F'(He~I) = F(He~I) + EW_{abs}(SB~99)*F (CONT)$,  where  $F(He~I)$ and  $F'(He~I)$ are the "raw'' and the corrected fluxes, respectively, $EW_{abs}(SB~99)$ is the stellar absorption equivalent width from the SB~99 models, and $F (CONT)$ is the flux in the continuum measured from the spectra.

\subsection{Reddening correction}

For H~II regions, the reddening was derived from the H$\delta$/H$\alpha$,  H$\gamma$/H$\alpha$ and  H$\beta$/H$\alpha$ ratios assuming  the 
\cite{cardelli89} extinction law with $R_V = 3.05$, according to the formula: 

\begin{equation}\label{reddening}
E(B-V)= \frac {\log_{10} [(FH_{\lambda1}/FH_{\lambda2})_o/(FH_{\lambda1}/FH_{\lambda2})_t]}{0.4\times R_V \times [A_{\lambda2}/A_V - A_{\lambda1}/A_V ]}
\end{equation}

where $\lambda1$ and $\lambda2$ are the wavelengths of the two Balmer lines,  $(FH_{\lambda1}/FH_{\lambda2})_o$ and $(FH_{\lambda1}/FH_{\lambda2})_t$ are, respectively, the observed and theoretical Balmer emission line ratios,  
and the magnitude attenuation ratio $A_{\lambda}/A_V$ is that from Cardelli' s law.
We adopted theoretical Balmer ratios of   $(FH_{\delta}/FH_{\alpha})_t=0.090$, $(FH_{\gamma}/FH_{\alpha})_t=0.163$, and $(FH_{\beta}/FH_{\alpha})_t=0.350$ from \cite{storey95}  for case B recombination assuming $T_e=10,000$ K and $n_e=100 \ cm^{-3}$, and $A_{H\delta}/A_V\sim1.45$,  $A_{H\gamma}/A_V\sim1.36$,  $A_{H\beta}/A_V\sim1.17$ and  $A_{H\alpha}/A_V\sim0.81$ from Cardelli' s extinction curve. 
For each Balmer line ratio, the error in $E(B-V)$ was obtained  by propagating the emission flux errors into Eq.~(\ref{reddening}).  
It can be easily verified from  Eq.~(\ref{reddening}) and from the  $A_{\lambda}/A_V$ values reported above that, for equal emission flux errors, the reddening uncertainty is higher when using Balmer line ratios with closer wavelength spacing, minimizing the  $A_{\lambda2} - A_{\lambda1}$ difference: for instance,  a 
 $\sim$5\% flux error provides an error $\sigma_{E(B-V)}$ in the range $\sim$0.04$-$0.07 mag if the reddening is estimated from  the H$\delta$/H$\alpha$,  
 H$\gamma$/H$\alpha$ and  H$\beta$/H$\alpha$ ratios (as in our case), while using e.g. H$\delta$/H$\gamma$ implies $\sigma_{E(B-V)}$ as high as $\sim$0.28 mag.  
  
For H~II regions, the reddening was obtained by averaging the results from the H$\delta$/H$\alpha$, H$\gamma$/H$\alpha$ and  H$\beta$/H$\alpha$ ratios, 
and its uncertainty was computed as the standard deviation; typically, the  $E(B-V)$ values derived from the three different Balmer ratios  turned out to be consistent with each other, within the errors. Differences in the $E(B-V)$ values obtained from different Balmer lines may arise from the fact that redder lines, affected by a lower extinction, probe larger optical depths of the nebula \citep{calzetti96}.
For PNe, we used instead only the H$\beta$/H$\alpha$ ratio, due to the faintness of the other Balmer lines. 
For H~II regions, the derived  $E(B-V)$ values are  in the range $\sim0.10\pm0.01-0.24\pm0.03$ mag, while for PNe they are in the range $\approx$0$-$0.4 mag  
with a typical uncertainty of $\sim$0.2 mag. 

The emission line fluxes are corrected for reddening according to the formula :

\begin{equation}
F_c = F_o \times 10^{0.4\times A_{\lambda}},
\end{equation}

where $F_o$ and $F_c$ are the observed and the extinction-corrected fluxes, respectively, and  $A_{\lambda} = (A_{\lambda}/A_V)\times R_V \times E(B-V)$.
The derived E(B$-V$) values and the reddening-corrected emission-line fluxes for H~II regions and PNe are given in Tables~\ref{h2_flux} and ~\ref{pn_flux}, respectively.

\section{Temperatures, Densities and Chemical Abundances}

Temperatures, densities, and chemical abundances for H II regions and PNe were derived using  the  {\it getCrossTemDen}  and {\it getIonAbundance} options in the 
1.0.1 version of the PyNeb code \citep{pyneb}, which is based upon the FIVEL program  developed  by  \cite{fivel} and \cite{abund}. The {\it getCrossTemDen}  task  simultaneously derives   
electron densities ($n_e$) and temperatures ($T_e$)  through an iterative process  assuming a density-sensitive and a temperature-sensitive diagnostic line ratio:  
the quantity (density or temperature) derived from one emission line ratio is inserted into the other, and the process is iterated until the two temperature-sensitive and density-sensitive diagnostics give self-consistent results.

Once the physical conditions are known, the {\it getIonAbundance} task computes the ionic abundance of a given ion relative to $H^+$ from the observed emission line intensities relative to H$\beta$. We ran PyNeb with the default data-set for  line emissivities, collision strengths, and radiative transition probabilities; the atomic data set sources for the various ions are provided in Table~\ref{atomic_set}. Notice that the adopted emissivities for  $He^{+}$ are those of  \cite{porter12,porter13}, which 
include collisional excitation.

\begin{center}
\begin{deluxetable*}{lll}
\tablecaption{Used Atomic Data Set \label{atomic_set}}
\tablehead{
\colhead{{\bf Ion}} &  \multicolumn{2}{c}{{\bf Emissivities}} 
}
\startdata
$H^+$ & \multicolumn{2}{c}{ \cite{storey95} }\\
$He^+$ & \multicolumn{2}{c}{\cite{porter12,porter13}}\\
$He^{+2}$ & \multicolumn{2}{c}{ \cite{storey95} }\\
\hline
& & \\
\multicolumn{1}{l}{ } &  \multicolumn{1}{l}{{\bf Transition Probabilities}} & \multicolumn{1}{l}{{\bf Collision Strengths}} \\ 
\hline
$N^+$ & \cite{galavis97}, \cite{wfd96}  & \cite{tayal11}  \\
$O^+$ & \cite{zei82}, \cite{wfd96} & \cite{pra06}, \cite{tayal07} \\
$O^{+2}$ & \cite{sz00}, \cite{wfd96} & \cite{ak99} \\
$S^+$ & \cite{pkw09}  &  \cite{tz10} \\
$S^{+2}$ &  \cite{pkw09} &  \cite{tg99}\\
$Ne^{+2}$ & \cite{galavis97}  & \cite{mb00} \\
$Ar^{+2}$ &  \cite{mend83}, \cite{ks86} & \cite{galavis95} \\
\enddata
\end{deluxetable*}
\end{center}

\subsection{H II regions}

For H II regions,  $n_e$  and $T_e$ values were derived using the density-sensitive [S II] $\lambda6716/\lambda6731$ diagnostic line ratio, and three sets of temperature-sensitive line ratios: 
[O~III]$\lambda4363/\lambda4959+\lambda5007$, [S~III]$\lambda6312/\lambda9069+\lambda9532$ 
and  [O~II]$\lambda7320+\lambda7330/\lambda3726+\lambda3729$. We found $n_e \lesssim 100 \ cm^{-3}$, and  $T_e$ in the range $9000 - 11000 \ K$. Density and 
temperature values for the individual H~II regions are given in Table~\ref{h2_abund}. The associated errors were derived by inputting into the {\it getCrossTemDen} 
task the $\pm 1 \sigma$ interval for each diagnostic line flux ratio.

\begin{figure}
\epsscale{1.2}
\plotone{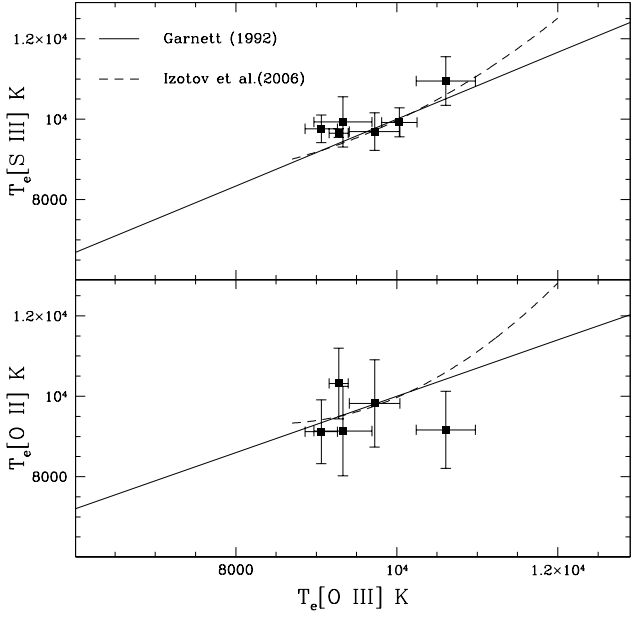}
\caption{Correlations between electron temperatures derived for H II regions in NGC~4449 through different diagnostics: 
[O III]$\lambda$4363/[O III]$\lambda\lambda$4959,5007  for $T_e$[O III],  
[S~III]$\lambda$6312/[S~III]$\lambda\lambda$9069,9532  for $T_e$[S~III], and [O~II]$\lambda\lambda$3726,29/[O II]$\lambda\lambda$7320,30 for $T_e$[O~II]. 
The solid and dashed lines are the predicted correlations based on photoionization models from \cite{garnett92} 
and from \cite{izotov06}, respectively. 
\label{temps}}
\end{figure}

The availability of multiple $T_e$ measurements in H~II regions allowed us to investigate the comparison between temperatures measured for different ions 
($O^+$, $O^{+2}$, $S^{+2}$). The behaviour of  $T_e$[S~III]  against $T_e$[O~III], and of  $T_e$[O~II] against $T_e$[O~III] for our H~II regions is shown in Figure~\ref{temps}, 
together with the predicted correlations from \cite{garnett92} and \cite{izotov06} based on photoionization models.  The derived temperatures are consistent, within the errors, with the predictions from the models, although they do not exhibit clear correlations, probably because of the small temperature range sampled by our data. In particular, we notice that  the data points in the $T_e$[O~II] versus $T_e$[O~III] diagram exhibit a large scatter around the theoretical relations, an effect that was also found  and  discussed by other authors \citep[e.g.][]{k03,bresolin09a,berg15}. A detailed discussion of the possible theoretical and observational causes of this disagreement (e.g. recombination contribution to the [O~II]$\lambda\lambda7320,30$ lines, radiative transfer and shocks affecting the [O II]~lines, reddening uncertainties)  can be found in these studies.
We notice that our [O~II] temperatures are affected by large observational errors, both because of the uncertain flux calibration below $\sim$4000 \AA, as discussed in Section~2, and because of the large uncertainty in the extinction-corrected [O~II] ratios, due to the large wavelength difference between the  [O~II]$\lambda\lambda3726,29$ doublet and the [O~II]$\lambda\lambda7320,30$ complex.\footnote{The  [O~II]$\lambda\lambda7320,30$ complex consists in fact 
 of the blend of  the two [O~II]$\lambda\lambda$7319,20 lines and of the blend of the two  [O~II]$\lambda\lambda$7330,31 lines.}

 Chemical abundances were derived assuming a three-zone model for the electron temperature structure: 
the [O~III] temperature was adopted for the highest-ionization zone ($O^{+2}$, $Ne^{+2}$, $He^{+}$, $He^{+2}$) , the [S~III] temperature for the intermediate-ionization zone ($S^{+2}$, $Ar^{+2}$), and the [O~II] temperature for the low-ionization zone ($O^{0}$, $O^{+}$, $N^{+}$, $S^{+}$, $Fe^{+2}$).
For $T_e$[O~III]  and $T_e$[S~III], we used the temperatures directly derived from the [O~III]$\lambda4363/\lambda\lambda4959,5007$ and  
[S~III]$\lambda6312/\lambda\lambda9069,9532$ line ratios. For the [O II] temperature, instead, in view of the problems observed in the T$_e$[O~II]  
versus T$_e$[O~III]  plane, we used the relation from  \cite{garnett92}:

\begin{center}
\begin{deluxetable*}{lcccccc}
\tabletypesize{\scriptsize}
\tablecaption{Derived properties for H~II regions in NGC~4449 \label{h2_abund}}
\tablehead{
\colhead{Property} & \colhead{H~II-1} &  \colhead{H~II-2}  &  \colhead{H~II-3}  &  \colhead{H~II-4} &  \colhead{H~II-5} &  \colhead{H~II-6} 
}
\startdata
R.A.[J2000] & 12:28:12.626 & 12:28:09.456 & 12:28:17.798 & 12:28:16.224 & 12:28:13.002 & 12:28:13.925 \\
Dec.[J2000] & +44:05:04.35 & +44:05:20.35 & +44:06:32.49 & +44:06:43.32 & +44:06:56.38 & +44:07:19.04 \\
$R/R_{25}$ & 0.20 & 0.13 & 0.49 & 0.46 & 0.44 & 0.57 \\
\hline
$n_e [cm^{-3}]$ &    $50_{-30}^{+30}$ &   $120_{-90}^ {+100}$ &   $40_{-40}^{+70}$ &    $70_{-70}^{+100}$ &   $40_{-40}^{+100}$ &  $30_{- 30}^{+100}$ \\
$T_e (O^+)$ [K] &   10300$\pm$   900 &    9100 $\pm$  800 &    $-$ &    9000 $\pm$   1000 &    10000 $\pm$   1000 &    9000 $\pm$ 1000 \\
$T_e (O^{++})$ [K] &    9300 $\pm$  100 &    9100 $\pm$  200 &   10000 $\pm$  200 &    9300 $\pm$  400 &    9700 $\pm$  300 &   10600 $\pm$  400 \\
$T_e (S^{++})$ [K] &    9600 $\pm$ 100 &    9800 $\pm$  300 &    9900 $\pm$ 400 &    9900 $\pm$  600 &    9700 $\pm$  500 &   10900 $\pm$  600 \\
\hline
$(O^+/H^+)\times10^4$  &   0.50 $\pm$  0.04 &   0.88 $\pm$  0.09 &   1.3 $\pm$  0.2 &   1.0 $\pm$  0.1 &   0.9 $\pm$  0.1 &   1.2 $\pm$  0.2 \\
$(O^{++}/H^+)\times10^4$  &   2.0 $\pm$  0.1 &   2.0 $\pm$  0.2 &   0.75 $\pm$  0.06 &   1.4 $\pm$  0.2 &   1.3 $\pm$  0.1 &   0.9 $\pm$  0.1 \\
$12+ \log(O/H)$ &   8.40 $\pm$  0.02 &   8.46 $\pm$  0.03 &   8.32 $\pm$  0.03 &   8.39 $\pm$  0.04 &   8.34 $\pm$  0.04 &   8.32 $\pm$  0.04 \\
\hline
$(He/H)$ &    0.089 $\pm$   0.003 &    0.090 $\pm$   0.005 &    0.084 $\pm$   0.004 &    0.092 $\pm$   0.008 &    0.086 $\pm$   0.006 &    0.095 $\pm$   0.007 \\
\hline
$(N^+/H^+)\times10^6$  &   1.96 $\pm$  0.06 &   3.1 $\pm$  0.3 &   5.2 $\pm$  0.5 &   3.7 $\pm$  0.6 &   3.1 $\pm$  0.4 &   6.6 $\pm$  0.8 \\
$12+ \log(N/H)$ &   6.99 $\pm$  0.01 &   7.04 $\pm$  0.04 &   6.99 $\pm$  0.04 &   7.02 $\pm$  0.07 &   6.95 $\pm$  0.05 &   7.15 $\pm$  0.05 \\
\hline
$(Ne^{++}/H^+)\times10^5$  &   4.8 $\pm$  0.3 &   3.9 $\pm$  0.4 &   1.5 $\pm$  0.1 &   3.4 $\pm$  0.6 &   2.3 $\pm$  0.3 &   3.0 $\pm$  0.4 \\
$12+ \log(Ne/H)$ &   7.74 $\pm$  0.02 &   7.71 $\pm$  0.04 &   7.52 $\pm$  0.04 &   7.71 $\pm$  0.08 &   7.53 $\pm$  0.06 &   7.76 $\pm$  0.06 \\
\hline
$(S^{+}/H^+)\times10^7$  &   4.1 $\pm$  0.1 &   3.9 $\pm$  0.4 &  14 $\pm$  1 &   7 $\pm$  1 &   7.2 $\pm$  0.9 &  20 $\pm$  2 \\
$(S^{++}/H^+)\times10^6$  &   3.00 $\pm$  0.09 &   3.3 $\pm$  0.3 &   2.7 $\pm$  0.3 &   2.9 $\pm$  0.5 &   3.4 $\pm$  0.3 &   2.5 $\pm$  0.3 \\
$12+ \log(S/H)$ &   6.68 $\pm$  0.01 &   6.64 $\pm$  0.04 &   6.60 $\pm$  0.03 &   6.58 $\pm$  0.06 &   6.64 $\pm$  0.04 &   6.65 $\pm$  0.04 \\
\hline
$(Ar^{++}/H^+)\times10^7$  &   8.8 $\pm$  0.5 &   9.3 $\pm$  0.6 &   6.5 $\pm$  0.4 &   7.4 $\pm$  0.9 &   9.3 $\pm$  0.9 &   6.6 $\pm$  0.6 \\
$12+ \log(Ar/H)$ &   5.98 $\pm$  0.02 &   5.99 $\pm$  0.03 &   5.88 $\pm$  0.03 &   5.90 $\pm$  0.06 &   6.00 $\pm$  0.04 &   5.88 $\pm$  0.04 \\
\hline
$(Fe^{++}/H^+)\times10^7$  &   0.74 $\pm$  0.06 &   1.8 $\pm$  0.2 &   4.0 $\pm$  0.5 &   1.7 $\pm$  0.2 &   2.4 $\pm$  0.5 &   5.6 $\pm$  0.7 \\
$12+ \log(Fe/H)$ &   5.69 $\pm$  0.03 &   5.92 $\pm$  0.04 &   5.97 $\pm$  0.06 &   5.79 $\pm$  0.05 &   5.94 $\pm$  0.09 &   6.18 $\pm$  0.06 \\
\enddata
\end{deluxetable*}
\end{center}

\begin{equation}
T_e[O~II] = 0.70 \times T_e [O~III] + 3000 \ K,
\end{equation}

 an approach that is widely applied in the literature to reduce the uncertainty in the [O~II] temperature determination \citep[e.g.][]{k03,bresolin11,berg15}.
However, we caution that the errors on the [O~II] temperature derived with this model-based relation  are formal errors obtained by propagating the  $T_e$[O~III] uncertainties,  without assigning any error to the temperature calibration itself; as a consequence, the errors on $T_e$[O~II] may be underestimated, a problem that has been discussed by previous studies \citep[e.g.][]{hagele06}.

To determine the abundances of the various ions, we used the extinction-corrected fluxes (listed in Table~\ref{h2_flux}) for the following lines:  He~I~$\lambda$4471, 
He~I~$\lambda$5876,  and He~I~$\lambda$6678 for $He^{+}$, He~II~$\lambda$4686 (when available) for $He^{+2}$, [N II]$\lambda$6548,6584 for $N^+$,  [O III]$\lambda$4959,5007 for $O^{+2}$,
[O~II]$\lambda\lambda$3726,29 and [O~II]$\lambda\lambda$7320,30 for  $O^{+}$, [O~I]$\lambda$6300 and [O~I]$\lambda$6364 for  $O^{0}$,  
[Ne~III]$\lambda$3869 for  $Ne^{+2}$, [S~II]$\lambda\lambda$6716,31 for  $S^{+}$,  [S~III]$\lambda$6312 and [S~III]$\lambda9069+\lambda9531$ for  $S^{+2}$, 
[Ar~III]$\lambda$7136 and   [Ar~III]$\lambda$7751 for  $Ar^{+2}$, and [Fe~III]$\lambda\lambda4986,87$ for $Fe^{+2}$. 
The PyNeb code adopts the He~I emissivities of \cite{porter12,porter13} including collisional excitation, so no correction to the emission line fluxes for this effect 
\citep{clegg87} needs to be applied. The He~I~$\lambda$7065 line, which has a strong contribution from collisional excitation, was not included in the 
computation of the  $He^{+}$; in fact, the uncertainties on the derived $n_e$ values translate into large errors on the  $He^{+}$ abundance due to the 
 strong dependence of the  He~I~$\lambda$7065 emissivity on density  \cite[see e.g. Fig.~4 of] []{porter12}.

To get an estimate of the ion abundance uncertainties, we ran  
the  {\it getIonAbundance} task for  ranges of temperatures, densities, and flux ratios within the $\pm 1 \sigma$ levels, and conservatively adopted the maximum excursion around the nominal abundance value as our error. 
When multiple sets of lines were available for a single ion
(i.e. $He^{+}$, $O^{+}$, $O^{0}$, $S^{+2}$, $Ar^{+2}$), its abundance was computed by averaging all the abundances from the various lines (or line complexes). 
Typically, the standard deviation around the mean abundance from the different lines is lower than or comparable with the error obtained by propagating the individual abundance 
uncertainties; conservatively, we adopted the largest value as our uncertainty on the abundance determinations.

Total element abundances were derived from the abundances of ions seen in the optical spectra using ionization correction factors (ICFs).
For Oxygen, the total abundance was computed as  $O/H= (O^+ + O^{+2})/H^+$. 
From  \cite{izotov06}, the contribution of $O^{+3}$ to the total oxygen abundance is expected to  be  $<$1\%, since $O^+ /(O^+ + O^{+2}) >0.1$ in our H II regions. 
We did not add the contribution from $O^0$ because  it is associated to neutral hydrogen, and almost all the emission in the  [O~I]~$\lambda$6300, 6364 lines comes from photodissociation regions \citep{abel05}.

To compute the abundances of the other elements, we adopted the ICFs from  \cite{izotov06} for the ``high'' Z regime ($12 + \log O/H \ge 8.2$): 
 
 \begin{equation}
 ICF(N^+)=  -1.476v +1.752 + 0.688/v,
  \end{equation}

  \begin{equation} 
 ICF(Ne^{+2}) =  -0.591w+0.927+0.546/w, 
  \end{equation}
  
 \begin{equation}
 ICF(S^+ + S^{+2})=0.178v+ 0.610+0.153/v,
  \end{equation}
 
 \begin{equation}
 ICF(Ar^{+2})= 0.517v+0.763+0.042/v,
 \end{equation}
 
 \begin{equation}
 ICF(Fe^{+2})=-1.377v+1.606+1.045/v
 \end{equation}

 where
 
 \begin{equation}
 v = O^+/(O^+ + O^{+2}),  w = O^{+2}/(O^+ + O^{+2}).
  \end{equation}

In H~II-6, where the He~II~$\lambda4686$ nebular emission line was clearly detected  (see Fig.~\ref{slit10}), the He abundance was computed as $He^+ + He^{+2}$
(with $He^{+2}$ contributing for $\sim$1\% to the total He abundance), while we neglected  the He$^{+2}$ contribution in all the other H~II regions. We notice that a modest nebular He~II $\lambda4686$ emission could be present in  H~II-1, 
superimposed  upon a much stronger Wolf-Rayet broad  emission component (see Section~5 and Table~\ref{h2_flux}); however, since this nebular  He~II 
contribution turns out very small, and is furthermore affected by large uncertainties due to the dominating WR component, we decided to neglect it in the computation of the total He abundance for  region H~II-1.

To derive the total abundance of He,  one should in principle account for the ionization structure of the nebula. In fact, the radius of the $He^+$ zone 
can be smaller than the radius of the $H^+$ zone in the case of soft ionizing radiation, or larger in the case of hard radiation. 
In the former case, a correction for unseen neutral helium needs to be applied, resulting in a ionization correction factor $ICF(He^+ + He^{+2})>1$ \citep{izotov07}.
\cite{izotov13} ran photoionization models to investigate the behavior of $ICF(He^+ + He^{+2})$ as a function of metallicity and excitation parameter $w$. According to their 
``high'' Z models, $ICF(He^+ + He^{+2})$ approaches unity for large $w$ values and  $ICF \sim$1.03 for $w\sim$0.3. Since $w>0.3$ in our H~II regions, 
it is reasonable to neglect this correction.

\subsection{Planetary Nebulae \label{pn_section}}

\begin{center}
\begin{deluxetable*}{lcccc}
\tabletypesize{\scriptsize}
\tablecaption{Derived properties for PNe in NGC~4449 \label{pn_abund}}
\tablehead{
\colhead{Property} & \colhead{PN-1} &  \colhead{PN-2}  &  \colhead{PN-3}  &  \colhead{PN-5} 
}
\startdata
R.A. [J2000] & 12:28:04.126 & 12:28:03.540 & 12:28:03.972  & 12:28:13.950  \\
Dec. [J2000] & +44:04:25.14 & +44:04:34.80 & +44:05:56.78 & +44:07:45.29 \\
$R/R_{25}$ & 0.57 & 0.56 & 0.43 & 0.71  \\
\hline
$n_e [cm^{-3}]$ &   $600_ {-400}^{+500}$ &   $300_ {-300}^{+400}$ &  $1300_{-700}^{+1000}$ &  $-$ \\
$T_e (O^{++})$ [K] &   12200 $\pm$  900 &   14000 $\pm$   1000 &   13000 $\pm$  1000 &   13000 $\pm$  1000 \\
\hline
$(O^+/H^+)\times10^5$  &   0.4 $\pm$  0.1 &   0.9 $\pm$  0.3 &   1.2 $\pm$  0.4 &   0.07 $\pm$  0.02 \\
$(O^{++}/H^+)\times10^4$  &   2.2 $\pm$  0.5 &   1.6 $\pm$  0.4 &   1.7 $\pm$  0.4 &   1.9 $\pm$  0.4 \\
$12+ \log(O/H)$ &   8.3 $\pm$  0.1 &   8.3 $\pm$  0.1 &   8.4 $\pm$  0.1 &   8.3 $\pm$  0.1 \\
\hline
$He/H$ &    0.08 $\pm$   0.02 &    0.09 $\pm$   0.02 &    0.08 $\pm$   0.01 &    0.10 $\pm$   0.02 \\
\hline
$(N^+/H^+)\times10^6$  &   3.4 $\pm$  0.9 &   3.4 $\pm$  0.9 &   7 $\pm$  2 &   1.2 $\pm$  0.3 \\
$12+ \log(N/H)$ &   8.2 $\pm$  0.1 &   7.9 $\pm$  0.1 &   8.2 $\pm$  0.1 &   8.5 $\pm$  0.1 \\
\hline
$(Ne^{++}/H^+)\times10^5$  &   4 $\pm$  1 &   2.9 $\pm$  0.9 &   2.7 $\pm$  0.8 &   3 $\pm$  1 \\
$12+ \log(Ne/H)$ &   7.6 $\pm$  0.1 &   7.5 $\pm$  0.1 &   7.6 $\pm$  0.1 &   7.5 $\pm$  0.1 \\
\hline
$(S^{+}/H^+)\times10^7$  &   4 $\pm$  1 &   2.9 $\pm$  0.8 &   3.2 $\pm$  0.9 & $-$ \\
$(S^{++}/H^+)\times10^6$  &   1.8 $\pm$  0.7 &   2.0 $\pm$  0.8 &   1.6 $\pm$  0.7 & $-$ \\
$12+ \log(S/H)$ &   6.8 $\pm$  0.1 &   6.6 $\pm$  0.2 &   6.6 $\pm$  0.2 &  $-$ \\
\hline
$(Ar^{++}/H^+)\times10^7$  &   4.5 $\pm$  1.3 &   8 $\pm$  2 &   4 $\pm$  1 & $-$ \\
$12+ \log(Ar/H)$ &   5.9 $\pm$  0.1 &   6.2 $\pm$  0.1 &   5.9 $\pm$  0.1 &  $-$ \\
\enddata
\end{deluxetable*}
\end{center}

Densities and temperatures of PNe were derived using the density-sensitive [S~II]$\lambda6716/\lambda6731$ line ratio, and 
the temperature-sensitive [O~III]$\lambda4363/\lambda\lambda4959+5007$ ratio. 
Figure~\ref{pne_2d} shows that the [O~III]$\lambda$4363 line was detected in all five PNe. We excluded PN~4 from our study since its 
2D spectra appeared highly contaminated from diffuse emission due to nearby H~II regions. 
For all the other PNe, we obtained [O~III] temperatures in the range 12,000 - 14,000 K. Densities were derived  for PN~1, PN~2 and PN~3 with large errors 
($n_e = 600^{+500}_{-400}$,  $300^{+400}_{-300}$,  $1300^{+1000}_{-700} \ cm^{-3}$),   while  for PN~5, where the [S~II]$\lambda\lambda6716,31$ lines were not detected, 
we assumed  $n_e =$ 1000 cm$^{-3}$.  
Following the same approach adopted by many studies in the literature \citep[e.g.][]{stasi13,idiart07}, we assumed that the temperature of all the ions is equal to $T_e [O III]$. 
In fact,  empirical relations between $T_e$[O III]  and $T_e$[ N II] derived in the literature for PNe \citep{kaler86,kb94,wl07} show important spreads and have different trends.

For all the PNe but PN~4, we derived the abundances  of  $He^+$, $N^+$,  $O^{+}$,  $O^{+2}$, and $Ne^{+2}$ from the He~I~$\lambda$5876, 
 [N~II]$\lambda$6548,84, [O~II]$\lambda\lambda$3726,29,
[O~III]$\lambda$4959, 5007 and  [Ne~III] $\lambda$3869 lines;   $S^{+}$,  $S^{+2}$, and $Ar^{+2}$ abundances were derived for PN~1, PN~2, and PN~3 from the 
[S~II]$\lambda\lambda$6716,31, [S~III]$\lambda9069$, and [Ar~III]$\lambda$7136 lines;  $He^{+2}$ abundances were obtained only 
for PN~2 and PN~3 from the He~II~$\lambda$4686 line. For these two PNe, the total He abundance was computed as  $He^+ + He^{+2}$, while the 
$He^{+2}$ contribution was omitted for PN~1 and PN~5.

To derive total element abundances, we used the ICFs from \cite{kb94}, hereafter KB94. 
For oxygen, the correction due to unseen $O^{+3}$ is:

\begin{equation}
ICF(O^+ + O^{++}) = \left( \frac{He^+ + He^{+2}}{He^+}\right)^{2/3}.
\end{equation}

The absence of He~II lines in PN~1 and PN~5 indicates negligible $He^{+2}$ abundances, and thus we do not expect an important amount of  $O^{+3}$ in these two objects. 
On the other hand, we derived  $ICF(O^+ + O^{++})\sim1.1$ and $\sim$1.4 for  PN~2 and PN~3, respectively.

For the other elements, the KB94 ICFs are:

\begin{equation}
ICF(N^+)=\frac{O}{O^{+}}, 
\end{equation}

\begin{equation}
ICF(Ne^{++})=\frac{O}{O^{+2}}, 
\end{equation}

\begin{equation}
ICF(Ar^{++})=1.87, 
\end{equation}

\begin{equation}
ICF(S^+ + S^{++})=  \left[ \left( 1 - \frac{O^+}{O} \right)^3 \right]^{-1/3}
\end{equation}

The derived PN abundances are provided in Table~\ref{pn_abund}. 

 Recently, a new set of ICFs was presented by \cite{delgado14} (hereafter DMS14). The new ICFs from DMS14 are based on a large grid of photoionization models and provide significant improvement with respect to previous ICFs for PNe. We present in the  Appendix a description of the new ICFs  and evaluate the effect on the derived PN abundances. We find that oxygen is very  little affected by the new ICFs, with abundance differences of only a few percent in dex. For the other elements, i.e. N, Ne, S and Ar, 
the difference in abundance is larger than for O, but always within $\sim$0.1 dex, comparable to the errors associated with our derived abundances.

\section{Wolf-Rayet features \label{section_wr}}

According to stellar evolution models, the most massive stars \citep[$M\gsim20 M_{\odot}$, for a solar metallicity model with rotation, see][]{mm05} 
evolve into the Wolf-Rayet (WR) phase $\approx$2-5 Myr after their birth.  A WR star is  a bare stellar core that has lost the main part of its H-rich envelope via strong winds \citep[e.g.][]{maeder91,maeder95}, or by mass transfer through the Roche Lobe in close binary systems \citep[e.g.][]{chiosi86}.
The  characteristic features of WR stars are broad emission lines of helium, nitrogen, carbon and oxygen formed in the high-velocity wind region 
surrounding the hot stellar photosphere. 
In the optical, two main emission features can be identified: the so-called {\it blue bump} around 4600 - 4700 \AA, and the 
 {\it red bump} centered around 5800 \AA, usually fainter than the blue bump. The blue bump is due to the blend of a broad He~II~$\lambda$4686 \AA  \  emission feature 
 with  N~III~$\lambda$4640 \AA \ (WN subtype) or with C~III$~\lambda$4652 \AA \ (WC subtype). The red bump is due to the  C~IV~$\lambda$5808 \AA \ emission in WC stars, and is more rarely observed than the blue bump.  The WN and WC subtypes represent an evolutionary sequence since the ejection process is believed to occur in succession, first exposing 
the surface mainly composed of the nitrogen-rich products of the CNO cycle (WN stars), and later the carbon-rich layer due to He-burning (WC and WO)  
\citep[][and references therein]{dray03b}.

 \begin{figure}
\epsscale{1.2}
\plotone{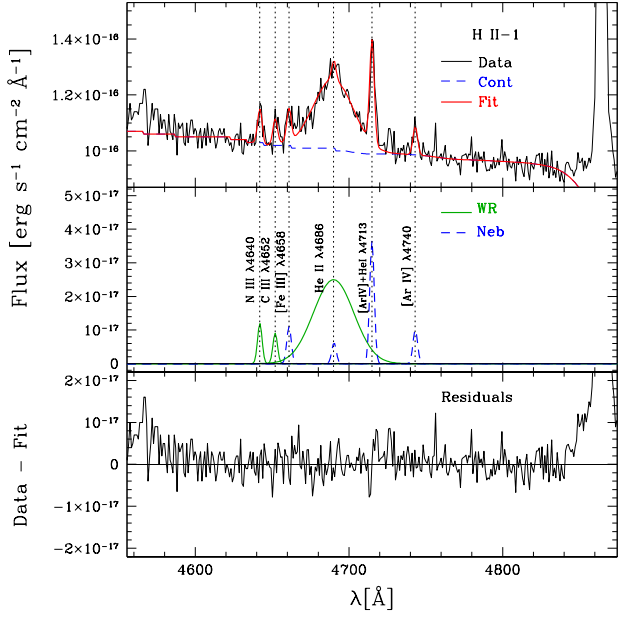}
\caption{Portion of the spectrum of H~II-1 around the region of the Wolf-Rayet blue bump at $\sim$4690 \AA.   {\bf Top panel:} Observed spectrum (thin black line) 
and total (continuum plus emission lines) fit (red thick line).  The continuum has been modelled with a  Z$=$0.004, 3-4 Myr old SSP, normalized at 
$4770-4840$ \AA (see Section~\ref{section_wr} for details).
 {\bf Middle panel:} fitted emission lines (solid green line for WR features, dashed blue line for nebular  narrow emission lines). {\bf Bottom panel:} residual after subtracting the best-fit model. 
\label{wr3}}
\end{figure}

\begin{figure}
\epsscale{1.2}
\plotone{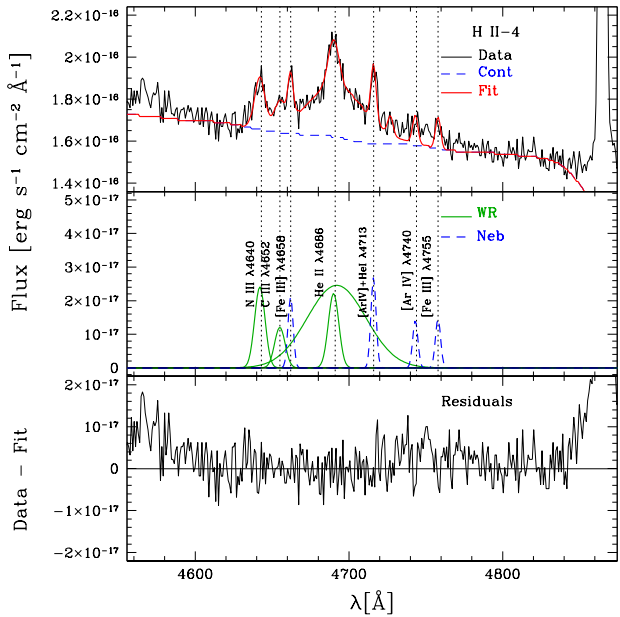}
\caption{Same as in Figure~\ref{wr3} but for H~II-4.  
\label{wr8}}
\end{figure}

\begin{figure}
\epsscale{1.}
\plotone{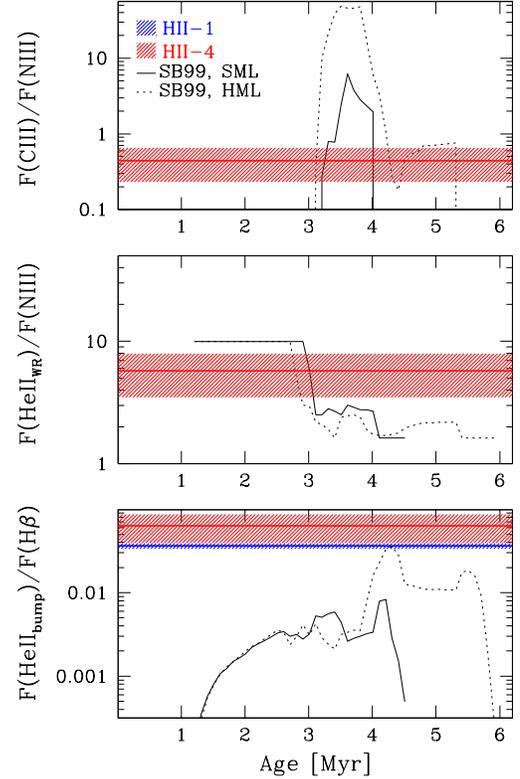}
\caption{Wolf-Rayet features detected in regions H~II-1 and H~II-4 against Starburst 99 (SB99) models. The models are based on the Geneva tracks with metallicity $Z=0.008$ with a standard mass-loss prescription (solid line) and with a high mass-loss prescription (dotted line). From top to bottom: F(C~III)/F(N~III), F(He~II$_{WR}$)/F(N~III), and 
F(He~II$_{WR}$)/F(H$\beta$) ratios  versus age. Notice that the F(He~II$_{WR}$) flux refers to the broad WR component without the contribution from the nebular He~II emission
line.  
 \label{wr_sb99}}
\end{figure}

We detected the blue bump in two (H~II -1 and H~II -4) out of the six H~II regions studied in NGC~4449. On the other hand, the wavelength range of the red bump falls close to the region of low sensitivity of the blue and red MODS detectors, preventing us to draw conclusions about the presence of this feature.  
 The blue bump spectral region in  H~II-1 and H~II-4 was modeled  by simultaneously fitting the WR features (N~III~$\lambda$4640, C~III~$\lambda$4652, He~II~$\lambda$4686) and the nebular emission lines ([Fe~III] $\lambda$4658, He~II~$\lambda$4686, [Ar~IV]$+$He~I~$\lambda$4713, [Ar~IV]$\lambda$4740) following an  approach similar to that of  \cite{brinch08}. To evaluate the underlying stellar continuum, we performed a spectral energy distribution (SED) fit to the 4000-7000 \AA  \ range (avoiding  the regions contaminated by nebular emission lines) using SSP models from the Padova group \citep{marigo08,chavez09}.
The SED of H~II -1 and H~II -4 turned out to be best reproduced by a Z$=$0.004, 3-4 Myr old population; this result, which we expect to be highly  affected by the age-metallicity degeneracy, is not intended to draw conclusions on the physical properties of the underlying stars, but has the mere purpose of providing a reliable continuum below the bump. 
The fits to the blue bump in regions  H~II-1 and H~II-4  are shown in Figures~\ref{wr3} and ~\ref{wr8}, respectively. We fixed the nebular emission lines to have the same Gaussian widths as the other emission lines in the 4000-6000 \AA\ spectral range (FWHM$\sim$4 \AA), while the FWHMs of the WR features were allowed to vary as free parameters. 

For region H~II-1, the width of the broad He~II~$\lambda$4686  feature is best fitted with a Gaussian FWHM of $\sim$30 \AA, corresponding to a velocity 
$\sigma\sim$800 km s$^{-1}$ \footnote{computed as ${\rm \sigma = \frac{ \sqrt{FWHM^2 -FWHM_{instr}^2 }}{2.35}\times \frac{c}{\lambda}}$, where FWHM and FWHM$_{instr}$ are the measured and the instrumental widths, respectively.};  the presence of a nebular contribution to this line is not well constrained given the large errors (see Table~\ref{h2_flux}). 
Surprisingly, the N~III~$\lambda$4640 and  C~III~$\lambda$4652 features show widths comparable to those of the nebular emission lines. A similar result was found by \cite{smith16} 
when fitting the WR blue bump for cluster \#5 in the blue compact dwarf galaxy NGC~5253; as they noticed, the narrow widths derived for N~III and C~III would suggest that these lines are likely to be nebular in origin, although detecting these transitions is unusual. 
 
 For region H~II-4, it was necessary to assume two broad emission components to obtain a satisfactory fit to the He~II~$\lambda$4686  feature: our best fit provides two Gaussians  with FWHMs  of $\sim$8 \AA \ and $\sim$ 45 \AA, corresponding to velocities of $\sim$200 km s$^{-1}$  and $\sim$1200 km s$^{-1}$,  while the N~III and C~III features are reproduced by two Gaussians with FWHM$\sim$8 \AA. Notice that wind velocities derived in WR stars can be as high as $\sim$2500 km s$^{-1}$ \citep[e.g.,][]{wind}.

We show in Figure~\ref{wr_sb99} a comparison of the WR features in  H~II-1 and H~II-4 against the Starburst99 instantaneous burst models \citep{sb99}.
 For region H~II-1, we do not show the ratios  involving the  C~III and N~III emission because, as previously discussed, the narrow widths derived for these lines 
would suggest that they are nebular in origin and not due to WR stars.
The plotted models were computed with the Geneva $Z=0.008$ stellar tracks \citep{geneva,meynet94} assuming either a standard mass-loss (SML) or a high mass-loss (HML) rate.  It is well known that models of massive stars suffer uncertainties due to rotation \citep[e..g][]{mm05} and to possible binary evolution \citep[e.g.][]{eldridge08,vanb07}. 
Models including rotation have been computed by the Geneva group for some metallicities, but are not available for $Z=0.008$.

Figure~\ref{wr_sb99} shows that WR features (mainly C~III, N~III and He~II  $\lambda$4686) are visible during a limited age range, from $\approx$1 Myr to $\approx$4.5 Myr in the case of SML models, and up to $\approx$6 Myr for HML models. While the N~III emission is always present during the WR phase, the C~III emission due to the later appearance of WC stars is observed only after $\approx$3 Myr (top panel of Fig.~\ref{wr_sb99}). This holds both for the SML and for the HML models, although we notice that the latter ones imply significantly higher F(C~III)/F(N~III) ratios.  The presence of broad  C~III emission in H~II-4 indicates the existence of a WR population at least $\approx$ 3 Myr old. The behaviour of the  F(He~II  $\lambda$4686)/F(N~III) ratio is displayed in the middle panel of Fig.~\ref{wr_sb99}. The SML and HML models predict a moderate difference in this ratio. 
The  F(He~II)/F(N~III) ratio is as high as $\sim$10 in the earliest phases, and then rapidly decreases after $\sim$3 Myr reaching down to $\sim$2. For H~II-4, we derive a ratio 
of $\sim 6 \pm 2$, compatible with an age of $\sim$3 Myr. 
Finally, the bottom panel of Fig.~\ref{wr_sb99} shows the  evolution of  the F(He~II  $\lambda$4686)/F(H$\beta$) ratio, which is proportional to the number of WR stars over the number of ionizing OB stars. Here the difference between the two sets of models is striking: while the SML largely under-predict the number of WR over OB stars, the HML provide a very satisfactory match for ages older than $\approx$4 Myr  for both  H~II-1 and H~II-4. This is in agreement with past studies in  the literature showing that the observed properties of WR stars require, in absence of stellar models with rotation,  the inclusion of an enhanced mass loss \citep[e.g.][]{geneva,schaerer92}. 
 However, we caution that the difficulty of the models  in reproducing the strength of the blue bump could be due to the presence of stars other than ``classical'' WR, such as  
massive, mass losing core-hydrogen burning stars close to the main sequence, a stellar phase not  yet accounted for in the evolution models \citep{iizw40}.
Using the HML models we derive, from the observed He~II~$\lambda$4686 flux,  a number of WR stars of $\approx4$ and $\approx8\pm2$ in regions  H~II-1 and H~II-4, respectively.

\section{Results on the Chemical Abundances and Abundance Ratios}

\begin{figure*}
\epsscale{1}
\plotone{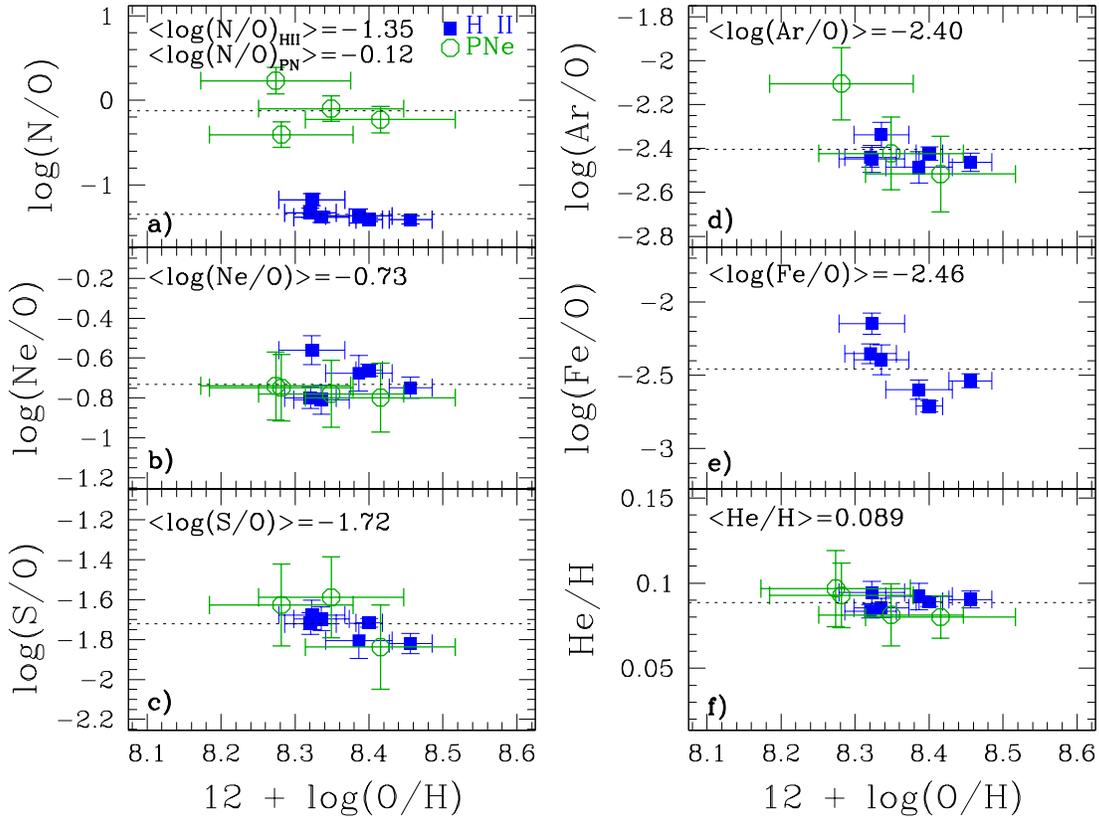}
\caption{Abundance ratios as a function of total oxygen abundance. Solid and open symbols are for H~II regions and PNe, respectively. 
Within each panel, the label and the dotted horizontal line indicate the average value for the combined H~II region and PN data; only in panel a), separate $\log(N/O)$ mean values for H~II regions and PNe are provided.  
\label{ratios}}
\end{figure*}

Element abundances derived in H~II regions and PNe are given in Tables~\ref{h2_abund} and \ref{pn_abund}.  
The oxygen abundance interval  spanned by our sample is ${\rm 8.3 \pm 0.1 \leq12 + \log(O/H)\leq8.46\pm0.03}$, with average O abundances for 
H~II regions and PNe of 8.37 $\pm$ 0.05 and   8.3 $\pm$ 0.1, respectively. The H~II region results are consistent with previous literature measurements  based on the direct temperature method:  \cite{talent80}, see also \cite{skillman89}, derived an average oxygen abundance of  $12 + \log(O/H)=8.31 \pm 0.07$ for H~II regions in NGC~4449; later, \cite{berg12} obtained new MMT spectra for the bright H~II knot located a few arcsec to the south of our H~II-6 region (slit~10 in  Figure~\ref{image}), and found  an average abundance of $12 + \log(O/H)=8.20 \pm 0.08$, consistent with our value of  $12 + \log(O/H)=8.32 \pm 0.04$ within the errors.

The trend of H~II region and PN abundance ratios versus total oxygen abundance is illustrated in Figure~\ref{ratios}. 
We find that, within the oxygen interval spanned by our data, the Ne/O, S/O and Ar/O ratios 
are similar for the H~II region and PN sub-samples, and are compatible, within the errors, with a constant trend as a function of oxygen 
(Figs~\ref{ratios}b,~\ref{ratios}c and ~\ref{ratios}d). This behaviour is consistent with what is commonly observed in other studies 
\citep[e.g.,][]{richer07,bresolin10,stasi13,anni15}. The explanation is that  $\alpha$-elements are all synthesized by massive stars on similar timescales, thus their abundance variations  occur in lockstep, maintaining the corresponding ratios constant. The similarity between the Ne/O, S/O and Ar/O ratios measured in PNe and H~II regions is not surprising 
since $\alpha$ - elements are not significantly affected during the evolution of low and intermediate mass stars. 
In NGC~4449, the  average values of $<\log(Ne/O)>\simeq-0.73$, $<\log(S/O)>\simeq-1.72$, and $<\log(Ar/O)>\simeq-2.40$ are consistent with typical abundance ratios derived in H~II regions of star-forming dwarf galaxies \citep[see Figure~13 of][for NGC~1705]{anni15}.

The major abundance difference between  H~II regions and PNe is observed for Nitrogen (Figure~\ref{ratios}a), showing a dichotomy in the N/O distribution.
Our H~II regions exhibit an average $<\log(N/O)_{H~II}>\simeq-1.35$, comparable to values measured in H~II regions of luminous dwarf galaxies \citep[e.g.][]{ks96,berg12} 
and of spirals for similar oxygen abundances as in NGC~4449 \citep[e.g.][]{bresolin09a,berg15,croxall16}. 
On the other hand, our PNe are more than $\simeq$1 dex enhanced in N with respect to H~II regions, with an average $<\log(N/O)_{PNe}>\simeq-0.12$. 
 This is not unusual since previous studies have shown that PNe in nearby galaxies are enriched in N with respect to H~II regions; there is a large scatter in the amount 
of the enrichment, with N/O ratios from close to those measured in  H~II regions up to $\approx$ 1 dex higher \citep[e.g.][]{pena07,richer07,richer08,bresolin10,stasi13,garcia16}.
Highly N-enriched PNe are found both in star-forming galaxies and in quiescent early-type galaxies, where star formation ceased a long time ago \citep[e.g.][]{richer08}.
However, we notice that our PNe, despite their significant N enrichment, do not appear to be enhanced in He (Fig.~\ref{ratios}f); to our knowledge, there are no models that can simultaneously increase the N abundance by a factor of 10 and leave He unchanged \citep[see e.g.][]{karakas07}. A possible explanation is that our derived PN He abundances are
uncertain because the detected  He~I $\lambda5876$ line is significantly fainter than two nearby sky lines at $\lambda\sim5867$ \AA \ and  $\lambda\sim5890$ \AA.

  From the theoretical point of view, the N/O enhancement in PNe is the natural consequence of nitrogen being mostly synthesized 
 in intermediate mass stars, that are the PN progenitors, and brought to the stellar surface during dredge-up episodes occurring in the RGB and AGB phases;
 a significant  N production may also occur in the most  massive and luminous AGB stars through HBB (see Section~1). 
PNe exhibiting the most extreme  N (and He) abundances, classified as type~I, 
are thought to be the descendants of massive ($>$ 3 \MSUN), relatively young  (age$\lesssim$400 Myr) AGB stars experiencing HBB \citep[e.g.][]{stanghellini06,corradi95}.
\cite{tpp97} proposed $\log(N/O)>-0.42$ and He/H$>0.105$ as an empirical criterion to select type I  systems; three PNe out of four in our sample satisfy the condition in N/O, but their helium abundances are similar to those of H~II regions around He/H$\simeq$0.09 (see Figure~\ref{ratios}f),  which can not be explained with existing models.
  Although the reliability of the derived He abundances for our PNe could be questioned as discussed before, a strong selection bias should be invoked to explain why a fraction as high as 3/4 of our sample derives from massive star progenitors.

\begin{figure*}
\epsscale{1}
\plotone{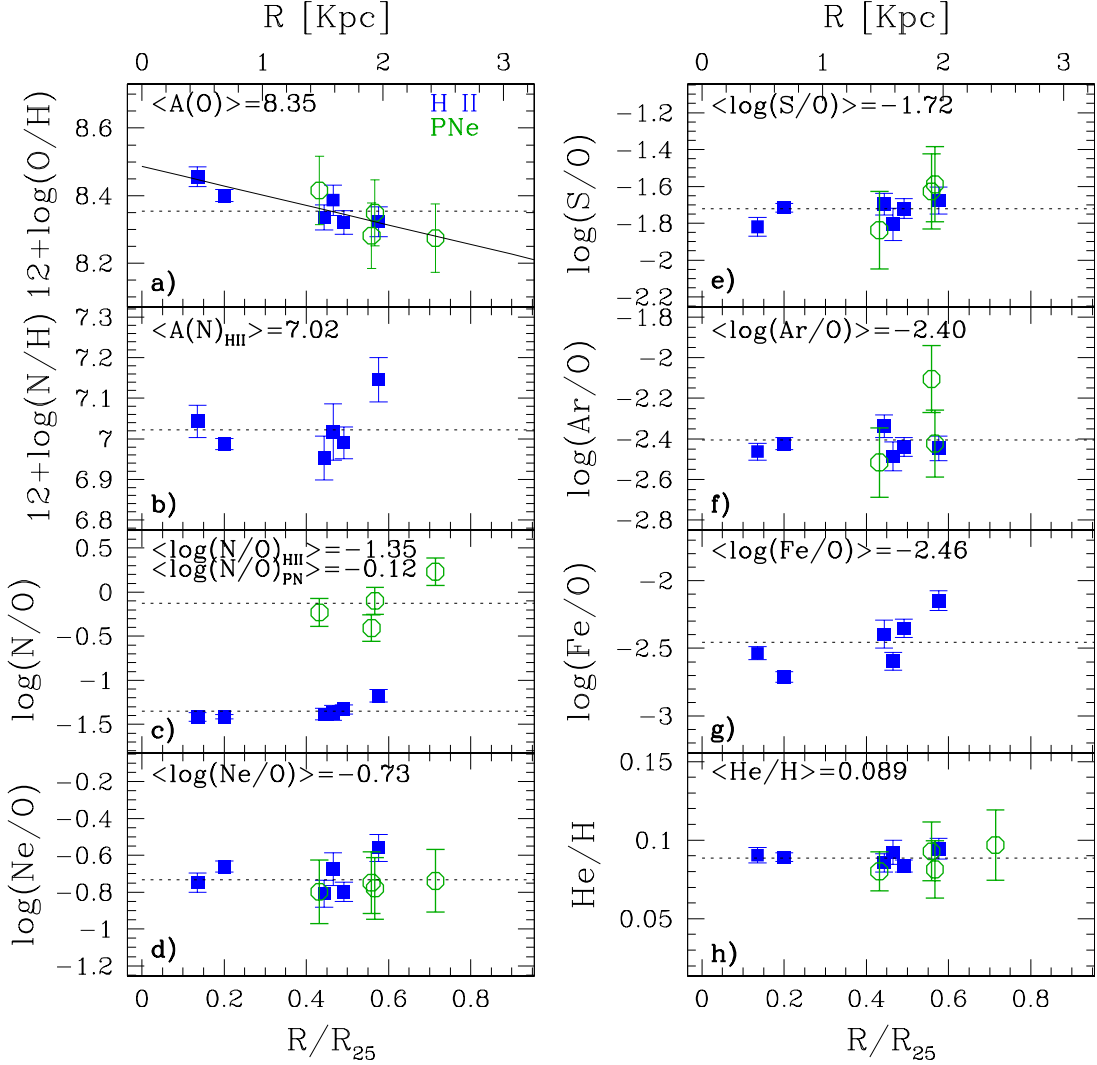}
\caption{Element abundances and abundance ratios as a function of galacto-centric distance $R/R_{25}$, where $R_{25}=3.4$ kpc. The linear galacto-centric scale in kpc is also indicated on top. Solid and open symbols are for H~II regions and PNe, respectively. Within each panel, the label and the dotted horizontal line indicate the average value for the combined H~II region and PN data; only in panel c), separate $\log(N/O)$ mean values for H~II regions and PNe are provided. In panel b), in order to better visualise
the range of nitrogen variation as a function of radius  for H~II regions, we do not include PNe, whose N abundance is so high (see Table~\ref{pn_abund}) that the ordinate scale would be too compressed. 
\label{grad}}
\end{figure*}

We find no significant trends in the N/O vs O/H distribution of Figure~\ref{ratios}a. 
 Historically, the absence of a trend in N/O vs O/H for low-metallicity systems was taken as the first indication that nitrogen cannot be a pure secondary element.  Primary elements (like He, C and O)  are those whose production can start already in stars with primordial initial chemical composition. Secondary elements are those that can be synthesized only if the star already contains their seed elements at birth or at the evolutionary phase when the physical conditions allow the element to be synthesized. As a natural consequence, the abundance of secondary elements is predicted to increase as the square of primary ones \citep{tinsley80}. Hence, had N been of fully secondary nature, its ratio to oxygen should have been proportional to the O abundance. Since N/O is instead always found to be quite independent of O/H in the nebulae of individual galaxies \citep[see, e.g.][]{diaz86}, a significant fraction of N must be of  primary  origin. In practice, its nature depends on whether the C used to synthesize N was produced by previous nuclear reactions in the same star or was already present in its initial chemical composition. An inspection of the chemical yields computed for low- and intermediate-mass stars \citep[see e.g.][]{karakas10,ventura13,vincenzo16} shows that N is mainly of secondary origin above a metallicity $\approx$ half of solar, and is mainly produced by M$\gsim 3.5 M_{\odot}$ stars experiencing HBB  \citep[see also Figs 1 and 2 in][]{romano10}. In massive stars N can have a significant primary  origin if their metallicity is low and they rotate sufficiently fast \citep{mm02}.

Finally, the distribution of the Fe abundance,  derived only for H~II regions, is shown in  Figure~\ref{ratios}e. 
The Fe/O ratio shows no clear trend with oxygen: in fact, although there is a hint for an Fe/O decrease with increasing O abundance, in agreement with the behaviour revealed 
by other studies \citep[e.g.,][]{izotov06,guseva11,delgado11} and commonly interpreted as Fe depletion into dust grains, the range in oxygen abundance probed by our data is likely too small  to claim a clear trend. In particular, we notice that the data in the Fe/O versus O plane shown by \cite{delgado11} span an oxygen interval of  almost $\sim$2 dex, compared to a range of only $\sim$0.2 dex for the NGC~4449 data. Our Fe and O abundances  nicely fall upon the region occupied by  H~II regions in the  \cite{delgado11} plot. 

\section{Results on the abundance spatial distributions}

The behaviour of  element abundances as a function of galacto-centric distance is shown in Figure~\ref{grad}.
For immediate comparison between NGC~4449 and other literature studies, the radial distance is expressed in terms of $R/R_{25}$, where the optical isophotal radius $R_{25}=3.1'$ is taken from \cite{pilyugin15}, and corresponds to $\approx$3.4 kpc at NGC~4449's distance of $\approx$ 3.8 Mpc. 

 Figure~\ref{grad}a shows that H~II regions and PNe exhibit similar oxygen abundances in the galacto-centric distance range of overlap, despite the fact that they 
represent different evolutionary stages of the galaxy. The same result was found for other star-forming dwarf and spiral galaxies by previous studies 
reporting similar abundances for H~II regions and bright PNe  \citep[e.g.][]{richer93,magrini05,richer07,bresolin10,stasi13}. 
 From the analysis of the CMD of the resolved stars, we know that NGC~4449 has been actively forming stars over the last 1 Gyr \citep{anni08,mcquinn10,sacchi17}; therefore we would expect a significant chemical enrichment since the PN progenitors were formed, i.e. since $\sim$100 Myr ago or more.
On the other hand, the similarity in oxygen abundance between H~II regions and PNe suggests that this is not the case.
The galactic outflow observed in NGC~4449  \citep{dceca97,summers03,bomans14} may have played an important role, 
expelling the recently produced $\alpha$-elements. Accretion of metal poor gas or acquisition of smaller gaseous systems may have also contributed to dilute the metals in the ISM.

Figure~\ref{grad}a illustrates that both H~II regions and PNe show a well-defined oxygen  gradient. We thus combined H~II and PN data to infer a global relation from a linear least-squares fit:  
\begin{equation}
{\rm  12 + \log(O/H)= -0.29 (\pm 0.06) \times R/R_{25}+ 8.49 (\pm 0.030)}. 
 \end{equation}
 
\noindent  Hence our gradient is ${\rm  (-0.29\pm0.08) \ dex \ R_{25}^{-1}}$ \footnote{or ${\rm (-0.09\pm0.05) \ dex \ R_{Kpc}^{-1}}$ if the actual galactocentric distance R in Kpc is considered.}, in good agreement with the value of ${\rm (-0.23\pm 0.03) \ dex \ R_{25}^{-1}}$ obtained 
 by  \cite{pilyugin15}, once we correct his O/H gradient  for the larger distance adopted in his work (${\rm D_{NGC~4449}\approx4.1 \ Mpc)}$. On the other hand, we notice that the central extrapolated oxygen abundance derived by  \cite{pilyugin15} is  $12 + \log(O/H)=8.26\pm0.01$,  more than 0.2 dex lower than ours. 
 
The presence of metallicity gradients in late-type dwarf galaxies has been widely discussed in the literature. 
For a long time, dIrrs and BCDs have been considered to have nearly constant radial trends, at least within the observational uncertainties 
\citep[e.g.,][]{kobul97,croxall09,lagos13,haurberg13}. Two possible explanations have been proposed for this behaviour:   (a) the ejecta from stellar winds and supernovae are dispersed and mixed across the ISM on short timescales, of the order of $\lesssim10^7$ year; (b) freshly synthesized elements remain unmixed with the surrounding ISM and reside in a hot $10^6$ K phase or a cold, dusty, molecular phase \citep{kobul97}. However, detections of negative metallicity gradients from stars and H~II regions have been reported in the literature for the dIrr NGC 6822  \citep{venn04,lee06},  and spectroscopic studies of individual RGB stars have shown slightly negative gradients in [Fe/H] for the SMC, the LMC, and the dIrr WLM \citep[][and references therein]{leaman14}. Very recently,  our study of  the BCD NGC~1705 \citep{anni15} and the study by  \cite{pilyugin15} 
showed that negative nebular metallicity gradients  are indeed  present in late-type dwarf galaxies; our results for NGC~4449 reinforce this scenario.  
We suspect that these recent studies were able to reveal gradients  previously undetected thanks to the high-quality data implied, allowing for much smaller uncertainties on the element abundance determinations.

Figures~\ref{grad}d, \ref{grad}e, and \ref{grad}f show a constant trend of the Ne/O, S/O, and Ar/O abundance ratios with galacto-centric distance, 
which, as described in Section~6, is expected because $\alpha$-elements are all synthesized by massive stars on similar timescales and vary in lockstep. 
Also the He abundance remains constant with galacto-centric distance (Figure~\ref{grad}h), in agreement with the absence of any trend of He with oxygen in Figure~\ref{ratios}f. 
On the other hand, Fe/O decreases with increasing galacto-centric distance (Figure~\ref{grad}g), which is expected from the presence of an oxygen radial gradient and from the fact that Fe/O decreases with increasing O  in Figure~\ref{ratios}e.

 Figure~\ref{grad}b and ~\ref{grad}c  show the radial distribution of the N abundance and of the N/O ratio, respectively. 
The behaviour of nitrogen in Figures~\ref{grad}b and ~\ref{grad}c deserves particular discussion. In fact, despite the existence of a well-defined oxygen metallicity gradient in  
Figure~\ref{grad}a, no clear trend of the N abundance with galacto-centric distance is observed for H~II regions.  This behaviour is surprising, since we expect that the present ISM in NGC~4449 contains a significant component of secondary N   
\citep[see e.g.][]{vincenzo16}, implying that an oxygen metallicity gradient should be accompanied by a nitrogen gradient at least as steep \citep[see for instance studies of spiral galaxies, e.g.][]{diaz86,bresolin09a,croxall15}. 

Given the evidence for a conspicuous  population of WR stars in NGC~4449 
\citep[][see also Section~5]{mk97,bietenholz10,Srivastava14,sokal15}, local pollution from WR ejecta enriched in N is an attractive possibility to explain the observed  behaviour. Although significant amounts of N and C  are expected to be injected by WR stars on theoretical grounds \citep[e.g.][]{chiosi86,dray03a,mm05},  
observational results have been discrepant so far, suggesting either the absence \citep{ks96,kob99} or the presence  \citep{k97,lopez07,lopez11,james11,james13} of localised metal enrichment by massive star ejecta.  
A proposed explanation for these ambiguous results is that an N/O excess is observed only after a recently completed WR phase, when the WR features are weak and the ejecta have cooled. In this picture, strong WR features trace very young regions where the stellar ejecta are still in a hot (T$\approx 10^6$ K) phase and do not show up in optical spectroscopy of H~II regions \citep{tenorio96,wofford09}. This scenario is supported by a study of a large galaxy sample from the SDSS \citep{brinch08} where an excess in N/O is found for WR galaxies with EW(H$\beta$)$<$100 \AA  \ (i.e. burst ages $\gsim$6 Myr), while WR and non-WR galaxies do not show difference in N/O for EW(H$\beta$)$>$100 \AA \ (i.e. burst ages $\lesssim$6 Myr). Indeed, the two regions with strong WR features in our sample (H~II-1 and H~II-4) do not exhibit particularly high N/O values, while H~II-6, which does not have WR features 
but is located in a region of very active SF, presents an N/O excess. Whether there is a correlation in NGC~4449 between the age of the burst and the N abundance for the individual H~II regions will be investigated in a forthcoming paper \citep{sacchi17} based on UV LEGUS data \citep{legus}.

\section{Distinctive properties of the PN population  \label{pn_prop} }

Further insights on the overall evolutionary status of the PNe in NGC~4449 can be attained 
by considering the full sample of 28 {\it bona fide} candidates examined either with
spectroscopic or with photometric observations (see Table~\ref{pne_total}).
The statistics is affected by a selection bias, since our PN detection is restrained 
only to the most active objects with prominent [O~III] emission at 5007~\AA\ (for the 
target to clearly stand out in the $V$ or F502N band frames, compared to $B$ and $I$ imaging). 
Notice that younger stars do not produce brighter [O~III] planetary nebulae; 
however, at  NGC~4449's metallicity and lower, the highest luminosity that a PN can attain 
increases with increasing oxygen abundance \citep{dopita92,richer93}. Therefore, we 
can not exclude that our selection criterium has picked-up only the most oxygen-rich PNe 
in NGC~4449.

Within the biased and limited size of our sample, a preliminary, yet useful, estimate
of the luminosity-specific PN number density $\alpha = N_{PN}/L_{gal}$ 
\citep{jacoby80} may be attempted. This parameter directly relates the 
amount of light in a stellar system to be associated to any observed PN sample, 
and it closely traces the distinctive evolutionary properties of the underlying 
stellar population in the parent galaxy.
For this task we first require an estimate of NGC~4449 bolometric luminosity,
followed by a quantitative assessment of the completeness factor of our PN counts.

\begin{center}
\begin{deluxetable*}{lccc|lccc}
\tabletypesize{\scriptsize}
\tablecaption{The total sample of 28 PNe.\label{pne_total}}
\tablehead{
\colhead{ID}   & \colhead{m$_{F555W}$}  & \colhead{m$_{F502N}$} & \colhead{$(M-M^*)_{[OIII]}$} &
\colhead{ID}   & \colhead{m$_{F555W}$} &  \colhead{m$_{F502N}$}  & \colhead{$(M-M^*)_{[OIII]}$} \\
\colhead{} & \colhead{[Vega mag]} &  \colhead{[Vega mag]}  & \colhead{[mag]} &    
\colhead{} & \colhead{[Vega mag]}  & \colhead{[Vega mag]}   & \colhead{[mag]} \\    
}
\startdata
   {\bf PN-1} & 23.86 & $-$ & 0.82 & {\bf PN-15} & 25.41 & 22.85 & 2.36\\ 
   {\bf PN-2} & 24.39 & $-$ & 1.01 & {\bf PN-16} & 25.49 & 23.10 & 2.44\\ 
   {\bf PN-3} & 24.46 & 22.03 & 1.25 & {\bf PN-17} & 24.57 & 22.00 & 1.52\\ 
   {\bf PN-4} & 24.78 & 22.6 & 0.57 & {\bf PN-18} & 23.82 & 21.85 & 0.77\\ 
   {\bf PN-5} & 24.26 & $-$  & 1.01 & {\bf PN-19} & 23.38 & 21.83 & 0.33\\ 
   {\bf PN-6} & 24.33 & 22.02 & 1.28 & {\bf PN-20} & 23.63 & 21.60 & 0.58\\ 
   {\bf PN-7} & 24.47 & 22.38 & 1.42 & {\bf PN-21} & 24.27 & 22.69 & 1.22\\ 
   {\bf PN-8} & 25.11 & 22.72 & 2.06 & {\bf PN-22} & 24.89 & 22.19 & 1.84\\ 
   {\bf PN-9} & 24.62 & 22.48 & 1.57 & {\bf PN-23} & 24.34 & 21.89 & 1.29\\ 
  {\bf PN-10} & 24.51 & 22.18 & 1.46 & {\bf PN-24} & 23.29 & 21.30 & 0.24\\ 
  {\bf PN-11} & 24.20 & 21.85 & 1.15 & {\bf PN-25} & 25.15 & 23.16 & 2.10\\ 
  {\bf PN-12} & 23.96 & 21.56 & 0.91 & {\bf PN-26} & 24.70 & 21.99 & 1.65\\ 
  {\bf PN-13} & 23.93 & 21.57 & 0.88 & {\bf PN-27} & 23.09 & 20.75 & 0.04\\
  {\bf PN-14} & 25.59 & 22.93 & 2.54 & {\bf PN-28} & 24.97 & 22.30 & 1.92\\ 
\enddata
\tablecomments{Apparent m$_{F555W}$ and m$_{F502N}$ magnitudes derived for our sample of 28 PNe from 
HST/ACS data. The $(M-M^*)_{[OIII]}$ magnitude difference for PN-1 to PN-5 is 
obtained from the observed $[OIII]$ fluxes listed in Table~\ref{pn_flux} 
(corrected for the distance modulus), by assuming $M^*_{[OIII]} = -4.36$ as  
the PN cutoff-magnitude. For the remaining objects, we nominally assume 
$m_{[OIII]} = m_{F555W}+0.5$, as explained in the text.}
\end{deluxetable*}
\end{center}

\begin{figure}
\includegraphics[width=\hsize]{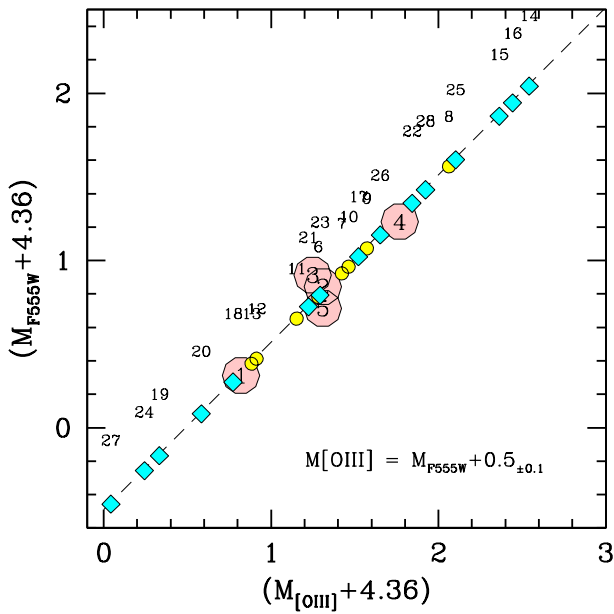}
\caption{The $M_{F555W}$ magnitudes of the spectroscopic PN sample
are compared to the corresponding $M_{[O~III]}$ values (big dot markers).
All magnitudes have been corrected for the distance modulus and offset by a 
value of  $+4.36$ mag, corresponding to the PNLF  bright  cutoff magnitude ($M^*_{[O~III]}$).
A straight relationship is in place with $M_{[O~III]} - M_{F555W} = 0.5\pm0.1$. 
When applied to the total photometric PN sample 
(diamond markers), this offset allows us to assess the $M_{[O~III]}$ distribution 
of the whole sample of 28 PNe showing that our observations actually probed the bright 
tail of NGC~4449 PNLF, down to $(M_{[O~III]}-M^*_{[O~III]}) \sim 2.5$~mag.
In the plot each nebula is labeled according to its entry ID of Table~\ref{pne_total}.
\label{pne_o3vmag}}
 \end{figure}

The apparent integrated magnitude $V^{N4449}_o = 9.47 \pm 0.3$ and the color
$(B-V)^{N4449}_o = 0.36 \pm 0.07$ of NGC~4449 are taken from the corresponding 
RC3 \citep{rc3} and \citet{gronwall04} entries, assuming a Galactic foreground 
reddening of $E(B-V) = 0.019$ \citep{schlegel98}. A match of these figures with the 
\citet{buzzoni05} Im template galaxy model (see Table~A7 therein)  suggests a bolometric 
correction in the range $(Bol-V) = -0.84 \pm 0.02$, from which an absolute value of 
$M_{bol} = -19.3 \pm 0.3$ and a total bolometric luminosity of 
$L = (4.0 \pm 1.0)\,10^9~L_\odot$ can be obtained, once 
accounting for the galaxy distance modulus, and 
assuming for the Sun $M^{bol}_\odot = +4.72$ \citep{lang80}.

\begin{figure*}
\epsscale{0.8}
\plotone{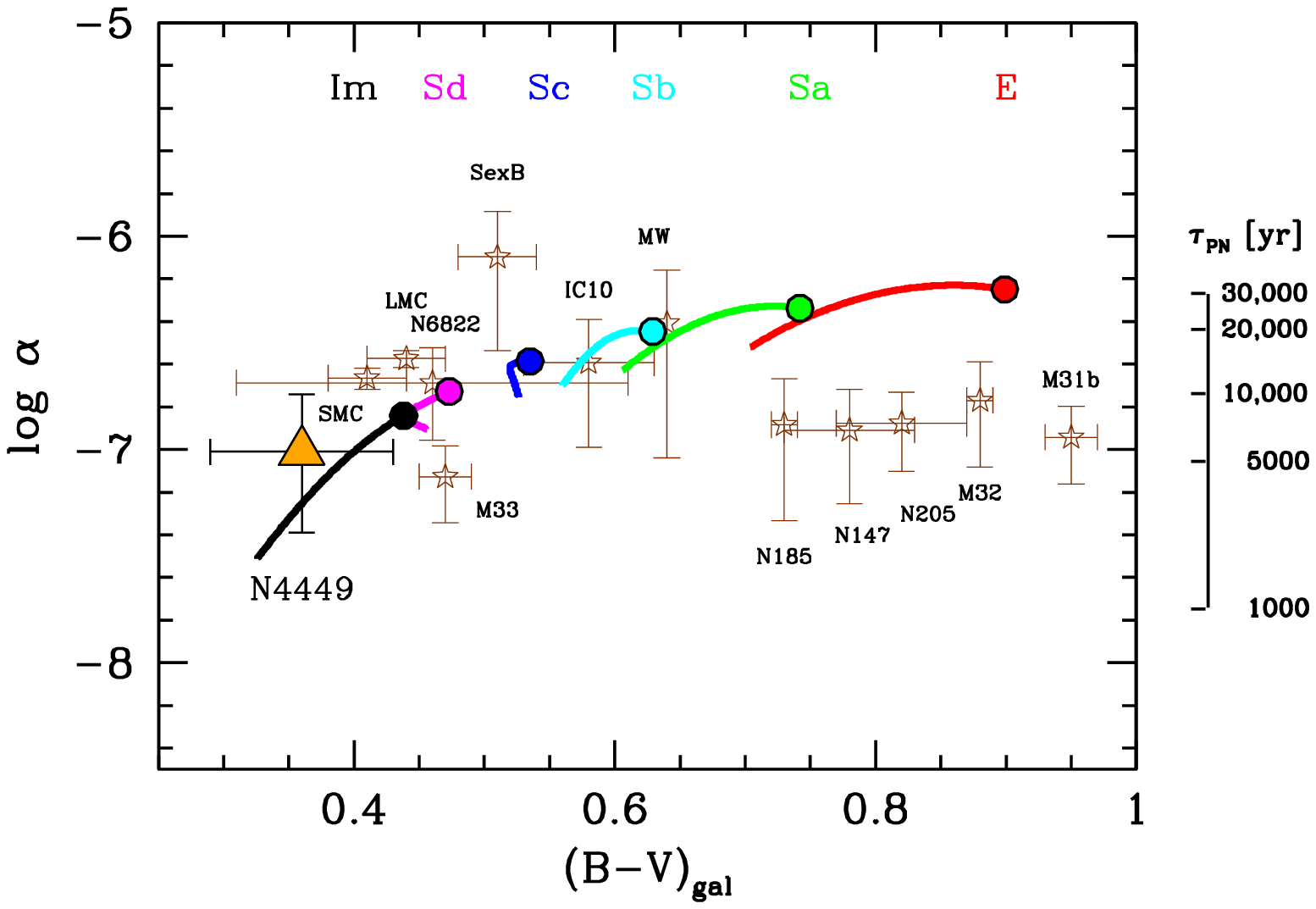}
\caption{A comprehensive overview of the derived luminosity-specific PN number density 
$\alpha$ for NGC~4449, compared with Local Group galaxies (star markers) as from \citet{buzzoni06}.
Also superposed are the \cite{buzzoni05} template galaxy models along the 
whole morphological sequence from 2 to 15 Gyr, with the latter limit
marked by the big solid dots.  The \cite{weidemann00} empirical scheme is adopted for 
properly assessing stellar mass loss in the galaxy models.
An indicative estimate of the mean representative PN visibility timescale (in years) is sketched on the right scale, as discussed in the text.
\label{pn_models}}
 \end{figure*}

Our PN completeness level can be estimated by relying on the classical \citet{henize63}
PN luminosity function (PNLF), in the form of an exponential curve with a sharp truncation 
designed to accomodate the bright end  \citep[cf. e.g.][]{ciardullo89,jacoby02}.
An absolute $M_{[O~III]}$ magnitude (at 10~pc) can be derived for each nebula in our 
spectroscopic sample, according to the observed [O~III] flux of Table~\ref{pn_flux_raw}, 
corrected for the distance modulus of $(m-M)_0\sim27.9$ \citep{anni08},  as 
$M_{[O~III]} = -2.5\,\log F_{[O~III]} - 13.74$. These figures can be contrasted
with the bright cut-off magnitude (M$^*_{[O~III]}$) of  \cite{ciardullo89} PNLF 
that, for the NGC 4449 metallicity, can be set at $M^*_{[O~III]} = -4.36$ 
 \citep{ciardullo02}. Our results are summarized in Table~\ref{pne_total} and 
Fig.~\ref{pne_o3vmag}.

The figure also  provides a mapping between $M_{F555W}$ magnitudes and
$M_{[O~III]}$ magnitudes for the five planetary nebulae with spectroscopy.
For this task we relied on the observed $m_{F555W}$ 
magnitudes, after correcting the Table~\ref{pne_total} entries for distance modulus.
As expected, the  $M_{F555W}$ magnitude of the five spectroscopic nebulae happens to be a 
quite confident proxy of the corresponding  $M_{[O~III]}$, 
with  $M_{[O~III]} = M_{F555W} + 0.5\pm0.1$ (see Fig.~\ref{pne_o3vmag}).
 When applied to the remaining photometric targets,
one can therefore conclude that our observations sample the bright  
tail of NGC~4449 PNLF, down to  $(M_{[O~III]}-M^*_{[O~III]}) \sim 2.5$~mag.

Adopting the standard PNLF, as scaled for instance from the M31 \citep{ciardullo89}
or LMC observations \citep{reid10} down to $\sim 8$~mag fainter than the bright cut-off
limit $M^*_{[O~III]}$, we obtain a total expected number of  $280 \pm 53$ (Poissonian rms) PNe for our field. 
A lower value of $\sim 145 \pm 27$ would be obtained assuming instead the SMC PNLF, as 
from the deep [O~III]$_{5007}$ observations of \cite{jacoby02}. These values 
need to be corrected for the fact that we are missing the PNe in the most crowded, central 
$R/R_{25} \lesssim 0.1$ galaxy regions: adopting the  galaxy surface brightness profile of \cite{rich12},   
 we estimate a $\sim$40 \% correction to the PN number, which translates into $392 \pm 74$ for an assumed 
 \cite{ciardullo89} PNLF (or $203 \pm 38$ PNe assuming the SMC PNLF).  We caution that these estimates are highly uncertain because our extrapolation 
 assumes that our sample  is proportional to the complete sample from the brightest PN down to $\sim$2.5 mag below the PNLF cutoff; indeed a robust determination of the PN completeness as a function of magnitude would require artificial star tests performed on 
 the images, which is beyond the scope of this paper.


With these figures, our estimate of the $\alpha$ parameter leads eventually to
\begin{equation}
\log \alpha = \log \left[ \frac{392_{\pm74}}{4.0_{\pm 0.1}\,10^9}\right] = -7.01^{+0.27}_{-0.38}, 
\end{equation}

to which, in addition to the Poissonian error from the counts, we attached  an 
uncertainty range that either assumes a 50\% fraction of PNe in excess due to a more conservative 
completeness limit at $\sim1.5$ mag below the bright cutoff or, conversely, a halved reduction of the PN number, as 
from the \citet{jacoby02} SMC PNLF.

A comparison with other Local Group galaxies is shown in Fig.~\ref{pn_models}, following 
\citet{buzzoni06}. In the plot we also report the theoretical predictions for 
\citet*{buzzoni05} template galaxy models, between 2 and 15 Gyr, according to 
\citet*{weidemann00} stellar mass-loss scheme.
 The models predict that $\alpha$ decreases in young and/or star-forming galaxies, 
compared to more ÔquiescentÕ early-type systems as a consequence 
of a smaller population of PNe embedded in a higher galaxy luminosity per unit mass (i.e. a lower M/L ratio).
 We notice that the luminosity-specific PN number in NGC~4449, as well as the values derived for other late-type systems, 
 agree quite well with the \cite{buzzoni05} models (on the other hand, the models are less successful for the earliest galaxy types). 

 From a theoretical point of view, the luminosity-specific PN number density easily relates 
with the reference visibility lifetime ($\tau_{PN}$) of the PN events, being
$\alpha = {\cal{B}} \times \tau_{PN}$, with $\cal{B}$ the so-called specific evolutionary
flux of a stellar population \citep[see][for the theoretical background]{buzzoni06}.
The  PN visibility lifetime $\tau_{PN}$ depends both on the chemical and dynamical properties 
of the ejected material, and on the stellar core evolution during the PAGB phase.
In general, for young and intermediate-age SSPs, the PAGB timescale  ($\tau_{PAGB}$) is shorter than the  dynamical time-scale 
for the nebula evaporation, so that the PN lifetime might likely be shorter for more massive, and younger, stellar progenitors. 
This trend is schematically sketched in Fig.~\ref{pn_models} relying on \cite{buzzoni06} theoretical framework.
According to these models,  the typical visibility lifetime for the PNe in NGC 4449 is predicted to be 
a few thousand years.

 \section{Conclusions}

 We presented new deep multi-object spectroscopy with LBT/MODS of H~II regions and PNe in the starburst irregular galaxy NGC~4449, at a 
 distance of $\approx$3.8 Mpc from us.  The  [O~III]$\lambda4363$ auroral line was detected in all spectra, allowing for a direct determination of the O$^{+2}$ 
 temperature. For the H~II regions, the O$^{+}$ and  S$^{+2}$ temperatures were also derived from the [O~II]$\lambda7320+\lambda7330/\lambda3726+\lambda3729$ and 
 [S~III]$\lambda6312/\lambda9069+\lambda9532$ ratios.  Using the ``direct'' method, we  
 derived the abundance of He, N, O, Ne, Ar, and S for six H~II regions and, for the first time, for four PNe in NGC~4449. 
 Iron abundances were also derived for the H~II regions, but this element is notoriously highly affected by depletion into dust grains.  
 The combined H~II region and PN sample covers a galacto-centric distance range of 
 $\sim$2 kpc, corresponding to $\approx$70\% of the $R_{25}$ isophotal radius. 
 Our main results are: 
 
 \begin{enumerate}
 
 \item The total H~II region $+$ PN sample spans $\sim$0.2 dex in oxygen abundance, with average $12+\log(O/H)$ values of  
 8.37 $\pm$ 0.05 and   8.3 $\pm$ 0.1 for H~II regions and PNe, respectively. The results for the H~II regions are consistent, within the errors, 
 with previous literature estimates  based on the direct temperature method. 
 
 \item We find a well defined  trend of decreasing oxygen abundance with increasing galacto-centric distance: 
  $12+\log(O/H) = -0.29(\pm0.06) \times R/R_{25}+8.49(\pm0.030)$, with H~II regions and PNe exhibiting similar oxygen abundances at the same galacto-centric distance. 
    This result, coupled with our previous finding of a negative metallicity gradient for H~II regions in the blue compact dwarf NGC~1705 
  \citep{anni15} and with the recent results by \cite{pilyugin15}, suggests that {\it metallicity gradients do exist in irregular galaxies}, at odds 
  with what was previously believed \citep[e.g.][]{kobul97,croxall09,haurberg13,lagos13}.

 \item Despite the presence of a negative oxygen gradient, 
 nitrogen does not exhibit any well-defined radial trend. This is unexpected, since an important component of
 secondary nitrogen should exist in the present-day ISM of NGC~4449.  
Building on previous literature studies showing evidence  for N/O inhomogeneities in Wolf-Rayet galaxies, we suggest that the anomalous nitrogen behaviour may be due to local enrichment by the conspicuous Wolf-Rayet population in NGC~4449.

 \item  The studied PNe exhibit a significant nitrogen enhancement with respect to H~II regions ($\gsim$1 dex); this behaviour is in agreement with previous chemical abundance studies of PNe in galaxies of different morphological types. On the other hand, we also find that the PN helium abundances are similar to those of NGC~4449' s H~II regions,  around He/H$\simeq$0.09 (although we caution that  our PN He estimates are very uncertain because the detected  He~I $\lambda5876$ line is significantly fainter than two nearby sky lines at $\lambda\sim5867$ \AA \ and  $\lambda\sim5890$ \AA). From the theoretical point of view, we expect both N and He to be enhanced in PNe because they are both synthesized and brought to the stellar surface through dredge-up episodes occurring in the RGB and AGB phases of intermediate mass stars. We are not aware of any model producing a factor $\sim$10 enhancement in N while leaving He unchanged.

\item Our  PN  oxygen (and $\alpha$-element, more in general) abundances are, on the other hand, similar to those of H~II regions  in the galacto-centric distance range of overlap. This indicates that the NGC~4449's ISM has not been significantly enriched in metals  since the progenitors of the PNe were formed (i.e., since $\sim$ 100 Myr ago or more).
Recently produced $\alpha$-elements may have been 
 expelled from NGC~4449 by the galactic outflow, or may still reside in a hot phase \citep[see e.g.][for NGC~1569]{martin02}; also, acquisition of metal poor gas may have diluted 
 the metals in the ISM.

\item  The derived luminosity-specific PN number density ($\alpha=N_{PN}/L_{gal}$) in NGC~4449 agrees quite well with the \cite{buzzoni05} template galaxy models 
that predict the behaviour of $\alpha$ as a function of galaxy morphological type and color; according to these models, the $\alpha$ value derived   in NGC~4449 
translates into a typical visibility lifetime for the PN population  of a few thousand years.

 \item Two out of the six studied H~II regions show broad emission features associated with Wolf-Rayet stars of WN and WC subtypes. From a comparison with population 
  synthesis models, we infer that a WR population at least 3$-$4 Myr old must be present in NGC~4449.

 \end{enumerate}

\acknowledgments

F. A. thanks C. Leitherer and E. Skillman for useful discussion, V. Luridiana for support with the Pyneb code, and G. Delgado-Inglada for suggestions 
related to the use the new ionization correction factors.
We are grateful to the referee, Michael Richer, for his careful reading of the manuscript and for the useful, precise and constructive comments which helped 
us to significantly improve the paper. 
F. A. and this work have been supported by PRIN MIUR through grant 2010LY5N2T\_006.
D. R. benefited from the International Space Science Institute (ISSI) in Bern, thanks to the funding of the team ``The formation and evolution of the galactic halo'' (PI D. Romano). 
This research was partly supported also by the Munich Institute for Astro- and Particle Physics (MIAPP) of the DFG cluster of excellence ``Origin and Structure of the Universe''.
This work was based on LBT/MODS data. 
The LBT is an international collaboration among institutions in the United States, Italy and Germany. LBT Corporation partners are: the University of Arizona on behalf of the Arizona Board of Regents; Istituto Nazionale di Astrofisica, Italy; LBT Beteiligungsgesellschaft, Germany, representing the Max-Planck Society, the Leibniz Institute for Astrophysics Potsdam, and Heidelberg University; the Ohio State University, and the Research Corporation, on behalf of the University of Notre Dame, University of Minnesota and University of Virginia.
We acknowledge the support from the LBT-Italian Coordination Facility for the execution of observations, data distribution and reduction.

\newpage 
\appendix{}

\section{H~II regions and PNe spectra}

\begin{figure*}[h]
\epsscale{0.75}
\plotone{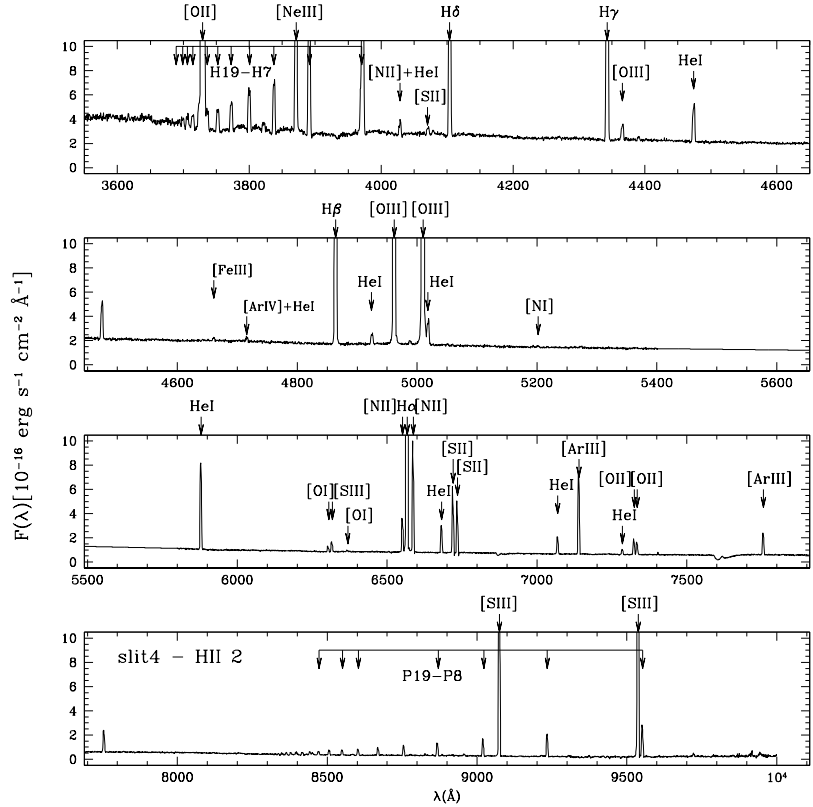}
\caption{LBT/MODS spectra in the blue and red channels for H~II-2 in NGC~4449 with indicated all the identified emission lines. 
The spectra have been scaled such that the details are evident. 
 \label{slit4}}
\end{figure*}

\begin{figure*}[h]
\epsscale{0.75}
\plotone{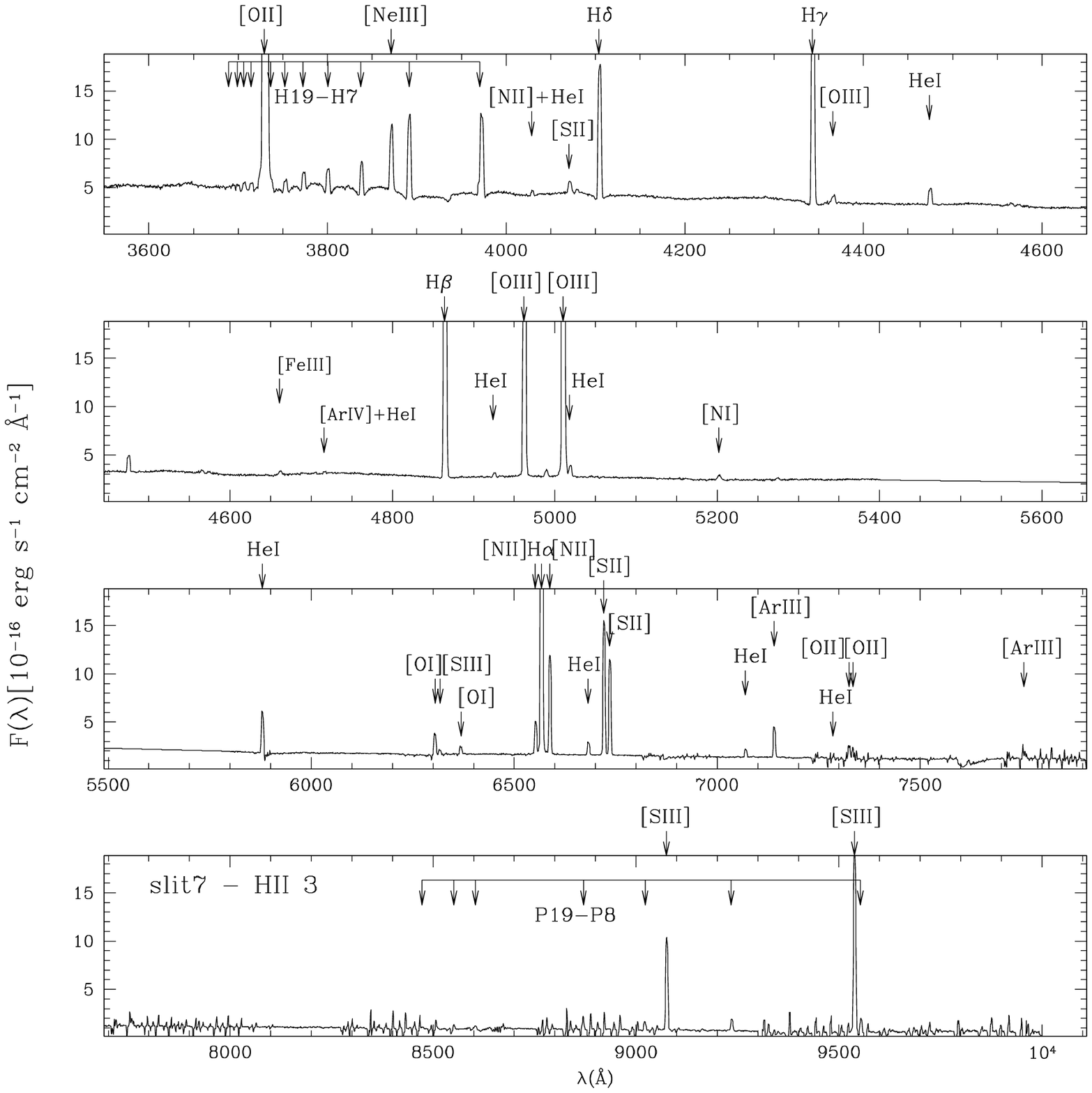}
\caption{Same as Fig.~\ref{slit4} but for  H~II-3. 
 \label{slit7}}
\end{figure*}

\begin{figure*}[h]
\epsscale{0.75}
\plotone{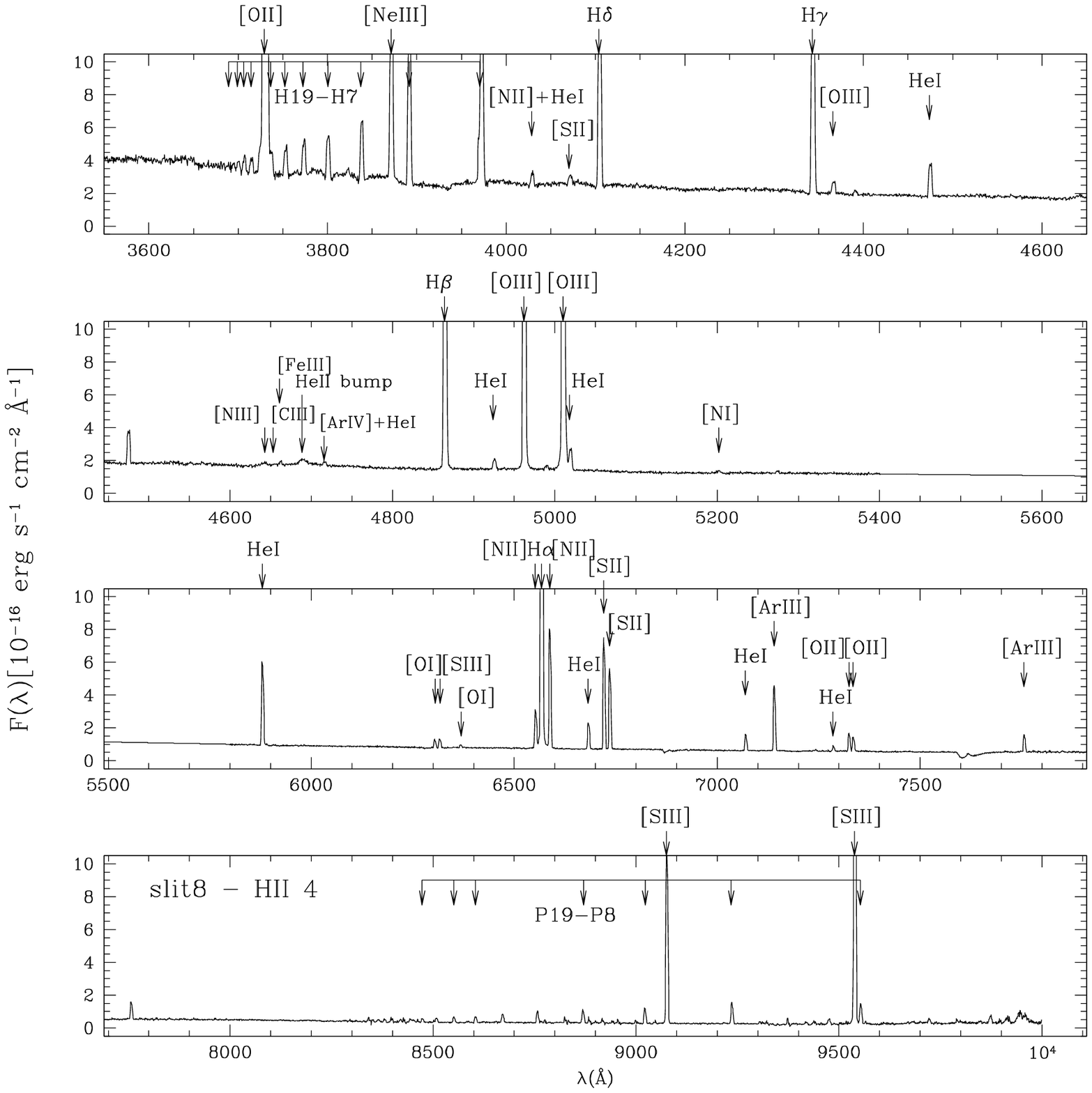}
\caption{Same as Fig.~\ref{slit4} but for  H~II-4. 
 \label{slit8}}
\end{figure*}

\begin{figure*}[h]
\epsscale{0.75}
\plotone{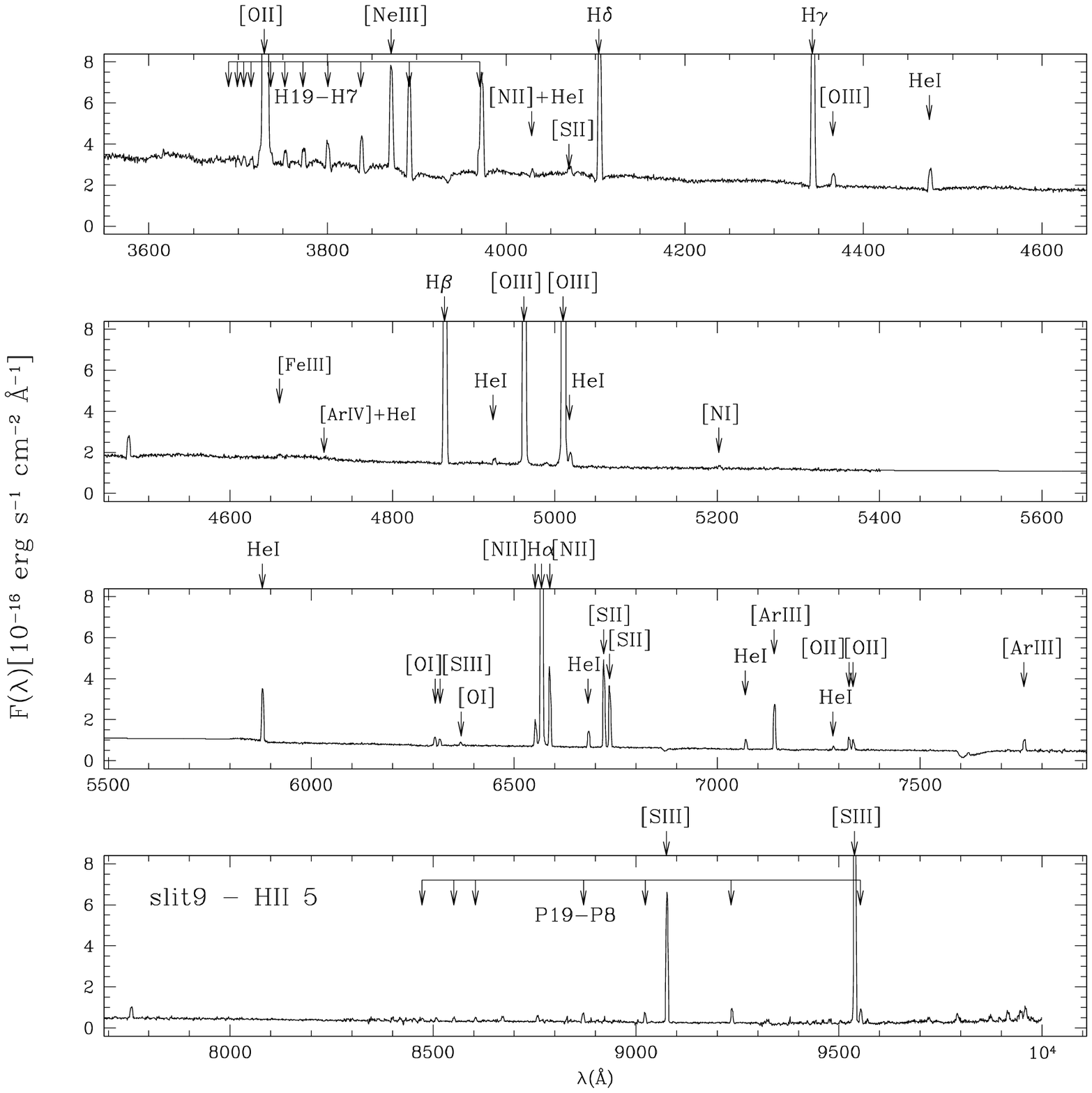}
\caption{Same as Fig.~\ref{slit4} but for  H~II-5. 
 \label{slit9}}
\end{figure*}

\begin{figure*}[h!]
\epsscale{0.7}
\plotone{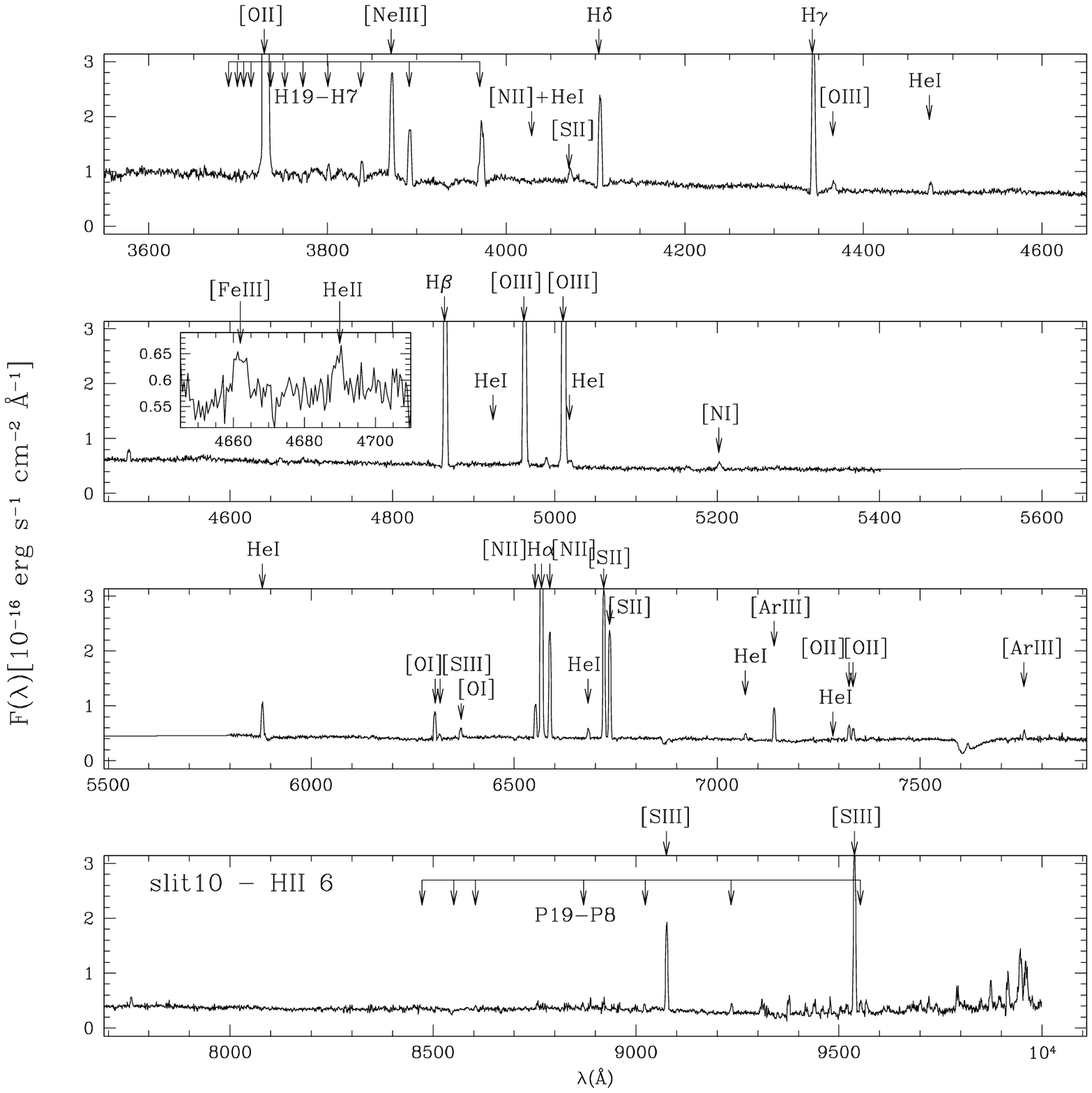}
\caption{Same as Fig.~\ref{slit4} but for  H~II-6. The small insertion provides a zoom into the $\sim$4640 - 4710 \AA \ wavelength range to highlight the faint [Fe~III]$\lambda$4658 and  He~II$\lambda$4686 lines.
 \label{slit10}}
\end{figure*}

\begin{figure*}[h!]
\epsscale{0.75}
\plotone{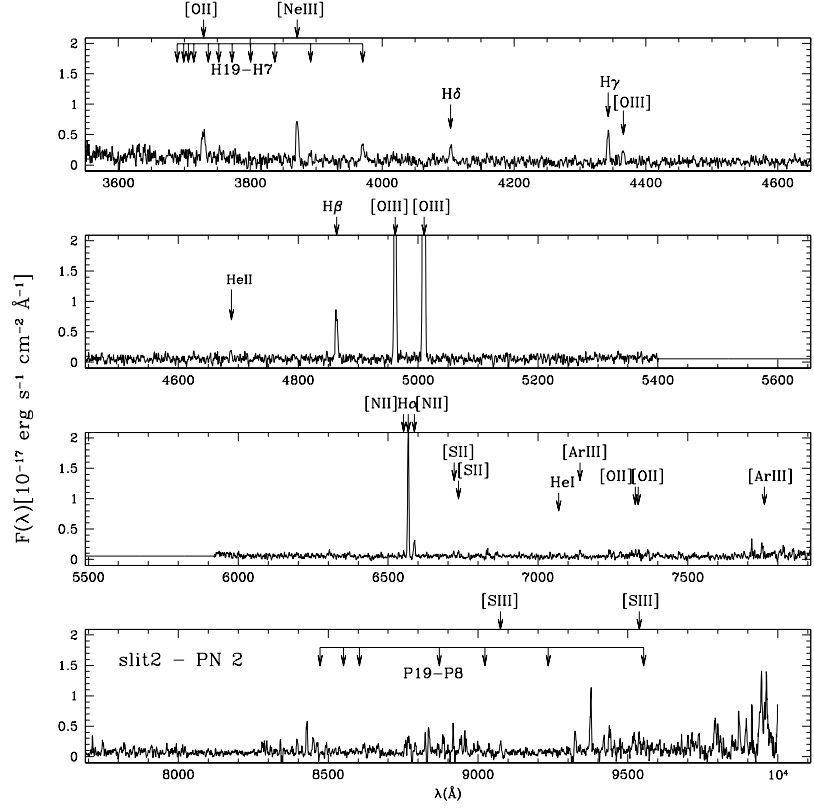}
\caption{LBT/MODS spectra in the blue and red channels for PN-2 in NGC~4449 with indicated all the identified emission lines. 
A  $\sim$1~\AA \ boxcar filter smoothing was applied to the spectrum to better highlight the low singal-to-noise features.  
The spectra have been scaled such that the details are evident. 
 \label{slit2}}
\end{figure*}

\begin{figure*}[h!]
\epsscale{0.7}
\plotone{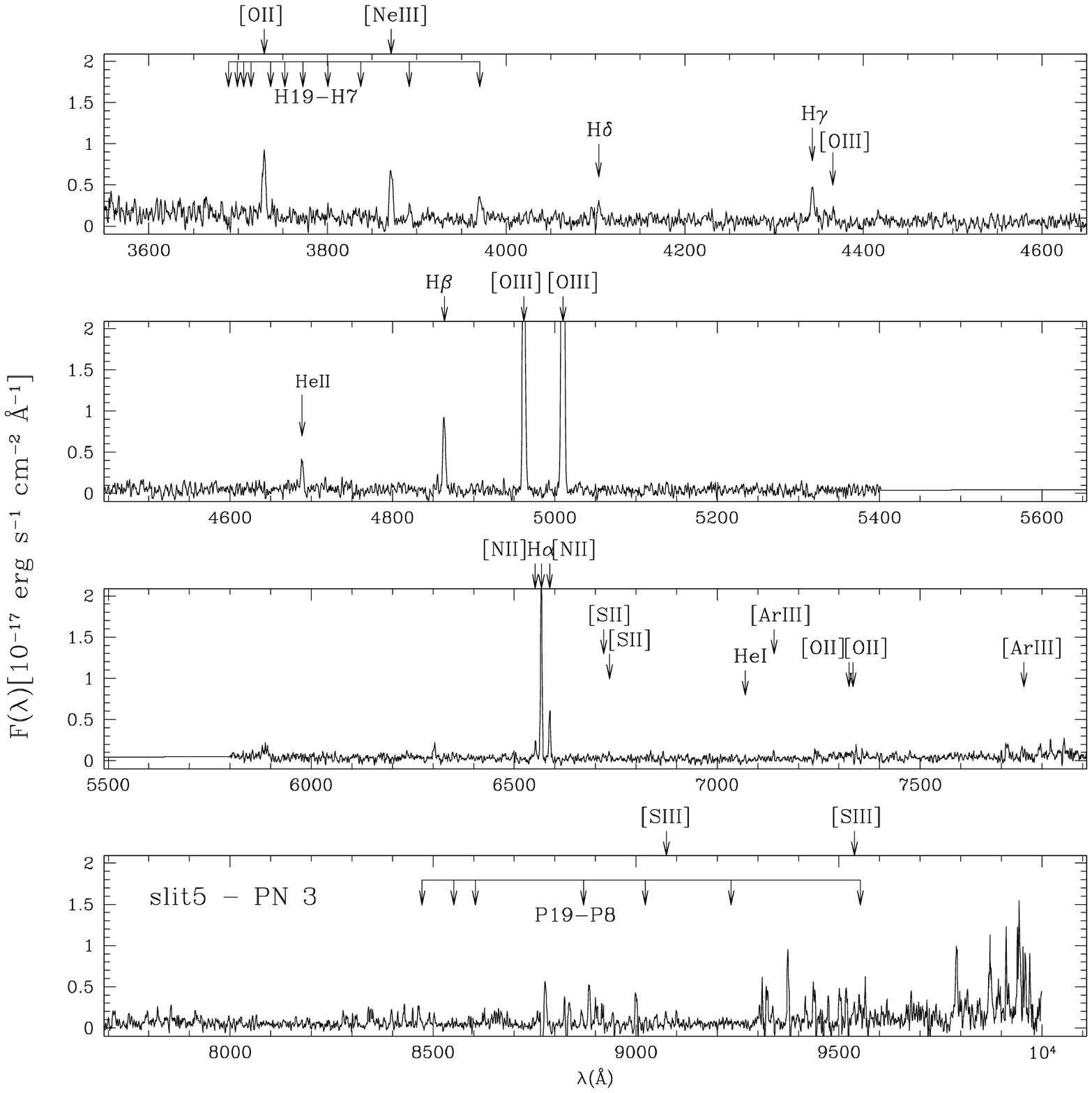}
\caption{Same as Fig.~\ref{slit2} but for  PN-3. 
 \label{slit5}}
\end{figure*}

\begin{figure*}[h!]
\epsscale{0.7}
\plotone{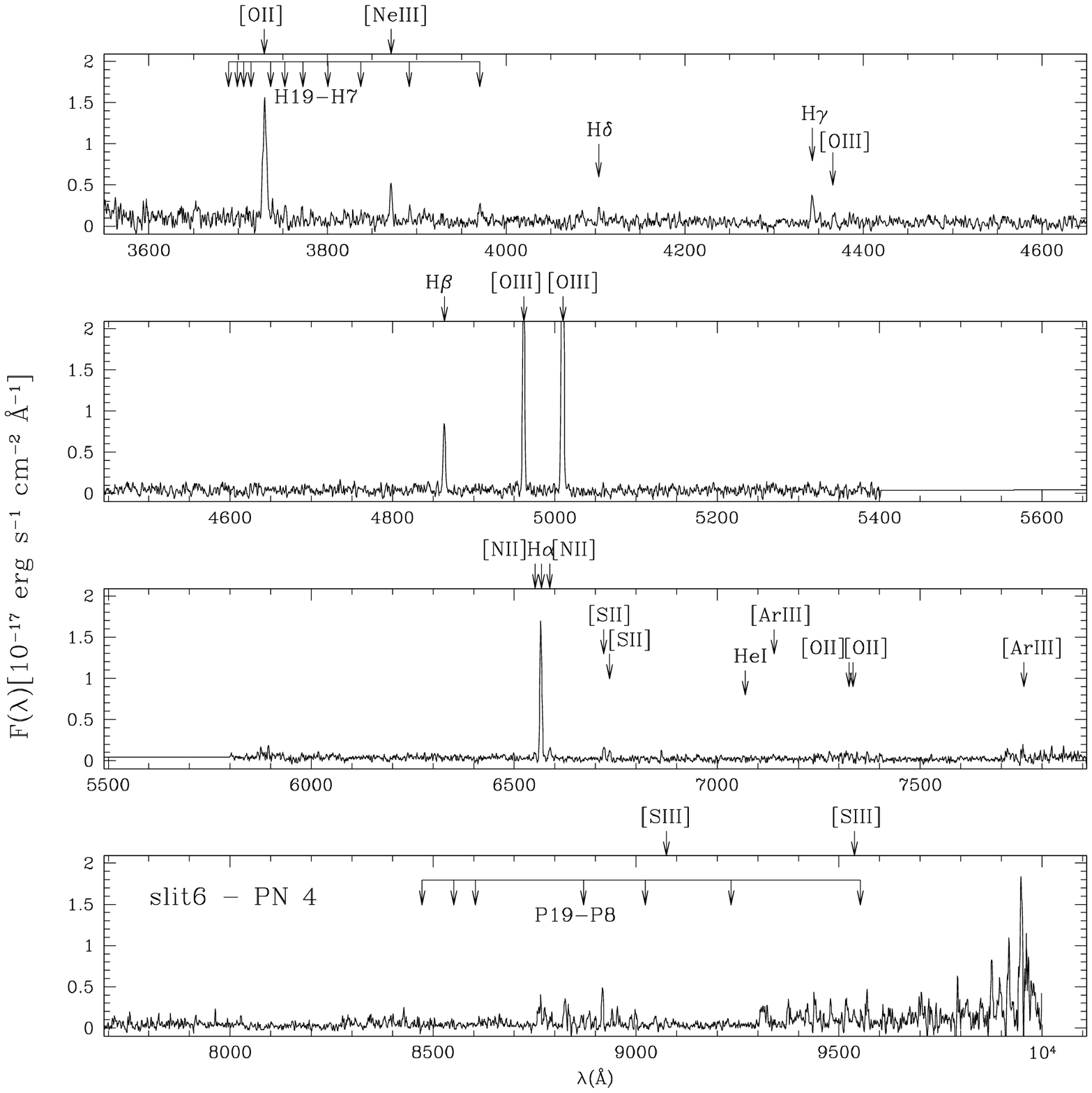}
\caption{Same as Fig.~\ref{slit2} but for  PN-4.  
\label{slit6}}
\end{figure*}

\begin{figure*}[h!]
\epsscale{0.7}
\plotone{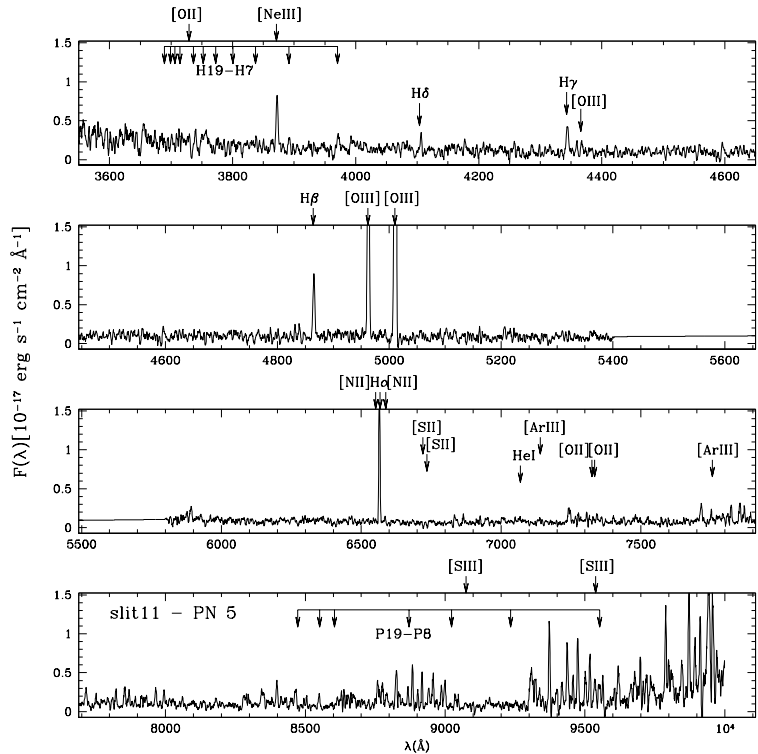}
\caption{Same as Fig.~\ref{slit2} but for  PN-5. 
 \label{slit11}}
\end{figure*}

We present the LBT/MODS spectra of H~II regions H~II-2,  H~II-3,  H~II-4,  H~II-5, H~II-6 in Figures~\ref{slit4},~\ref{slit7},~\ref{slit8},~\ref{slit9},~\ref{slit10}, respectively, 
and of planetary nebulae PN-2, PN-3, PN-4, PN-5 in Figures~\ref{slit2},~\ref{slit5},~\ref{slit6},~\ref{slit11}, respectively.

\newpage

\section{``Raw'' emission line fluxes}

We provide in Tables~\ref{h2_flux_raw} and ~\ref{pn_flux_raw} the measured emission line fluxes, with no reddening correction applied, for our studied H~II regions and 
PNe in NGC~4449. The reported flux values were obtained by averaging the results from different measurements, as outlined in Section~3; the associated uncertainties 
were simply obtained as the  standard deviation of the different measurements. Notice that these errors do not account for additional uncertainties due to e.g. flux calibration. 

\setcounter{table}{1}

\begin{deluxetable}{lcccccc}
\tabletypesize{\scriptsize}
\tablecaption{Observed emission fluxes for H~II regions in NGC~4449 \label{h2_flux_raw}}
\tablehead{
\colhead{Line} & \colhead{H~II-1} &  \colhead{H~II-2}  &  \colhead{H~II-3}  &  \colhead{H~II-4} &  \colhead{H~II-5} &  \colhead{H~II-6} 
}
\startdata
{[O II]} $\lambda$3727 &    23.69 $\pm$    0.02 &    42.85 $\pm$    0.04 &    75 $\pm$    2 &    43.62 $\pm$    0.04 &    24.70 $\pm$    0.03 &   10.2 $\pm$   0.1 \\
H10 $\lambda$3978 &   1.41 $\pm$  0.01 &   1.41 $\pm$  0.02 &   1.19 $\pm$  0.01 &   1.17 $\pm$  0.01 &   0.60 $\pm$  0.01 &   0.13 $\pm$  0.01 \\
He I $\lambda$3820  &  0.210 $\pm$ 0.001 &  0.163 $\pm$ 0.001 & $-$ &  0.110 $\pm$ 0.001 &  $-$ & $-$ \\
H9$+$He II $\lambda$3835 &    1.99 $\pm$   0.01 &    1.907 $\pm$   0.006 &    1.65 $\pm$   0.01 &    1.63 $\pm$   0.01 &    0.837 $\pm$   0.002 &    0.195 $\pm$   0.001 \\
{[Ne III]} $\lambda$3869 &     7.22 $\pm$    0.02 &     5.32 $\pm$    0.01 &     2.64 $\pm$    0.01 &     4.08 $\pm$    0.01 &     1.96 $\pm$    0.01 &     0.75 $\pm$    0.01 \\
H8$+$He I $\lambda$3889 &     4.33 $\pm$    0.02 &     4.96 $\pm$    0.01 &     3.81 $\pm$    0.01 &     3.90 $\pm$    0.01 &     2.19 $\pm$    0.02 &     0.45 $\pm$    0.01 \\
H$\epsilon$ $+$ He I $+$[Ne III] $\lambda$3970 &     5.72 $\pm$    0.01 &     5.44 $\pm$    0.01 &     4.08 $\pm$    0.01 &     4.84 $\pm$    0.04 &     2.56 $\pm$    0.03 &     0.64 $\pm$    0.01 \\
He I $\lambda$4026 &    0.462 $\pm$   0.006 &    0.420 $\pm$   0.004 &    0.18 $\pm$   0.01 &    0.284 $\pm$   0.001 &    0.105 $\pm$   0.004 &  $-$ \\
{[S II]} $\lambda$4068 &    0.168 $\pm$   0.005 &    0.184 $\pm$   0.002 &    0.50 $\pm$   0.01 &    0.177 $\pm$   0.002 &    0.143 $\pm$   0.004 &    0.088 $\pm$   0.001 \\
{[S II]} $\lambda$4076 &    0.049 $\pm$   0.008 &    0.095 $\pm$   0.001 &    0.155 $\pm$   0.002 &  $-$ &  $-$ &    0.034 $\pm$   0.002 \\
H$\delta$ $\lambda$4101 &     6.08 $\pm$    0.01 &     6.96 $\pm$    0.01 &     5.96 $\pm$    0.04 &     6.10 $\pm$    0.03 &     3.36 $\pm$    0.04 &     0.69 $\pm$    0.01 \\
H$\gamma$ $\lambda$4340 &    10.93 $\pm$    0.01 &    12.54 $\pm$    0.01 &    10.46 $\pm$    0.01 &     9.75 $\pm$    0.01 &     5.74 $\pm$    0.04 &     1.25 $\pm$    0.01 \\
{[O III]} $\lambda$4363 &    0.51 $\pm$   0.02 &    0.460 $\pm$   0.005 &    0.289 $\pm$   0.004 &    0.310 $\pm$   0.003 &    0.222 $\pm$   0.004 &    0.059 $\pm$   0.001 \\
He I $\lambda$4389 &    0.137 $\pm$   0.003 &    0.098 $\pm$   0.003 &  $-$ &    0.096 $\pm$   0.002 &  $-$ &  $-$ \\
He I $\lambda$4471 &     1.03 $\pm$    0.01 &     1.15 $\pm$    0.01 &     0.70 $\pm$    0.01 &     0.81 $\pm$    0.01 &     0.39 $\pm$    0.01 &     0.07 $\pm$    0.01 \\
{[N III](WR)} $\lambda$4640 &    0.046 $\pm$   0.002 &  $-$ &  $-$ &    0.23 $\pm$   0.01 & $-$ &  $-$ \\
{[C III](WR)} $\lambda$4652 &    0.028 $\pm$   0.007 &  $-$ &  $-$ &    0.10 $\pm$   0.03 & $-$0 &  $-$ \\
{[Fe III]} $\lambda$4658 &    0.034 $\pm$   0.006 &    0.078 $\pm$   0.001 &    0.186 $\pm$   0.004 &    0.090 $\pm$   0.007 &    0.060 $\pm$   0.001 &    0.036 $\pm$   0.004 \\
He II (WR) $\lambda$4686 &    0.89 $\pm$   0.04 &  $-$ & $-$ &    1.33 $\pm$   0.09 &  $-$ &  $-$ \\
He II  $\lambda$4686 &    0.10 $\pm$   0.08 & $-$ &  $-$ &  $-$ &  $-$ &    0.022 $\pm$   0.001 \\
{[Ar IV]}$+$He I $\lambda$4713 &    0.12 $\pm$   0.01 &    0.107 $\pm$   0.001 &  $-$ &    0.11 $\pm$   0.01 &  $-$ &  $-$ \\
{[Ar IV]} $\lambda$4740 &    0.041 $\pm$   0.003 &  $-$ &  $-$ &    0.064 $\pm$   0.009 &  $-$ &  $-$\\
H$\beta$ $\lambda$4861 &    25.02 $\pm$    0.02 &    29.06 $\pm$    0.01 &    22.77 $\pm$    0.04 &    21.76 $\pm$    0.03 &    12.42 $\pm$    0.01 &     2.72 $\pm$    0.01 \\
He I $\lambda$4922 &     0.30 $\pm$    0.01 &     0.35 $\pm$    0.01 &     0.17 $\pm$    0.01 &     0.23 $\pm$    0.02 &     0.10 $\pm$    0.01 &   $-$\\
{[O III]} $\lambda$4959 &   36.190 $\pm$   0.005 &   38.200 $\pm$   0.008 &   16.37 $\pm$   0.02 &   22.95 $\pm$   0.06 &   14.010 $\pm$   0.005 &    2.913 $\pm$   0.006 \\
{[Fe III]} $\lambda$4986 &    0.049 $\pm$   0.004 &    0.11 $\pm$   0.01 &    0.268 $\pm$   0.005 &    0.084 $\pm$   0.001 &    0.073 $\pm$   0.003 &    0.054 $\pm$   0.001 \\
{[O III]} $\lambda$5007 &   106.40 $\pm$    0.01 &   113.50 $\pm$    0.01 &    48.22 $\pm$    0.03 &    67.35 $\pm$    0.02 &    40.37 $\pm$    0.01 &     8.45 $\pm$    0.01 \\
He I $\lambda$5015  &    0.624 $\pm$   0.004 &    0.804 $\pm$   0.003 &    0.50 $\pm$   0.01 &    0.52 $\pm$   0.02 &    0.273 $\pm$   0.002 &    0.062 $\pm$   0.005 \\
{[N I]} $\lambda$5199 &    0.054 $\pm$   0.001 &    0.041 $\pm$   0.001 &    0.243 $\pm$   0.006 &    0.059 $\pm$   0.002 &    0.044 $\pm$   0.001 &    0.064 $\pm$   0.001 \\
He I $\lambda$5876  &     3.14 $\pm$    0.01 &     4.04 $\pm$    0.01 &     2.72 $\pm$    0.04 &     3.06 $\pm$    0.01 &     1.57 $\pm$    0.01 &     0.40 $\pm$    0.01 \\
{[OI]}  $\lambda$6302 &  0.322 $\pm$ 0.002 &  0.231 $\pm$ 0.001 & $-$ &  0.283 $\pm$ 0.002 &  0.247 $\pm$ 0.002 &  0.314 $\pm$ 0.001 \\
{[S III]} $\lambda$6314 &    0.307 $\pm$   0.001 &    0.471 $\pm$   0.001 &    0.309 $\pm$   0.008 &    0.321 $\pm$   0.001 &    0.189 $\pm$   0.001 &    0.054 $\pm$   0.001 \\
{[OI]} $\lambda$6365  &    0.102 $\pm$   0.001 &    0.060 $\pm$   0.004 &    0.526 $\pm$   0.001 &    0.102 $\pm$   0.004 &    0.087 $\pm$   0.002 &    0.104 $\pm$   0.001 \\
{[NII]} $\lambda$6548  &     0.80 $\pm$    0.01 &     1.53 $\pm$    0.02 &     2.4 $\pm$    0.1 &     1.42 $\pm$    0.01 &     0.71 $\pm$    0.01 &     0.40 $\pm$    0.01 \\
H$\alpha$ $\lambda$6563  &    79.61 $\pm$    0.01 &   110.00 $\pm$    0.05 &    82.6 $\pm$    0.2 &    82.70 $\pm$    0.01 &    44.28 $\pm$    0.01 &    10.55 $\pm$    0.01 \\
{[N II]} $\lambda$6584  &     2.24 $\pm$    0.03 &     4.59 $\pm$    0.02 &     6.8 $\pm$    0.1 &     4.18 $\pm$    0.01 &     2.07 $\pm$    0.01 &     1.22 $\pm$    0.01 \\
He I $\lambda$6678   &    0.965 $\pm$   0.001 &    1.257 $\pm$   0.004 &    0.89 $\pm$   0.03 &    0.984 $\pm$   0.004 &    0.477 $\pm$   0.001 &    0.122 $\pm$   0.001 \\
{[S II]} $\lambda$6716  &     2.29 $\pm$    0.01 &     2.67 $\pm$    0.01 &     9.09 $\pm$    0.05 &     3.93 $\pm$    0.01 &     2.36 $\pm$    0.01 &     1.79 $\pm$    0.01 \\
{[S II]} $\lambda$6731  &     1.67 $\pm$    0.01 &     2.08 $\pm$    0.01 &     6.53 $\pm$    0.05 &     2.90 $\pm$    0.01 &     1.70 $\pm$    0.01 &     1.28 $\pm$    0.01 \\
He I $\lambda$7065  &    0.544 $\pm$   0.001 &    0.802 $\pm$   0.002 &    0.534 $\pm$   0.001 &    0.582 $\pm$   0.004 &    0.278 $\pm$   0.001 &    0.066 $\pm$   0.001 \\
{[Ar III]} $\lambda$7136  &    2.388 $\pm$   0.001 &    3.668 $\pm$   0.005 &    2.014 $\pm$   0.002 &    2.223 $\pm$   0.005 &    1.331 $\pm$   0.001 &    0.349 $\pm$   0.001 \\
He I $\lambda$7281  &  0.164 $\pm$ 0.002 &  0.235 $\pm$ 0.001 &  $-$ &  0.167 $\pm$ 0.001 &  0.097 $\pm$ 0.001 & $-$ \\
{[O II]} $\lambda$7320  &  0.327 $\pm$ 0.002 &  0.694 $\pm$ 0.002 & $-$ &  0.624 $\pm$ 0.001 &  0.353 $\pm$ 0.001 &  0.152 $\pm$ 0.001 \\
{[O II]} $\lambda$7330  &  0.276 $\pm$ 0.002 &  0.592 $\pm$ 0.003 & $-$ &  0.508 $\pm$ 0.001 &  0.287 $\pm$ 0.001 &  0.123 $\pm$ 0.002 \\
{[Ar III]} $\lambda$7751  &   0.66 $\pm$  0.01 &   1.00 $\pm$  0.01 & $-$ &   0.58 $\pm$  0.01 &   0.38 $\pm$  0.01 &   0.08 $\pm$  0.01 \\
P10 $\lambda$9017  &    0.582 $\pm$   0.005 &    0.808 $\pm$   0.006 &  $-$ &    0.57 $\pm$   0.02 &    0.280 $\pm$   0.009 &    0.055 $\pm$   0.002 \\
{[S III]} $\lambda$9069  &    6.298 $\pm$   0.002 &    9.749 $\pm$   0.009 &    6.36 $\pm$   0.03 &    6.76 $\pm$   0.02 &    3.823 $\pm$   0.008 &    0.965 $\pm$   0.002 \\
P9 $\lambda$9229   &  0.692 $\pm$ 0.008 &  0.989 $\pm$ 0.003 &  0.800 $\pm$ 0.004 &  0.785 $\pm$ 0.003 &  0.426 $\pm$ 0.001 & $-$ \\
{[S III]} $\lambda$9532  &   13.200 $\pm$   0.005 &   24.34 $\pm$   0.04 &   13.64 $\pm$   0.03 &   14.67 $\pm$   0.03 &    8.78 $\pm$   0.02 &    2.03 $\pm$   0.02 \\
P8 $\lambda$9547  &    0.840 $\pm$   0.009 &    1.492 $\pm$   0.030 &  $-$ &    0.79 $\pm$   0.02 &    0.44 $\pm$   0.01 &  $-$ \\
F555W [Vega mag] & 18.1 &   17.6 & 18.2 &  18.3 &  18.5 &  19.4 \\ 
\enddata
\tablecomments{The fluxes, in units of $10^{-15} erg  \ s^{-1} cm^{-2} \AA^{-1}$, were not corrected for reddening. For each line, the reported uncertainty is  
the standard deviation from different measurements, as described in Section~3.}
\end{deluxetable}

\begin{deluxetable}{lccccc}
\tabletypesize{\scriptsize}
\tablecaption{Observed emission fluxes for PNe in NGC~4449 \label{pn_flux_raw}}
\tablehead{
\colhead{Line} & \colhead{PN-1} &  \colhead{PN-2}  &  \colhead{PN-3}  &  \colhead{PN-4} &  \colhead{PN-5} 
}
\startdata
{[O II]} $\lambda$3727 &    1.28 $\pm$   0.03 &    2.43 $\pm$   0.03 &    3.18 $\pm$   0.09 &    6.92 $\pm$   0.07 &  $-$ \\
{[Ne III]} $\lambda$3869 &    3.68 $\pm$   0.03 &    2.66 $\pm$   0.04 &    2.51 $\pm$   0.03 &    1.31 $\pm$   0.02 &    2.41 $\pm$   0.04 \\
H8$+$He I $\lambda$3889 &    0.55 $\pm$   0.01 &    0.71 $\pm$   0.02 &    0.69 $\pm$   0.02 &    0.40 $\pm$   0.01 &    0.49 $\pm$   0.02 \\
H$\epsilon$ $+$ He I $+$[Ne III] $\lambda$3970 &    1.55 $\pm$   0.02 &    1.34 $\pm$   0.01 &    1.37 $\pm$   0.06 &    0.60 $\pm$   0.01 &  $-$ \\
H$\delta$ $\lambda$4101 &    1.19 $\pm$   0.02 &    0.99 $\pm$   0.04 &    0.71 $\pm$   0.01 &  $-$ &  $-$ \\
H$\gamma$ $\lambda$4340 &    2.06 $\pm$   0.03 &    1.91 $\pm$   0.04 &    1.77 $\pm$   0.09 &    1.16 $\pm$   0.02 &    1.41 $\pm$   0.02 \\
{[O III]} $\lambda$4363 &    0.71 $\pm$   0.01 &    0.63 $\pm$   0.01 &    0.62 $\pm$   0.06 &    0.40 $\pm$   0.02 &    0.53 $\pm$   0.01 \\
He II  $\lambda$4686 &  $<0.15$ &    0.55 $\pm$   0.03 &    1.37 $\pm$   0.02 &  $<0.15$ &  $<0.2$ \\
H$\beta$ $\lambda$4861 &    5.21 $\pm$   0.03 &    3.36 $\pm$   0.03 &    3.50 $\pm$   0.07 &    2.44 $\pm$   0.04 &    3.10 $\pm$   0.02 \\
{[O III]} $\lambda$4959 &   20.29 $\pm$   0.02 &   13.84 $\pm$   0.02 &   13.9 $\pm$   0.1 &    8.25 $\pm$   0.01 &   13.63 $\pm$   0.01 \\
{[O III]} $\lambda$5007 &   59.06 $\pm$   0.05 &   40.30 $\pm$   0.03 &   40.15 $\pm$   0.07 &   24.2 $\pm$   0.2 &   38.08 $\pm$   0.02 \\
He I $\lambda$5876  &    0.59 $\pm$   0.01 &    0.38 $\pm$   0.03 &    0.24 $\pm$   0.01 &    0.34 $\pm$   0.02 &    0.47 $\pm$   0.03 \\
{[NII]} $\lambda$6548  &    0.54 $\pm$   0.02 &    0.34 $\pm$   0.02 &    0.83 $\pm$   0.01 &    0.40 $\pm$   0.01 &    0.16 $\pm$   0.02 \\
H$\alpha$ $\lambda$6563  &   14.50 $\pm$   0.03 &   10.12 $\pm$   0.02 &    9.19 $\pm$   0.01 &    9.96 $\pm$   0.02 &    9.77 $\pm$   0.02 \\
{[N II]} $\lambda$6584  &    1.31 $\pm$   0.02 &    1.30 $\pm$   0.01 &    2.49 $\pm$   0.01 &    0.84 $\pm$   0.01 &    0.37 $\pm$   0.01 \\
{[S II]} $\lambda$6716  &    0.54 $\pm$   0.03 &    0.41 $\pm$   0.02 &    0.30 $\pm$   0.01 &    0.98 $\pm$   0.01 &  $-$ \\
{[S II]} $\lambda$6731  &    0.55 $\pm$   0.04 &    0.37 $\pm$   0.02 &    0.38 $\pm$   0.01 &    0.66 $\pm$   0.01 &  $-$ \\
He I $\lambda$7065  &    0.74 $\pm$   0.03 &  $-$ &  $-$ &  $-$ &    0.25 $\pm$   0.01 \\
{[Ar III]} $\lambda$7136  &    0.40 $\pm$   0.01 &    0.63 $\pm$   0.02 &    0.30 $\pm$   0.01 &  $-$ &  $-$ \\
{[S III]} $\lambda$9069  &    0.92 $\pm$   0.02 &    0.90 $\pm$   0.01 &    0.65 $\pm$   0.03 &  $-$ &  $-$ \\
{[S III]} $\lambda$9532  &    1.71 $\pm$   0.08 &  $-$ &  $-$ &  $-$ &  $-$ \\
F555W [Vega mag] & 23.9 &   24.4 & 24.5 &  24.8 &  24.3 \\ 
\enddata
\tablecomments{The fluxes, in units of $10^{-17} erg  \ s^{-1} cm^{-2} \AA^{-1}$, were not corrected for reddening. For each line, the reported uncertainty is  
the standard deviation from different measurements, as described in Section~3.}
\end{deluxetable}

\newpage

\section{PN chemical abundances with new ionization correction factors \label{app1}}

\begin{figure*}
\epsscale{1}
\plotone{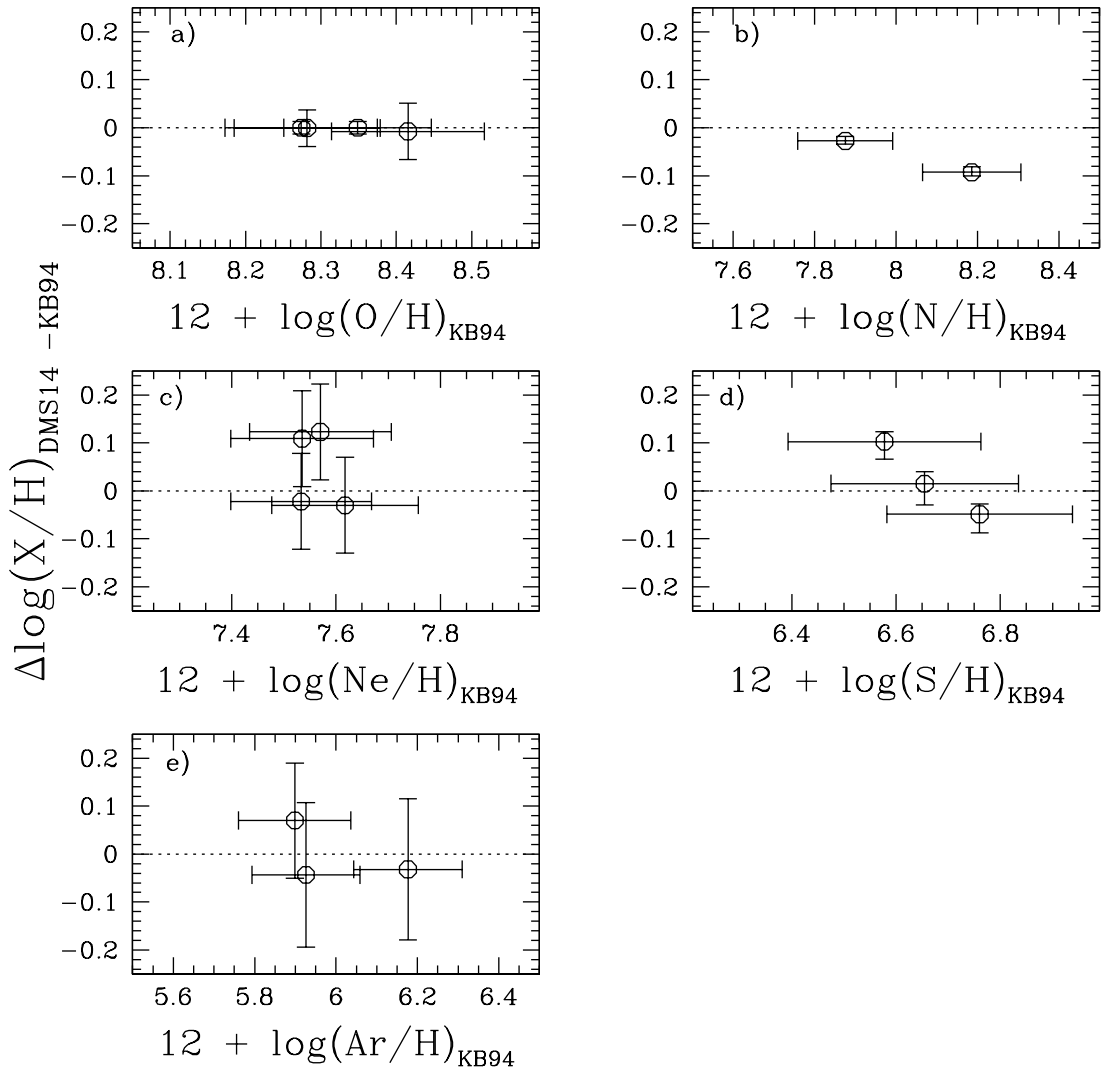}
\caption{Abundance difference due to the use of the new ICFs by DMS14 compared to the old ICFs by KB94, used  in Section~\ref{pn_section} of this paper 
(see  Appendix for details).
 \label{icf}}
\end{figure*}

Recently, \cite{delgado14}  (hereafter DMS14) presented new ionization correction factors (ICFs) for PNe using a large grid of photoionization models covering 
a wide range of physical parameters. Analytical expressions for the uncertainties associated with the new ICFs are also provided. 
We performed a comparison of the ICFs from \cite{kb94} (KB94), used in Section~\ref{pn_section} of this paper, with the new DMS14 recipes and evaluated the effect on the derived PN chemical abundances. Figure~\ref{icf} shows the difference in abundance due to the use of the new ICFs by DMS14 compared to the old ICFs by KB94.

For Helium, DMS14 suggest to calculate the He/H total abundance simply by adding He$^+$/H$^+$ and   He$^{++}$/H$^+$, i. e. neglecting any correction for neutral helium. 
This corresponds to the same approach that we adopted in Section~\ref{pn_section}, therefore no comparison needs to be made for He. 

For Oxygen, DMS14 propose: 

\begin{equation}
O = (O^+ + O^{++})  \times ICF(O^+ + O^{++}) 
\label{a0}
\end{equation}

where 

\begin{equation}
\log ICF(O^+ + O^{++}) = \frac{0.08 v + 0.006v^2}{0.34-0.27v},
\label{a1}
\end{equation}

and 

\begin{equation}
v = \frac{He^{++}}{(He^+ + He^{++})}; ~~~~~~~ w = \frac{O^{++}}{(O^+ + O^{++})}.
\label{a2}
\end{equation}

Equation~\ref{a1} is valid for $v\leq0.95$ and thus applies to all our 4 PNe. Panel a) of Figure~\ref{icf} shows that the difference in O abundance due to the use of the new ICFs is very small (a few percent in dex) and largely below the errors associated with the derived $12 + \log(O/H)$ values. 

For Nitrogen, DMS14 propose the following formulas valid only until $w=0.95$:

\begin{equation}
\frac{N}{O} = \frac{N^+}{O^+} \times ICF(N^+/O^+),
\label{a3}
\end{equation}

where 

\begin{equation}
\log ICF(N^+/O^+) = -0.16w(1+logv)
\label{a4}
\end{equation}

when He~II lines are detected, and 

\begin{equation}
\log ICF(N^+/O^+) = 0.64w
\label{a5}
\end{equation}

when He~II lines are not detected. However, Eq.~\ref{a5} may work fine for matter bounded nebulae but not for radiation bounded models and therefore the recommendation is to use the usual $N/O=N^+/O^+$ expression until the issue is further explored (Delgado-Inglada, private communication). 
In our sample, the condition $w\le0.95$ is only satisfied by  PN-2 and PN-3, with observed He~II lines, and therefore we used Eq.~\ref{a4} to compute the new abundances. 
Panel b) of Figure~\ref{icf} shows that the new N abundances tend to be lower than the previous ones;  however, the difference is within $\sim$0.1 dex, i. e. comparable to the uncertainty associated with the  $12 + \log(N/H)$ values. 

For Neon, the abundance is:

\begin{equation}
\frac{Ne}{O} = \frac{Ne^{++}}{O^{++}} \times ICF(Ne^{++}/O^{++}),
\label{a6}
\end{equation}

where 

\begin{equation}
ICF(Ne^{++}/O^{++}) = w + \left(\frac{0.014}{v'} + 2v'^{2.7}\right)^3 (0.7 + 0.2w -0.8w^2);
\label{a7}
\end{equation}

 $v'=0.01$ if no He~II lines are detected (PN-1 and PN-5) and $v'=v$ if $v\ge0.015$ (PN-2 and PN-3). Panel c) of Figure~\ref{icf} shows that the difference in Ne abundance is  
 within $\sim$0.1 dex, smaller than the errors for the $12 + \log(Ne/H)$ values; notice also the large uncertainties associated to the ICFs for Ne. 

For Sulfur, we measure both $S^+$  and $S^{++}$ in PN-1, PN-2 and PN-3; therefore DMS14 provides:

\begin{equation}
\frac{S}{O} = \frac{S^+ + S^{++}}{O^+} \times ICF((S^+ + S^{++})/O^+),
\label{a8}
\end{equation}

where 

\begin{equation}
\log ICF((S^+ + S^{++})/O^+) = \frac{-0.02 - 0.03w - 2.31w^2 + 2.19w^3}{0.69 + 2.09w-2.69w^2}
\label{a9}
\end{equation}

if $I(He~II)/I(H\beta)\ge0.02$ and 

\begin{equation}
ICF((S^+ + S^{++})/O^+) = 1
\label{a10}
\end{equation}

in all the other cases. We measure $I(He~II)/I(H\beta)\sim$0.2 and $\sim$0.4 in  PN-2 and PN-3, respectively, while for PN-1 we have only an upper limit of 
$I(He~II)/I(H\beta)<0.03$ from Table~\ref{pn_flux_raw}; therefore  
Eq.~\ref{a9} is applied to all three PNe.  
Panel d) of Figure~\ref{icf} shows no systematic trend of the new S abundances with respect to the old ones; the differences are within $\sim$0.1 dex, i.e. lower than the errors associated with the  $12 + \log(S/H)$ values. 

Finally, the  DMS14 recipe for Argon is:

\begin{equation}
Ar = Ar^{++} \times  \frac{O}{O^+ + O^{++}} \times ICF(Ar^{++}/(O^+ + O^{++})
\end{equation}

where, for $w>0.5$ (i.e. PN-1, PN-2, PN-3, PN-5):

\begin{equation}
\log ICF(Ar^{++}/(O^+ + O^{++}) = \frac{0.03w}{0.4-0.3w} - 0.05.
\end{equation}

Panel e) of Figure~\ref{icf} shows again differences in the Ar abundances within $\sim$0.1 dex, smaller than the uncertainty in the computed $12 + \log(Ar/H)$ values.

\end{document}